\documentclass[journal]{IEEEtran}

\usepackage{graphicx}
\usepackage{amssymb,amsmath,latexsym,amsfonts}
\usepackage{epsfig}
\usepackage{comment}
\usepackage{color}
\usepackage{algorithm}
\usepackage{algorithmic}

\newtheorem{definition}{Definition}%[section]
\newtheorem{theorem}{\noindent\textbf{Theorem}}%[section]
\newtheorem{lemma}{\noindent\textbf{Lemma}}

\newtheorem{assumption}{\noindent\textbf{Assumption}}

\def\done{\hspace*{\fill} \rule{1.8mm}{2.5mm}}

\begin{document}
%
% paper title5
%\title{Dynamic Adaptive Video Streaming over Multiple Interfaces Enabled Wireless Networks}
%\title{VARA: \underline{V}iew-time \underline{A}ware \underline{R}ate \underline{A}daptation 
%	for HTTP Adaptive Streaming}
%\title{Incentivizing Prompt and High-quality User Generated Content: A Game-Theoretic Approach}
\title{Can Early Joining Participants Contribute More?
\\ {\LARGE \it{-- Timeliness Sensitive Incentivization for Crowdsensing}} 
}
%\newline
%{\subtitle - Timeliness Sensitive Incentivization for Mobile Crowdsensing}
%\thanks{The second and third authors are ranked alphabetically and contribute equally.}}
%\subtitle{Timeliness Sensitive Incentivization for Mobile Crowdsensing}


\author{Yuedong~Xu,
        Yifan~Zhou,
        Yifan~Mao,
        Xu~Chen,
        and~Xiang~Li,
\IEEEcompsocitemizethanks{\IEEEcompsocthanksitem Yuedong Xu, Yifan Zhou, Yifan Mao and Xiang Li are 
with Research Center of Smart Networks and Systems, Fudan University, Shanghai, China. 
Xu Chen is with School of Data and Computer Science, Sun Yat-Sen University, Guangzhou, China. \protect\\
E-mail: \{ydxu, yfzhou13, 13307130211, lix\}@fudan.edu.cn; chenxu35@mail.sysu.edu.cn
\IEEEcompsocthanksitem Yifan Zhou and Yifan Mao contribute equally to this work.}% <-this % stops an unwanted space
}

%\footnote{The second and the third authors contributed equally.}.

%\begin{comment}
%\author{\authorblockN{Yuedong Xu$^*$, Yifan Zhou$^*$, Yifan Mao$^*$, Xu Chen$\dagger$, Xiang Li$^*$}
%\authorblockA{$^*$ Department of Electronic Engineering, Fudan University, Shanghai, China\\
%$\dagger$ School of Computer Science, Sun Yat-sen University, Guangzhou, China\\
%Email: \{ydxu,yfzhou13, 13307130211\}@fudan.edu.cn, chenxu35@mail.sysu.edu.cn, lix@fudan.edu.cn
%}
%}
%\end{comment}

%\author{\authorblockN{Yuedong Xu, Yifan Mao, Yifan Zhou, Xu Chen, Xiang Li}
%
%\thanks{ Yuedong Xu, Yifan Mao, Yifan Zhou and Xiang Li are with Department of Electronic Engineering, Fudan University, Shanghai, China. 
%Xu Chen is with Department of Computer Science, Gottingen University, Gottingen, Germany.
%
%Email: \{ydxu,13307130211,13307130190, lix\}@fudan.edu.cn, 
%
%xu.chen@cs.uni-goettingen.de}}

\maketitle

%\vspace{-0.5cm}
\begin{abstract} 

This paper investigates the incentive mechanism design from a novel and practically important 
perspective in which mobile users as contributors 
do not join simultaneously and a requester desires large efforts from early contributors. 
A two-stage Tullock contest framework is constructed: at the second stage the potential contributors compete 
for splittable reward by exerting efforts, %in crowdsensing, 
and at the first stage the requester 
can orchestrate the incentive mechanism to maximize his crowdsensing efficiency given the rewarding budget. 
A general reward discrimination mechanism is developed for timeliness sensitive crowdsensing
where an earlier contributor usually has a larger maximum achievable reward and thus allocates more efforts. 
Owning to the lack of joining time information, two practical implementations, namely  \emph{earliest-$n$} and 
\emph{termination time}, are announced to the contributors. 
For each of them, we formulate a Stackelberg Bayesian game in which the joining time of a contributor is his type and 
not available to his opponents. The uniqueness of Bayesian Nash equilibrium (BNE) is proved in each strategy. 
To maximize the requester's efficiency, we compute the optimal number of rewarded contributors in 
the earliest-$n$ scheme and the optimal deadline in the termination time 
scheme. Our contest framework is applicable not only to the closed crowdsensing with fixed number of contributors, 
but also to the open crowdsensing that the arrival of contributors is governed by a stochastic process. 
Extensive simulations manifest that with appropriate reward discriminations, the 
requester is able to achieve a much higher efficiency with the optimal selection of the number of 
rewarded contributiors and the termination time.
\end{abstract}

\IEEEpeerreviewmaketitle

\section{Introduction}
\label{sec:intro}

Recent years have witnessed the rapid proliferation of smartphones and wearable devices. 
They are equipped with a plethora of multi-modal sensors and they possess powerful computation as well as 
communication capabilities. With these technological features, ordinary mobile users can actively monitor their surrounding 
environments such as temperature, noise, vibration, network connectivity and geographic position without demanding sophisticated instruments. 
This leads to a new paradigm of problem-solving known as \emph{participatory crowdsensing}:  
a requester releases sensing tasks and collects contributed data from a number of mobile users 
that seek their individual benefits or the benefit of their community. Recent applications 
in  \cite{Mobisys10:Bao,Hondt,SenSys2014:Nawaz,noisetube} adopted crowdsensing 
to perform the indoor localization, collection of location information and noise monitoring, etc. 

The efficacy of a crowdsensing system heavily relies on the exerted efforts of mobile users. 
However, they are reluctant to share sensing capabilities due to the cost of energy, data traffic, time 
consumption, and risk of privacy leakage etc. 
A large body of studies have been devoted to developing efficient incentive mechanisms 
\cite{EC09:DiPalantino,Mobicom12:Yang,Infocom2013:IK,Mobihoc15:Jin,Infocom:Luo,Infocom:Luo2,Infocom2014:Feng} 
that can be roughly classified into two representative types: \emph{all-pay auction} \cite{EC09:DiPalantino,Infocom:Luo}  
and \emph{Tullock contest} \cite{Mobicom12:Yang,Infocom:Luo2}. 
%
%where (virtual) monetary reward is a commonly adopted incentive. 
%Mobile users compete for the reward by contributing high quality sensing results, and the requester designs  
%rules to assign the reward accordingly. 
%There exist two strings of representative incentive mechanism for 
%crowdsensing: %\emph{all-pay auction} \cite{EC09:DiPalantino,SODA12:Chawla,Infocom:Luo,TMC:Luo2} 
%\emph{all-pay auction} \cite{EC09:DiPalantino,Infocom:Luo}  
%and \emph{Tullock contest} \cite{Mobicom12:Yang,Infocom:Luo2}. 
In the former, the requester usually plays the role of 
auctioneer, and each contributor bids his efforts to the requester. 
The winning contributor acquires the entire reward, while 
the efforts of a losing contributor cannot be reimbursed. 
The latter approach, unlike all-pay auction, splits the total reward to all the contributors exerting positive efforts. 
Each contributor receives a fraction of the reward proportional to his efforts, and inversely proportional to the aggregate 
collected efforts. 

Incentive mechanisms based on auction theory are in general \emph{perfectly discriminatory}, i.e. 
the best bidder wins the competition while the others lose for sure \cite{Infocom:Luo2}. 
Owing to fear of sunk cost, all-pay auction is inclined to 
discouraging the participation of relatively weak contributors \cite{Franke}. 
%Franke et al. \cite{Franke} pointed out that under a complete information 
%setting, only two strongest contributors will engage in the auction at the equilibrium, and only one of them receive the positive 
%expected payoff. Obviously, this deviates from the goal of many crowdsensing applications: employing the wisdom of ordinary 
%(or even weak) contributors. 
In the light of its potential limitation, \cite{Mobicom12:Yang} 
proposed to utilize  Tullock contest to maximize the total sensing time of contributors. 
Tullock contest is \emph{partially discriminatory} so that each 
contributor gains a positive reward at the equilibrium if he exerts positive efforts. 
Though serving as a salient incentive mechanism 
for crowdsensing, the standard Tullock contest can be far away from optimality when the contributors are heterogeneous in their 
marginal costs of effort. Authors in \cite{Infocom:Luo2} presented an optimal discrimination strategy 
for Tullock contest when the marginal costs 
of the users are heterogeneous in crowdsensing. %, and each user is aware of his own marginal cost. 
The basic rationale is to enable a 
user of smaller marginal cost to exert more efforts under a Bayesian game framework. 

In this paper, we consider partially discriminatory incentive mechanisms at a new regime, 
namely \emph{timeliness sensitive crowdsensing contest (TSCC)}. This is motivated by versatile 
real-world crowdsensing applications where the timely sensing results are more valuable 
to the requester. For instance, CityExplorer \cite{cityexplorer} is a game-based crowdsensing system in which 
a winning player sets as many markers as possible in a city-wide game area within a finite 
time period by taking photos and providing concrete information. OpenSense \cite{opensense} 
enrols users to perform real-time air quality monitoring at different sites of a city. 
TruCentive \cite{trucentive}, CrowdPark \cite{crowdpark} and ParkNet \cite{parknet} collect   
the timely parking information from drivers and distributes to those in need of it. 
NoiseTube \cite{noisetube} is a participatory sensing framework for monitoring ambient noise in an area of 
several square kilometers. 

Though the above systems are designed for different purposes, they share 
a set of similar properties. Firstly, 
the contributors do not join \emph{instantly} and \emph{simultaneously}, 
while the requester desires large efforts from those who joins the sensing earlier.
The joining time can be the timestamp of a mobile user passing by a given sensing venue, 
assuming that the task is announced at time 0. 
Secondly, the contributors do \textbf{\emph{NOT}} exert efforts to change the joining times; 
in stead, they contribute efforts to execute the specific task. 
For instance, a driver will not drive his car out solely for the purpose of finding the availability of a parking lot. 
Executing a task such as taking and uploading photos consume a certain amount of resources. 
The cost of altering the joining time overwhelms that of task execution so that the joining time 
of the contributor is deemed as his property instead of his strategy to gain a larger share of the reward. 
Hence, the joining time and effort of a contributor are two 
orthogonal factors. 
%To clarify, the effort refers to energy and traffic expenditure or even monetary payment to execute a task which is 
%determined by the contributor endogenously, while his joining time is an exogenous variable. 
To incentivize more efforts from early contributors, a naive approach is to raise the reward, which is obviously 
not desired by the requester. From this angle, a fundamental question arises: \emph{can the requester 
incentivize more efforts from early joining contributors by designing appropriate 
discrimination mechanisms other than increasing his budget for crowdsensing?}

To answer this question, our first step is to examine in what form a feasible discrimination strategy should take 
so as to maximize the requester's efficiency under the Tullock contest model. 
Here, the efficiency is defined as the aggregate weighted efforts brought by per-unit of the requester's budget. This metric 
shares the similar principle as those in \cite{Mobicom12:Yang,Infocom:Luo2}, but is more general than a specific utility function.
We model the competition of contributors as a noncooperative game. Each contributor selfishly maximizes his 
payoff that is the difference between the received reward and cost of efforts.
Our analysis on the unique Nash equilibrium (NE) reveals  that the reward discrimination along with 
a ``virtual nature contributor'' can leverage the requester's high efficiency and the simplicity of mechanism design: 
the reward discrimination endows an early contributor the larger maximum achievable reward, and vice versa; 
the \emph{nature} player enables the requester to retain a fractional of reward when the number of contributors is small, 
thus improving his efficiency. 

Once Tullock contest structure has been determined for crowdsensing, our second step is to design practical 
incentives with timeliness sensitivity. 
Two practical approaches are proposed, the \emph{earliest-$n$ strategy} where 
only a subset of earliest contributors are rewarded, and the \emph{termination time strategy} 
that any contributor later than a ``deadline'' will be ruled out in this sensing task. 
In both strategies, they compete under an incomplete information scenario because each of them only knows his 
own joining time, while is unaware of those of the others, or is impossible to predict the joining times of future players. 
Hence, we formulate each competition as a two-stage Stackelberg Bayesian game, in which the 
requester announces the contest function to the contributors at Stage-I and they compete for the reward with the 
probability distribution of joining time at Stage-II. The existence and uniqueness of the 
Bayesian Nash equilibrium (BNE) are proved for each strategy. 
Based on these BNEs, the requester can compute the optimal number of rewarded contributors for the earliest-$n$ strategy 
and the optimal deadline for the termination time strategy.

Our major contributions are summarized below: 
\begin{itemize}

\item To the best of our knowledge, this is the first attempt to explore the design space for incentivizing more efforts 
from the early joining contributors.% with timeliness consideration. 

\item We show that the reward discrimination is suitable to elicit more efforts from early joining contributors. 
The NE of the Tullock contest is analyzed where a set of practical discrimination strategies are proposed 
to give preference to the early contributors. 

%analyze the NE of the general Tullock contest with aand find that  
%We propose the earliest-$n$ and termination time strategies to give preference to the early contributors. 
%This observation yields a simple Tullock contest framework with 
%insignificant compromise of the requester's efficiency.

\item We formulate Stackelberg Bayesian Nash games for the \emph{earliest-$n$} and the \emph{termination time} 
strategies. %In the closed crowdsensing system with fixed number of players, 
%The existence and uniqueness of the BNE for each strategy with incomplete 
%information on the joining times. 
The optimal number of rewarded contributors in the former and the 
optimal deadline in the latter are presented.

\item The Stackelberg Bayesian Nash game framework is generalized to the open crowdsensing system 
where the arrival of contributors is governed by a Poisson process.

\item Extensive simulations manifest that the proposed strategies can greatly improve the efficiency of the requester. 
In particular, the distribution of joining times for the closed system is derived from the WiFi access data of students in 
a campus building. 

\end{itemize}

The remainder of this paper is structured as follows. 
Second \ref{sec:model} presents the game model for participatory sensing 
with the timeliness consideration. A set of practical incentive strategies are proposed in section \ref{sec:multi_players}. 
Section \ref{sec:bayesian} analyzes the Bayesian Nash equilibria of incentive strategies and 
presents the optimal parameter configurations. 
Section \ref{sec:extension} extends the game framework to an open crowdsensing system with Poisson arrival of 
the contributors. The trace-driven experiments are performed in Section \ref{sec:perv}. 
Section \ref{sec:related} reviews the state-of-the-art work and Section \ref{sec:conclusion} concludes 
this paper.

\section{System Model}
\label{sec:model}

In this section, we present a suit of game models for participatory sensing that 
take into account the joining time of contributors. 

\subsection{Motivation and Basic Model}

In most of crowdsourcing and mobile sensing applications, the requester pursue not only high quality
but also timely efforts. However, recent research merely concentrates on the quality or effort while 
assuming that all the contributors participate in the crowdsensing simultaneously. 
In reality, the contributors more often join asynchronously. 
To harvest more efforts from the early joining contributors, 
the requester can increase the monetary reward, which is obviously undesirable.  
Our purpose here is to explore the design space for incentivizing timely and large efforts without 
increasing the requester's reward. 
The timeliness sensitive crowdsensing contest (TSCC) possesses two key factors, one is the \emph{joining 
time} and the other is the exerted \emph{effort}. 
The joining time is the duration between the instant of task distribution and that of task completion.  
We make an important assumption as the following. 

\begin{assumption}
{\it The joining time and effort of a contributor are perfectly complimentary factors in the crowdsensing contest.
The effort refers to various types of resources used to perform a sensing task, excluding those related to joining time. 
The joining time of a contributor is determined exogenously. }
\end{assumption}

This assumption manifests that the joining time and effort CANNOT be confused together. 
As an example, in mobile sensing applications, 
the joining time can be the timestamp of a mobile user passing by the given sensing site, and the effort 
refers to energy expenditure or even monetary cost to execute this task. 
A rational contributor will decide how many efforts he should spend on the sensing task, knowing 
his joining time. In what follows, we deliver the mathematical model for the incentive mechanism design. 

We consider a crowdsensing system with one requester and $N$ potential contributors. 
The requester releases a task at time 0 with a reward budget $B$, and the contributors compete for this reward. 
We denote by $e_i$ the effort made by contributor $i$, and by $t_i$ his joining time. Define two 
vectors $\mathbf{e}$ and $\mathbf{t}$ as $\mathbf{e} = \{e_i\}_{i=1}^N$ and $\mathbf{t} = \{t_i\}_{i=1}^N$.
The core of crowdsensing is the reward 
allocation mechanism that incentivizes the high effort from contributors, by taking the timeliness into consideration. 
We adopt a modified \emph{Tullock} contest success function (CSF) to characterize the competition among contributors. 
Define $\mathbf{e}_{-i} =  \{e_j\}_{j=1,\neq i}^N$ and 
$\mathbf{t}_{-i} =  \{t_j\}_{j=1,\neq i}^N$. 
Let $r_i([e_i, t_i], [\mathbf{e}_{-i}, \mathbf{t}_{-i}])$ be the reward obtained by the $i^{th}$ contributor, given the sets of 
joining time $\mathbf{t}$ and effort level $\mathbf{e}$. There has
\begin{eqnarray}
r_i([e_i, t_i], [\mathbf{e}_{-i}, \mathbf{t}_{-i}]) = \frac{e_i b(t_i)}{e_0 + e_i + \sum_{j{=}1,\neq i}^{N} e_j}, \quad \forall i,
\label{eq:basicincentive}
\end{eqnarray}
where $b(t_i)$, the maximum achievable reward of contributor $i$, is a function of his joining time $t_i$. 
In this Tullock crowdsensing contest, each contributor shares the total reward proportionally to his efforts, and 
inversely proportional to the aggregate efforts from all the contributors. 
The physical interpretation is that a contributor acquires a larger reward if he contributes more efforts, and 
a smaller reward if any of his opponent spends more. 

Two new features are introduced beyond the standard Tullock model. 
One is the ``reward discrimination'' on the joining times, that is, $b(t_i) \geq b(t_j)$ if $t_i < t_j$. An early joining contributor 
has a larger maximum achievable reward than a late one
To be noted, there are two other discrimination rules named ``weight discrimination'' and ``exponent discrimination''. 
However, only the reward discrimination is suitable for our problem where we leave the lengthy analysis in Appendix-II. 
The other feature is to introduce 
a constant $e_0$ that if it is positive, the reward will not be  completely assigned to the
contributors. If $e_0$ is 0, when there are only a couple of contributors, 
the requester has to assign the whole reward even though a very small amount of efforts are collected.  
Actually, a non-zero $e_0$ is equivalent to adding a \emph{NATURE} player, avoiding such adverse situations. 
A larger $e_0$ means that the requester retains a higher percentage of reward. 
Denote by $R$ the total reward paid to the contributors:
\begin{eqnarray}
R = \sum\nolimits_{i=1}^{N} r_i([\mathbf{e}, \mathbf{t}]).
\end{eqnarray}
In the incentive mechanism design, the requester needs to make the payment $R$ equal to the budget $B$ as the benchmark.
%To be noted, we also analyze two different discrimination rules named  
%in Appendix-I, and find that they fail to award more to the early joining contributors.   

%In comparison with standard Tullock model adopted in \cite{Mobicom12:Yang}, we introduce two modifications. 
%One is the discrimination on the joining times, that is, $b(t_i) \geq b(t_j)$ if $t_i < t_j$. An early joining contributor 
%has a larger maximum achievable reward than a late contributor. The other is to introduce 
%a constant $e_0$ that if it is positive, the reward will not be  completely assigned to the
%contributors. If $e_0$ is 0, when there are only a couple of contributors, 
%the requester has to assign the whole reward even though they spend negligible amount of efforts. 
%Actually, a non-zero $e_0$ is equivalent to adding a \emph{NATURE} player, avoiding such adverse situations. 
%A larger $e_0$ means that the requester retains a higher percentage of reward. 
%Denote by $R$ the total reward paid to the contributors that has $R = \sum\nolimits_{i=1}^{N} r_i([\mathbf{e}, \mathbf{t}])$. 
%In the incentive mechanism design, the requester needs to make the payment $R$ equal to the budget $B$ as the benchmark.

%\noindent\textbf{Remark 1:} Our modified Tullock contest model enables an early joining contributor to exert more 
%efforts and gain a higher payoff. Other discrimination strategies fail to reach both goals in which 
%detailed analyses can be found in the appendix-I. 

The payoff of contributor $i$, $\pi_i$, is denoted as the difference between his reward and the cost of efforts. Hence, there yields 
\begin{eqnarray}
\pi_i([e_i, t_i], [\mathbf{e}_{-i}, \mathbf{t}_{-i}]) = r_i([e_i, t_i], [\mathbf{e}_{-i}, \mathbf{t}_{-i}]) - e_i,
\end{eqnarray}
where the marginal cost of efforts is normalized as 1. When $t_i$ is not included in the CSF, 
%If $t_i$ is not inside the strategy profile, 
we can rewrite 
$r_i([e_i, t_i], [\mathbf{e}_{-i}, \mathbf{t}_{-i}])$ (resp.  $\pi_i([e_i, t_i], [\mathbf{e}_{-i}, \mathbf{t}_{-i}])$) as 
$r_i(e_i, \mathbf{e}_{-i})$ (resp. $\pi_i(e_i, \mathbf{e}_{-i})$) for simplicity.  
Concerning the requester's utility, an immediate contribution 
is no longer identically important to a late one, even if their efforts are the same. 
Inspired by this observation, we transform the importance of timeliness to the requester 
into a positive weight $w_i(t_i)$ (sometimes simplified as $w_i$) for contributor $i$. 
If $t_i < t_j$, there has $w_i \geq w_j$. Define $\mathcal{U}$ as the utility of the requester:
\begin{eqnarray}
\mathcal{U} = \sum\nolimits_{i=1}^{N}w_ie_i.
\end{eqnarray}

Define $\mathcal{E}$ as the requester's efficiency that is the utility brought by per-unit payment:
\begin{eqnarray}
\mathcal{E} = \mathcal{U}/R. %\frac{\mathcal{U}}{R}.
\end{eqnarray}
The efficiency $\mathcal{E}$ reflects the amount of utility brought by per-unit reward, and $\mathcal{E}$
serves as the metric of the requester to quantify the performance of incentive mechanisms. % in this work. 
Using $\mathcal{E}$ instead of $\mathcal{U}$ provides an intuitive understanding on how good an incentive mechanism 
can achieve. 
An auxiliary metric is named as ``discrimination gain'' denoted by 
$\mathcal{G}= \mathcal{E}_{d}/\mathcal{E}_{nd}$ where 
$\mathcal{E}_{nd}$ and  $\mathcal{E}_{d}$ indicate the requester's efficiencies without and with discrimination. 
In what follows, we formulate two different games to understand the competition of contributors, and explore 
the requester's utility maximization strategy. 

\noindent\textbf{Remark 1:} We consider a single requester because it is less likely that two requesters compete simultaneously 
over the same set of mobile users. 

\subsection{\textbf{G1}: Complete Information of Joining Times}

We formulate a noncooperative game to characterize the competition of contributors in which the joining time
is the common knowledge. In practice, when a contributor exerts certain efforts, he only knows his own joining time, while not 
those of his opponents. However, the complete information game allows us to gain important insights of 
the incentive mechanism design in a more tractable way. 
The game \textbf{G1} comprises three key elements:
\begin{itemize}

\item \emph{Players}: A set of $N$ potential contributors;

\item \emph{Strategies:} The action of player $i$ is $e_i$, $1\leq i\leq N$;

\item \emph{Payoffs:} The payoff of player $i$ is $\pi_i$, $1\leq i\leq N$.

\end{itemize}
Each player selfishly maximizes his individual payoff. The outcome of their competition is depicted by the famous \emph{Nash Equilibrium}
as the following. 
\begin{definition}
\emph{(\textbf{G1} Nash Equilibrium)}  The strategy profile $\mathbf{e}^*$ is a Nash equilibrium of \textbf{G1} if there exists 
\begin{eqnarray}
\pi_i(e_i^*, \mathbf{e}_{-i}^*)  \geq \pi_i(e_i, \mathbf{e}_{-i}^*), \quad \forall e_i \neq e_i^*, \; 1{\leq} i{\leq} N,
\end{eqnarray}
where $\mathbf{e}_{-i}^*$ is the set of efforts excluding $e_i^*$.
\end{definition}

\begin{definition}
\emph{(Individual Rationality)} Each player will receive a nonnegative payoff if he exerts a positive amount of efforts. 
\end{definition}

\subsection{\textbf{G2}: Incomplete Information of Joining Times}

The joining time is inherently a private information to the contributors. Especially, when the earliest contributor exerts his 
effort, he is by no means aware of the other's joining time that have not taken place. 
We hereby formulate a two-stage Stackelberg Bayesian game 
to characterize the crowdsensing contest where each player knows his exact joining time and the joining time distribution of 
the other contributors. In Stage-I, the requester announces the incentive mechanism 
and the joining time distribution so that his utility is optimized and the budget is balanced. 
In Stage-II, each contributor decides how many efforts to use to maximize his individual payoff. 

\noindent\textbf{Stage-II: Contributors' Bayesian Game.} Except that the players and the strategies are the same as those in \textbf{G1}, we 
append three different elements for this Bayesian game. % ($\forall 1\leq i\leq N$):
\begin{itemize}
\item \emph{Types:} The type of a contributor is the joining time $t_i$;

\item \emph{Probabilities:} The types of all the contributors are drawn from an i.i.d. priori distribution $F(t)$;

\item \emph{Payoffs:} The payoff of a contributor is the expectation $E[\pi_i([e_i, t_i], [\mathbf{e}_{-i}, \mathbf{t}_{-i}])]$ 
(occasionally written as $E[\pi_i]$). 
\end{itemize}
Each contributor chooses his effort $e_i$ to maximize his payoff with partial information. 
The equilibrium reached by the competing contributors is \emph{Bayesian Nash Equilibrium (BNE)} stated as the following. 
\begin{definition}
\emph{(Bayesian Nash Equilibrium)} The strategy profile $\mathbf{e}^*$ is a Bayesian Nash Equilibrium (BNE) of \textbf{G2} if 
for all $i\in\mathcal{N}$ and $e_i\neq e_i^*$, there has 
\begin{eqnarray}
E[\pi_i([e_i^*, t_i],[\mathbf{e}_{-i}^*, \mathbf{t}_{-i}])] \geq E[\pi_i([e_i, t_i],[\mathbf{e}_{-i}^*, \mathbf{t}_{-i}])]
\end{eqnarray}
where $\mathbf{e}_{-i}^*$ denotes the set of strategies excluding $e_i^*$, and $E(\cdot)$ denotes the expectation. 
\end{definition}

\noindent\textbf{Stage-I: Requester's CSF Choice.} 
At Stage-I, the requester can configure the parameter $e_0$ and the functions $b(t_i)$ for all $i$ 
so as to maximize his utility, given the budget $B$. Formally, the equilibrium $(e_0^*, b^*(\cdot))$ is 
the solution to the following optimization problem:
\begin{eqnarray}
 %&\max&\quad E[\mathcal{U}] \nonumber\\
 &\max&\quad E[\mathcal{E}] \nonumber\\
 &s.t.& \quad E[R] = B.
\end{eqnarray}
The requester's efficiency is optimized on the basis of the Bayesian Nash equilibrium at Stage-II. 
%The above optimization problem is also approximated by maximizing $E[\mathcal{E}]$ under 
%the same constraint so that both objectives are used interchangeably throughout the paper. 
%To iterate, our definition of requester's objective is more general than the difference between the 
%utility and budget that is usually adopted in the literature. 

\subsection{Comparison with Existing Models}

Our game framework is closely related to but different from the pioneering studies in \cite{Mobicom12:Yang} and \cite{Infocom:Luo2}. 
The major differences are summarized in three aspects:

\begin{itemize}

\item \textbf{Discrimination Rule.} There does not exist a discrimination rule in the crowdsensing contest of \cite{Mobicom12:Yang}. 
In \cite{Infocom:Luo2}, the marginal cost of a contributor is his private knowledge and the discrimination 
is based on the marginal cost. We investigate the discrimination on the joining time of the contributors where the 
origin is the different valuation of the requester on per-unit of effort at different joining times. 
The analytical framework in \cite{Infocom:Luo2} does not apply to our problem and hence a new framework is necessary. 

\item \textbf{Closed and Open Systems.} Only the closed system with a fixed number of contributors is studied in 
\cite{Mobicom12:Yang} and \cite{Infocom:Luo2}. Our game framework is applicable to not only this closed system but also the 
open system with the arrival of contributors governed by a stochastic process. The closed system usually exhibits the busty arrivals of 
the contributors while the open system has a more stable arrival rate. 
We are lucky to see that the incentive mechanisms can be orchestrated under the same framework in these two divergent systems. 

\item \textbf{Objective Function.} The objective of the requester in \cite{Mobicom12:Yang} and \cite{Infocom:Luo2} 
is to maximize his total utility on the efforts minus the payment to the contributors. We define a new metric for the requester, 
namely the crowdsensing efficiency. It is not constrained to the absolute net utility in a particular setting, but is to quantify 
the utility of the requester brought by per-unit of the budget. 

\end{itemize}

\section{Timeliness Sensitive Crowdsourcing Contest with Complete Information}
\label{sec:multi_players}

In this section, we analyze the competition of contributors with timeliness sensitivity, and present 
important insights in the design of practical incentive mechanisms. 

\subsection{Nash Equilibrium}

According to the individual rationality property, a contributor will not exert efforts if his participation cannot bring positive payoff. 
Hence, we need to scrutinize the participation of players on top of a NE. 
For simplicity of notations, we perform a change of variables by setting $E_{-i} = \sum\nolimits_{j=1,\neq i}^{N}e_j$ (esp. excluding $e_0$), $E = e_i + E_{-i}$ 
and $b(t_i) = b_i$. 
We observe that the utility function of each contributor $i$ is strictly concave in $e_i$. Then, the   
first order condition describes the global maximum of $\pi_i(e_i,\mathbf{e}_{-i})$ with respect to $e_i$,
\begin{eqnarray}
\frac{b_i(e_0+E_{-i})}{(e_0 + e_i + E_{-i})^2} - 1 \leq 0, \quad \forall i
\label{eq:n_ne_conditions}
\end{eqnarray}
where the equality holds upon $e_i {>} 0$. If this equality does not hold, the maximum is 
obtained at $e_i =0$, in which contributor $i$ does not ``participate'' in the crowdsensing. 
Therefore, the best response of the $i^{th}$ contributor to $E_{-i}$ is 
\begin{eqnarray}
e_i(\mathbf{e}_{-i}) := \max\Big\{  \sqrt{b_i (e_0+ E_{-i})} - (e_0+E_{-i}), \;0  \Big\}.
\label{eq:bestresponse}
\end{eqnarray}
The pure strategy Nash equilibrium must have $e_i^* = e_i(\mathbf{e}_{-i}^*)$ for each contributor. 
We have the following lemma with regard to the participation of contributors. 

\begin{lemma} (Principles of Participation)
In the standard contest function, the participation of a contributor satisfies:
\begin{itemize}
\item if $e_i^* > e_j^* >0$, there must have $b_i > b_j$; 

\item if $e_0\geq b_i$, the $i^{th}$ contributor does not participate in the contest;
when $e_0$ increases, the number of participating contributors decreases. 
\end{itemize}
\label{lemma:n_contributors}
\end{lemma}
\noindent\textbf{Proof:} Please refer to Appendix-I. \done % \ref{sec:appendix1}.

Based on the best response function in Eq.\eqref{eq:bestresponse}, 
there has 
\begin{eqnarray}
e_0+ E = \frac{(n-1)+\sqrt{(n-1)^2+4(\sum\nolimits_{j=1}^{n}\frac{1}{b_j})e_0}}{2(\sum\nolimits_{j=1}^{n}\frac{1}{b_j})},
\label{eq:sum_term}
\end{eqnarray}
when only the contributors from $1$ to $n$ exert positive effort at the NE. 
We hereby present a method to search the explicit NE within $N$ steps.
The first step is to compute $\mathbf{e}^*$ by assuming the participation 
of $n=N$ contributors at the NE.  
If $e_i^*$ is positive for all $i$, the NE is obtained. Otherwise, if any $e_i^*$ is negative, this means that some of the 
contributors do not participate at this NE. By removing the concurrent contributors with the smallest $b_i$, 
we proceed to search until all $e^*_i$ are positive for $1\leq i \leq n$. 
The NE strategy $e_i^*$ is subsequently computed by
\begin{eqnarray}
\left\{\begin{matrix}
e_i^* = (e_0+ E^*) {-} \frac{1}{b_i}(e_0+ E^*)^2,  &&\!\!\!\!\!\!\!\! \forall 1{\leq}i{\leq}n \\
e_i^*=0  && \!\!\!\!\!\!\!\! n{<}i {\leq} N 
\end{matrix}\right..
\label{eq:linearutility_ne}
\end{eqnarray}
The actual payment to the $i^{th}$ player is
 \begin{eqnarray}
r_i([\mathbf{e}^*_{i},\mathbf{t}^*_{i}]) = b_i - (e_0+E^*) \quad \textrm{if} \quad e_i^* \geq 0.
\label{eq:player_i_payment}
\end{eqnarray}
The detailed algorithm to find the unique NE is shown in Algorithm \ref{algo1}.

\begin{algorithm}
%\small
\caption{: Searching the NE}
\label{power_alc}
\begin{algorithmic}[1]
\renewcommand{\algorithmicrequire}{\textbf{Input: $\mathbf{b}$, $N$}; \quad \textbf{Output:} $\mathbf{e^{\star}}$}
\REQUIRE ~~ \\
%\ENSURE ~~\\
%\STATE //initialize the power of each antenna
\STATE \textbf{sort} $b_i\in\mathbf{b}$ in the descending order
\STATE  \textbf{for} $n = N$ to $1$ 
\STATE \ \ Compute $x_i^*$ using Eqs. \eqref{eq:sum_term} and \eqref{eq:player_i_payment}
\STATE \ \ \textbf{if} $x_i^{*} \geq 0$ for all $i=0,\cdots, n$
\STATE \ \ \ \ \textbf{exit}
\STATE \ \ \textbf{end}
\STATE \textbf{end}
\end{algorithmic}
\label{algo1}
\end{algorithm}

We next consider a special discrimination rule that the basic idea is to set the larger maximum rewards to 
the early contributors and 0 to late ones. Our purpose is to gain the important insights on the design of 
discrimination rules and on the parameter configuration. 

\medskip
\boxed{
	\begin{minipage}{3.1in}
		{\ensuremath{\mbox{\sc  Multi-Contributors:}}}
		$e_0 \geq 0$, $b_i =b \; \forall 1 \leq i \leq n$ and $b_{i} = 0 \; \forall n< i\leq N$; $w_i \geq w_j$ if $i< j$. 
	\end{minipage}
}
\medskip

When the maximum rewards of the early $n$ contributors are identical, and 
the remaining late ones are not rewarded, only the former ones participate at the NE. 
Their efforts at the NE are given by ($\forall i{\leq} n$)
\begin{eqnarray}
e_i^* = \big((n{-}1)b-2e_0n + \sqrt{(n{-}1)^2b^2 + 4e_0bn}\big)\big/(2n^2).
\label{eq:ne_nidentical_contributors}
\end{eqnarray} 
The total reward paid to the contributors is $R= \frac{bE^*}{e_0+E^*}$. 
Then, the NE efficiency of the requester is 
\begin{eqnarray}
\mathcal{E}(n) = \frac{\sum\nolimits_{i=1}^nw_i}{2n^2} \cdot \Big(  n-1 + \sqrt{(n-1)^2 + \frac{4}{b}e_0n} \Big) \label{eq:ncontributors_efficiency}
\end{eqnarray}
that yields $\frac{n-1}{n^2}\sum\nolimits_{i=1}^nw_i \leq \mathcal{E} < \frac{1}{n}\sum\nolimits_{i=1}^nw_i, \;\; \forall e_0\in[0, b).$
The upper and lower bounds are obtained or approximated when $e_0$ is 0 or is approaching $b$. 

\noindent\emph{Special case 1): the requester is not sensitive to the joining time (i.e. $w_i=w_j=w$ for all $1\leq i,j \leq N$).} 
The efficiency $\mathcal{E}$ is greater than $\frac{(n-1)w}{n}$ and any discrimination (i.e. $n<N$) yields a smaller efficiency. 

\noindent\emph{Special case 2): the requester is interested in only the earliest two contributions (i.e. $w_1=w_2=w >0$ and $w_i=0$ for all $i>2$).} The discrimination gain can be as high as $\frac{N^2}{4(N-1)}$ with the discrimination rule $n=2$. 

%The above example validates our intuition that the design of a discrimination rule is to find the appropriate $b_i$ for all $i$. 

\subsection{Understanding the function of $e_0$}

Introducing $e_0$ adds the complexity to the analysis and design of the incentive mechanism. 
We hereby investigate the role of $e_0$ and provide a simple guideline of configuring $e_0$ through the above example.

(1) A positive $e_0$ may prevent the participation of contributors with low maximum rewards. 
%If contributor $i$ arrives late, $b_i$ is then small at time $t_i$. 
When $b_i$ is below $e_0$ for a late contributor, he will 
not participate in the sensing at the NE. Hence, $e_0$ precludes the participation of late contributors 
so that the early contributors may exert more efforts. 

(2) A positive $e_0$ has a potential to improve the efficiency of the given discrimination rule.
According to Eq.\eqref{eq:ncontributors_efficiency}, the efficiency is an increasing function of $e_0$. 

(3) When the number of contributors, $n$, is large, the efficiency gain brought by $e_0$ becomes less 
remarkable. 

Given $n$ equal to 2, the efficiency is obtained by $\mathcal{E}(2) = \frac{w}{4}(1+\sqrt{1+\frac{8e_0}{b}})$. 
Then, $\mathcal{E}(2)$ is $\frac{w}{2}$ with $e_0 = 0$, and approaches $w$ as $e_0$ is sufficiently close to $b$. 
For $n$ equal to $100$, $\mathcal{E}(100)$ is $\frac{99w}{100}$ with $e_0 = 0$, and approaches $w$ as 
$e_0$ is sufficiently close to $b$.

(4) Choosing a large $e_0$ is beneficial to the requester, especially when only a few contributors participate in the 
competition. However, a large $e_0$ will reduce the number of participants at the NE. 
In realistic scenarios, the mobile users join the crowdsensing randomly so that 
the participation of contributors is very sensitive to the joining times in the presence of a large $e_0$. 
Hence, $e_0$ is suggested to be a fixed fraction of $b$ simply for the purpose of avoiding the scenarios with only 
a couple of participants.  

The KKT optimization is adopted for the optimal incentive mechanism design. Due to the page limit, we leave the detailed 
analysis in Appendix-I. 

\noindent\textbf{Remark 2:} The requester can orchestrate the reward discrimination scheme to maximize his crowdsensing 
efficiency, given the vector of the joining times. However, this is infeasible in reality because the 
joining time of a player is not known until he undertakes the 
sensing task. A practical incentive mechanism needs to be announced before the crowdsensing, and needs 
to be easily implemented. 
\vspace{-0.3cm}

\subsection{Practical Reward Discrimination Strategies}

We hereby turn the reward discrimination into reality, considering that 
each contributor is unaware of the joining time of other contributors. Three practical strategies are proposed. 

\emph{Earliest-$n$ Strategy:} Let $t_1$ until $t_N$ be the joining times of all the contributors. 
Let $T_1\leq T_2 \leq \cdots \leq T_N$ denote the ordered values of $t_1, t_2, \cdots, t_N$. Then, $\{T_i\}_{i=1}^N$ are 
called the order statistics of $\{t_i\}_{i=1}^{n}$. The earliest $n$ contributors will have a chance of being rewarded
(i.e. $b_i=b \; \forall i\leq n$), and all other late contributors will not be rewarded (i.e. $b_i = 0 \; \forall i> n$).

 \emph{Termination Time Strategy:} The requester sets up a termination time $T$ so that only contributions before $T$ 
are rewarded. Similarly, we denote by $t_1, \cdots, t_N$ the random joining time of all the contributors. 
If $t_i \leq T$, there has $b(t_i) = b$, and $b(t_i) = 0$ otherwise. 

\emph{Linearly Decreasing Strategy:} The maximum reward function is set to $b(t_i) = \max\{0, b - ht_i\}$ for every player $i$. 
The velocity $h$ can be tuned to further optimize the efficiency of the requester. 

In the earliest-$n$ strategy, a contributor is unaware of his ranking in terms of the joining time; in the termination time 
strategy, he is unaware of the number of opponents joining before the deadline; in the linearly decreasing strategy, 
he is unaware of the maximum achievable rewards of his opponents. 
These uncertainties lead to competitions with incomplete information. 

\noindent\textbf{Remark 3:} The linearly decreasing strategy is yet a different implementation of the earliest-$n$ 
strategy, which will be shown later on. 

%the linearly decreasing strategy is a 
%generalized version of the termination time strategy. 

\section{Stackelberg Bayesian Game with Timeliness Sensitivity}
\label{sec:bayesian}

In this section, we analyze the Stackelberg Bayesian Nash equilibrium (SBNE) of crowdsensing contests. 
Optimal strategies are proposed for the requester to maximize his efficiency.

%We provide 
%deep understanding on the interplay between timeliness and effort of the contributors, and propose optimal strategies 
%for the requester to improve his efficiency. 

\subsection{Earliest-$n$ Strategy}

In our context, all the contributors are not notified whether they are ranked as the earliest $n$ contributors. 
Hence, as the first step, we must compute the probability of being one of the earliest-$n$ contributors if 
the $i^{th}$ contributor joins at time $t_i$. Suppose that $t_i$ is the $j^{th}$ smallest joining time. Among all the contributors, 
there are $j-1$ contributors earlier than him and $N-j$ contributors later than him. Considering all the possible rankings of 
$t_i$ no larger than the $n^{th}$ place, the probability of being the earliest-$n$ contributor is given by
\begin{eqnarray}
P_{i\in\{n\}} = \sum_{j=1}^n {N-1\choose N-j} \cdot p^{j-1}(1-p)^{N-j},
\label{eq:earliest-n-prob}
\end{eqnarray}
where $p$ is the probability of a contributor joining before $t_i$. 
%The complementary probability is denoted by $P_{i\bar{\in}\{n\}} = 1- P_{i\in\{n\}}$. 
Note that in the earliest-$n$ strategy, the number of participated contributors can be at most $N$, though 
only the first $n$ of them are rewarded. 
Here, with certain abuse of notation, we can use a function $b(t_i)$ to denote $b P_{i\in\{n\}}$, 
and $b(t_i)$ is continuous, differentiable and strictly decreasing w.r.t. $t_i$. 
Then, the utility of contributor $i$ with joining time $t_i$ is simplified as
\begin{eqnarray}
%\mathbb{E}[\pi_i] = \frac{e_i b P_{i\in\{n\}} }{e_0+e_i+\sum\nolimits_{j\neq i}e_j} -  e_i. 
\pi_i([e_i, t_i], [\mathbf{e}_{-i}, \mathbf{t}_{-i}])= \frac{e_i b(t_i) }{e_0+e_i+E_{-i}} -  e_i. 
\label{eq:pi_earliest_n}
\end{eqnarray}
For contribution $i$, the joining time $t_i$ is his private information, while only the 
statistical joining time distribution of other contributions is known as a priori. 
In this situation, we model the crowdsourcing competition as a Bayesian game where 
the join time of a contributor is characterised as his \emph{type}. 
We will apply the backward induction principle to solve the proposed Bayesian game,
 i.e., first analyzing Stage-II and then determining the optimal policy at Stage-I accordingly.

\textbf{Stage-II: Finding Bayesian Nash Equilibrium.} 
For any contributor other than $i$, i.e. $j\in\mathcal{N}$ and $j\neq i$, since each 
$t_j$ is drawn from the common distribution $F(t)$, the expected utility of 
contributor $i$ on all the combinations is given by 
\begin{eqnarray}
\mathbb{E}[\pi_i([e_i, t_i], [\mathbf{e}_{-i}, \mathbf{t}_{-i}])] \!\!\!&=&\!\!\! {\int_{0}^{\infty}\cdots\int_{0}^{\infty}} \frac{e_ib(t_i)}{e_0+e_i + E_{-i}} \nonumber\\
&& \!\!\!\!\!\!\!\!\!\ \!\!\!\!\!\!\!\!\!\ \!\!\!\!\!\!\!\!\!\!\!\!\! \!\!\!\!\!\!\!\!\!\!\!\!\!\!\! \!\!\!\!\!\!\!\!\!\!\!\!\!\!\!
\cdot\prod\nolimits_{j=1,\neq i}^Nf(t_j)  dt_1 \cdots dt_{i{-}1}dt_{i{+}1}\cdots dt_N - e_i.
\label{eq:complicated_integral}
\end{eqnarray}
As the first step, we analyze the existence and uniqueness of this Bayesian Nash equilibrium. 
\begin{theorem} \emph{(Existence and Uniqueness of Earliest-$n$ BNE)}
The Earliest-$n$ incentive strategy has a unique Bayesian Nash equilibrium (BNE) in the contest. 
\label{lemma:E_U_earliest_n}
\end{theorem}
\noindent\textbf{Proof:} Please refer to Appendix-I. \done % \ref{sec:appendix1}.

We now turn to establish basic properties of pure-strategy Bayesian Nash equilibrium. 
As stated before, the Bayesian Nash equilibrium strategy $\mathbf{e}^*$ satisfies
\begin{eqnarray}
e_i^*(t_i) \in \arg\max_{e_i\geq 0} \mathbb{E}[\pi_i([e_i, t_i], [\mathbf{e}_{-i}, \mathbf{t}_{-i}])] - e_i, \;\;\; \forall i.
\end{eqnarray} 
Similar to the NE condition of the complete information game, for each contributor $i$ of the earliest-$n$ scheme, 
the equilibrium condition is given by 
\begin{eqnarray}
\mathbb{E}\Big[  \frac{e_0 + E_{-i}^*}{(e_0 + e_i^* + E_{-i}^*)^2} \Big] b(t_i) \leq 1, \quad \forall 1{\leq} i{\leq} N,
\label{eq:bayesian_NE_condition}
\end{eqnarray} 
with equality in the situation $e_i^* \geq 0$. Otherwise, the $i^{th}$ contributor does not participate.  
In general, there does not admit a close-form solution to this system of (in)equations. However, we can still 
infer interesting properties of the Bayesian Nash equilibrium. 

\begin{theorem} \emph{(Participation and Bound under Earliest-$n$ Strategy)} 
In the Bayesian Nash equilibrium of Earliest-$n$ scheme, contributor $i$'s strategy has the following property 
(with a meaningful condition $b>e_0$).
\begin{itemize}

\item If $n = N$, $e_i^*$ is positive with $b> e_0$ for all $i$.

\item If $1\leq n <N$, there exists a $\bar{t}$ such that $e_i^*$ is 0 for $t>\bar{t}$ while $e_i^*$ is positive and strictly decreasing for $t<\bar{t}$. Especially, $\bar{t}$ is no larger than $b^{-1}(e_0)$. 

\item The upper bound of $e_i^*(t_i)$ is $\frac{1}{4}b(t_i)$ for $e_0 \leq \frac{1}{4}b(t_i)$ and is 
$\frac{\frac{1}{4}b(t_i)^2 e_0}{(e_0+\frac{1}{4}b(t_i))^2}$ otherwise.  
\end{itemize}
\label{theorem:bounds}
\end{theorem}
\noindent\textbf{Proof:} Please refer to Appendix-I. \done % \ref{sec:appendix1}.

\noindent\textbf{Remark 4:} A contributor may not participate in the earliest-$n$ contest if his joining time is later than 
a certain threshold.

\textbf{Stage-I: Optimizing Requester's Efficiency.} 
In the earliest-$n$ scheme, the requester announces his reward allocation function before the contest takes place. 
As mentioned before, once $e_0$ is chosen to be a fraction of the maximum reward $b$, the efficiency is independent of 
$b$. The requester only needs to configure an optimal $n$ to optimize his efficiency. 

At the Bayesian Nash equilibrium, the amount of efforts exerted by contributor $i$ is $e_i^{*}(t_i)$. This contributor generates 
a utility of $w(t_i)e_i^*(t_i)$ to the requester. Since $t_i$ is unknown to 
the requester, his expected utility on contributor $i$'s effort is obtained by
\begin{eqnarray}
E[w(t_i)e_i^*(t_i)] = \int_{0}^{\infty} w(t_i)e_i^*(t_i) f(t_i) dt_i.
\end{eqnarray}
Because the joining time is i.i.d. at all the contributors (the assumption of identical distribution is not obligatory), 
the expected total utility of the requester is  
\begin{eqnarray}
\mathbb{E}[\mathcal{U}] = N\int_{0}^{\infty} w(t_i)e_i^*(t_i) f(t_i) dt_i. 
\end{eqnarray} 
For each set of joining time $\mathbf{t}$, the reward paid to all the contributors is
\begin{eqnarray}
R = \Big(b\sum\nolimits_{i=1}^{n}e_i^*(t_i)\Big)\Big/\Big(e_0 + \sum\nolimits_{i=1}^{N}e_i^*(t_i)\Big).
\end{eqnarray} 
at the BNE. Then, the expected reward paid by the requester takes the following integral form 
\begin{eqnarray}
\mathbb{E}[R] =\!\!\! \int_{0}^{\infty}\cdots\int_{0}^{\infty} \frac{b\sum\nolimits_{i=1}^{n}e_i^*(t_i)}{e_0 + \sum\nolimits_{i=1}^{N}e_i^*(t_i)}\prod_{i=1}^{N}f(t_i)dt_1\cdots dt_N. 
\end{eqnarray} 
The requester needs to configure an appropriate $b$ such that the expected reward apportioned to the contributors is 
equal to the budget $B$. 

For each vector of joining time $\mathbf{t}$, the efficiency of the requester is given by 
\begin{eqnarray}
\mathcal{E} = \frac{\big(\sum\nolimits_{i=1}^{N}w(t_i) e_i^*(t_i)\big)\big(e_0 + \sum\nolimits_{i=1}^{N}e_i^*(t_i)\big)}{b\sum\nolimits_{i=1}^{n}e_i^*(t_i)}. 
\end{eqnarray}
Then, the expected efficiency is 
\begin{eqnarray}
\mathbb{E}[\mathcal{E}] \!\!\!&=&\!\!\! \int_{0}^{\infty}\cdots\int_{0}^{\infty}\frac{\big(\sum\nolimits_{i=1}^{N}w(t_i) e_i^*(t_i)\big)\big(e_0 {+} \sum\nolimits_{i=1}^{N}e_i^*(t_i)\big)}{b\sum\nolimits_{i=1}^{n}e_i^*(t_i)} \nonumber\\
\!\!\!&&\!\!\!\prod\nolimits_{i=1}^{N}f(t_i)dt_1\cdots dt_N. 
\end{eqnarray}
The optimal $n^*$ can be searched for no more than $N$ times in the above formula. The main complexity stems from the 
multi-dimensional integral in computing the BNE at stage-II. This multi-dimensional integral can be substituted by 
the Monte Carlo sum via discretizing the integral zone into a large number of multi-dimensional cubes.

\subsection{Termination Time Strategy}

In the termination time strategy, the requester only allocates rewards to the contributors who join 
the contest before time $T$. Obviously, a contributor later than time $T$ will not participate in the contest. 
Given the distribution of joining times, a contributor arrives before time $T$ is computed as $p = F(T)$. 
In this game, the number of participating contributors is uncertain. 
Denote by $n_1$ the number of participating contributors in a pool of $n_2$ contributors in total. 
The probability that $n_1$ out of $n_2$ contributors join before the expiration time $T$ is given by
\begin{eqnarray}
\mathbb{P}(n_1, n_2) =  {n_2\choose n_1} p^{n_1}(1-p)^{n_2-n_1}
\label{eq:prob_termination_time}
\end{eqnarray}
due to their i.i.d. joining times. We analyze this Bayesian game using the backward induction as follows. 

\textbf{Stage-II: Finding Bayesian Nash Equilibrium.} 
We analyze the expected utility of the $i^{th}$ contributor here. If he participates, there will be at most $N-1$ other 
contributors.  To compute the expected utility, we exhaust all the possibilities in terms of the number of contributors in the contest. 
For a fixed maximum reward $b$, his expected utility is obtained by 
\begin{eqnarray}
\!\!\mathbb{E}[\pi_i] \!\!\!\!\!&=&\!\!\!\!\! \sum_{k=0}^{N-1}{N{-}1\choose k} p^k(1-p)^{N{-}1{-}k} \cdot \frac{e_ib}{\sum\nolimits_{j=0}^{k+1}e_j} - e_i \nonumber\\
\!\!\!\!\!\!\!&=&\!\!\!\!\! \sum_{k=0}^{N{-}1}\nolimits\mathbb{P}(k,N{-}1) \frac{e_ib}{\sum\nolimits_{j=0}^{k+1}e_j}  {-} e_i, \; \forall 1{\leq} i{\leq}k{+}1.
\label{eq:expected_utility_T}
\end{eqnarray} 
When $k$ is zero, the $i^{th}$ contributor only competes with the \emph{Nature} player. 
In this Bayesian game, each contributor aims to maximize his expected utility. To begin with, we show the 
existence of a unique Bayesian Nash equilibrium.
\begin{theorem} \emph{(Existence and Uniqueness of BNE for Termination Time Strategy)}
The crowdsensing contest using termination time strategy admits a unique symmetric Bayesian Nash equilibrium. 
\label{theorem:symmetric_BNE}
\end{theorem}
\noindent\textbf{Proof:} Please refer to Appendix-I. \done % \ref{sec:appendix1}.

The BNE is symmetric in which all the contributors joining before time $T$ exert the same amount of effort. 
According to the optimality conditions at the BNE, the following equations hold 
\begin{eqnarray}
\sum_{k=0}^{N-1}\nolimits\mathbb{P}(k,N{-}1) \frac{b(e_0 + ke_i^*)}{(e_0 + (k+1)e_i^*)^2} = 1, \;\;\; \forall i. 
\end{eqnarray}
The above equation intuitively shows that an increase of $b$ induces more effort from the contributors. 
For the special case $e_0 = 0$, $e_i^*$ admits a close form solution, that is, $e_i^* = \sum_{k=0}^{N-1}\nolimits\mathbb{P}(k,N{-}1)\frac{kb}{(k+1)^2}$. 

%\begin{lemma}
%There exists a unique symmetric  termination time strategy 
%\end{lemma}

\textbf{Stage-I: Optimizing Requester's Efficiency.} From the requester's perspective, contributor $i$ with the joining time 
$t_i$ ($t_i \leq T$) generates a utility $w(t_i) e_i^*$ to him. 
For the special case that no contributor arrives before $T$, the utility of the requester is 0 for sure, and his efficiency is 
set to 0 as a penalty of unsuccessful crowdsensing. 
Therefore, the expected total utility of the requester is obtained by
\begin{eqnarray}
\mathbb{E}[\mathcal{U}] = N \int_0^T w(t_i) e_i^*  f(t_i) dt_i = e_i^*N\int_0^T w(t_i) f(t_i) dt_i.
\end{eqnarray}
The number of contributors whose joining times before time $T$ ranges from 0 to $N$. The corresponding probability of 
having $k$ participants before $T$ is $\mathbb{P}(k, N)$. A fraction of the total reward allocated by the requester can be 
solved as
\begin{eqnarray}
\mathbb{E}[R] = \sum\nolimits_{k=1}^N \mathbb{P}(k, N) \frac{bke_i^*}{e_0 + ke_i^*}. 
\label{eq:expected_reward_T}
\end{eqnarray}
Here, $b$ is chosen to let $\mathbb{E}[R]$ be equal to the budget $B$. 
One should be informed that the summation begins from the subscript $k=1$, which is unlike \eqref{eq:expected_utility_T}. 
This is because Eq.\eqref{eq:expected_reward_T} counts the allocated reward when there is at least one contributors.

We next compute the efficiency of the requester. Denote by $\tilde{t}_i$ the joining time of contributor $i$ before termination 
time $T$. For each valid contributor, his joining time follows a conditional distribution of that of the original joining time, 
\begin{eqnarray}
\mathbb{P}(\tilde{t}_i \leq t| \tilde{t}_i \leq T) = \frac{F(t)}{F(T)} = \frac{F(t)}{p}.
\end{eqnarray}
The pdf of $\tilde{t}_i$ is subsequently obtained by
\begin{eqnarray}
f_{\tilde{t}_i} (t) = \frac{f(t)}{p}.
\end{eqnarray}
We then compute the efficiency of the requester for each vector of joining times 
\begin{eqnarray}
\mathcal{E} = \Big(\sum\nolimits_{l=1}^{k}w(\tilde{t}_l)(e_0+ke_i^*)\Big)\Big/(bk),\quad k> 0
\end{eqnarray}
when $k$ out of $N$ contributors are before $T$ and $\mathcal{E}=0$ for $k = 0$. 
The expectation of $\mathcal{E}$ incorporates two kinds of 
uncertainties; one is the number of valid contributors, and the other is the exact joining times of the valid contributors. 
With the detailed derivation in the Appendix-I, we have 
\begin{eqnarray}
\mathbb{E}[\mathcal{E}] %\!\!\!&=&\!\!\! \sum_{k=1}^{N}\mathbb{P}(k, N)  \frac{\mathbb{E}[\sum_{l=1}^{k}w(\tilde{t}_l)]e_i^*(e_0+ke_i^*)}{bke_i^*} \nonumber\\
%\!\!\!&=&\!\!\! 
= \big(\frac{e_0}{bp}(1-(1-p)^N) +  \frac{Ne_i^*}{b}\big)\int_0^T \!\!\!w(t) f(t)dt .
\label{eq:expected_eff_termination}
\end{eqnarray}

The optimal termination time $T$ is chosen to maximize the efficiency of the requester, that is, 

\medskip
\boxed{
	\begin{minipage}{3.3in}
		$T^*= \arg\max_{T} \big(\frac{e_0}{bp}(1{-}(1{-}p)^N) +  \frac{Ne_i^*}{b}\big)\int_0^T \!\!\!w(t) f(t)dt $
	\end{minipage}
}
\medskip
\vspace{-0.3cm}

\noindent where $p$ and $e_i^*$ are functions of $T$. The optimal $T^*$ can be approximated by enumerating a finite number of 
candidate termination times or by a bisection search. 

\subsection{Linearly Decreasing Strategy}

The requester sets $b(t)$ as a linearly decreasing function $b(t_i) = \max\{0, b - ht_i\}$
where the velocity $h$ is tunable to maximize his efficiency. 
For the fixed $h$, the Stackelberg Bayesian Nash equilibrium can be solved using the same approach as that of 
the earliest-$n$ strategy. In other words, $b(t_i)$ is $bP_{i\in\{n\}}$ in the earliest-$n$ strategy  
and is $\max\{0, b - ht_i\}$ in the linearly decreasing strategy. 
Another exception is that the earliest-$n$ strategy needs to try different $n$, while 
the linearly decreasing strategy needs to search the different velocity $h$. 
One can choose a finite set of candidate $h$ beforehand and try them one by one. 

\subsection{Comparisons}

%\noindent\textbf{Remark 4:} 

Both the earliest-$n$ and termination time strategies are practical simplifications to the original complicated 
reward allocation function. They constitute the contest with incomplete information model by using Bayesian game theory. 
Though their basic ideas are to incentivize the early joining contributors to exert more effort, they differ in 
several aspects. 
\begin{itemize}

\item The \emph{type} of the earliest-$n$ strategy is the joining time, while the \emph{type} is the number of 
contributors joining before time $T$ in the termination time strategy (implicitly determined by their joining time).

\item A contributor arriving after a certain time may not exert efforts in the earliest-$n$ strategy, while in the termination 
time strategy, all the contributors joining before $T$ exert the same amount of positive effort, and those arriving after $T$ will 
not spend any effort. 

\item The earliest-$n$ and the linearly decreasing strategies incur more complicated integral computations 
than the termination time strategy.

%\item The computation of the BNE is complicated in the earliest-$n$ strategy, while only a bisection search or a simple enumeration
%is needed in the termination time strategy. 

%\item The termination time strategy may risk the circumstance that very few even no contributors join before time $T$. 

\end{itemize}

%\subsection{Privacy Preservation}

\noindent\textbf{Remark 5:}
In the Bayesian game, each contributor is aware of the joining time distribution of all the others. 
When the joining time distributions of contributors are heterogeneous, announcing all these distributions may divulge 
the privacy of contributors, even though such information is anonymous. 
A feasible solution is to treat all of them homogeneous and to release the anonymous joining time distribution. 

\section{Open Crowdsensing System}
\label{sec:extension}

%In this section, we generalize our game framework to incorporate the privacy preservation of the heterogenous contributors
%and the open crowdsensing contest with external arriving contributors. 

In this section, we generalize our game framework to incorporate the open crowdsensing system with external arriving contributors.

\subsection{Modelling an Open Crowdsensing System}

Our previous models consider a fixed number of potential contributors, where the joining time of each contributor 
is characterized by a certain distribution. This is actually a \emph{closed} crowdsensing system. 
In some applications, the requester is \emph{open} to the arriving contributors without keeping the information of 
their personal joining time. Each arriving contributor decides the amount of efforts for crowdsensing, knowing the 
incentive mechanism and his joining time. 
Here, we suppose that the potential contributors join the contest at a Poisson rate denoted by 
$\lambda$. The \emph{public information} is thus this arrival rate, other than the joining time distributions of 
all the contributors. 

\begin{figure}[!htb]
\centering
\includegraphics[width=3.5in]{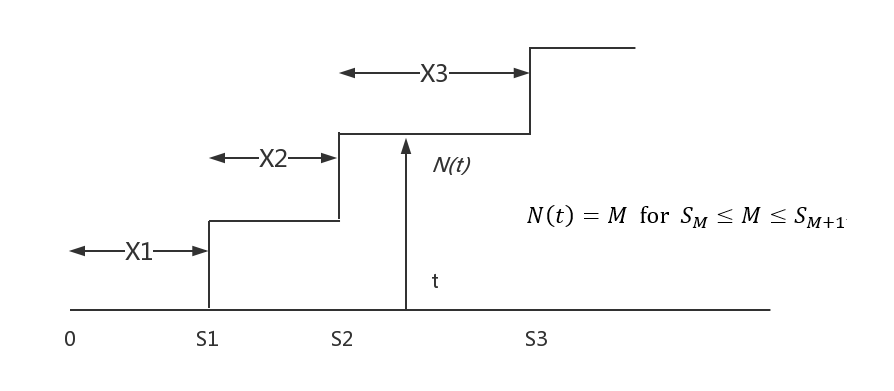}
\vspace{-0.8cm}
\caption{An arrival process and its arrival epochs $\{S_1, \cdots, S_M\}$, its inter-arrival intervals $\{X_1, \cdots, X_M\}$, and the 
counting process $N(t)$.}
\label{fig:poisson_arrival}
\vspace{-0.3cm}
\end{figure}

Figure \ref{fig:poisson_arrival} illustrates the counting process $N(t)$ 
where $S_m$ denotes the joining epoch of the $m^{th}$ contributor, and 
$X_m$ denotes the inter-arrival time between the $(m-1)^{th}$ contributor and the $m^{th}$ one. 
Note that the process starts at time 0 and that multiple arrivals cannot occur simultaneously (owing to the 
infinite divisibility of time). 
The arrival process can also be specified by two other stochastic processes. 
The first alternative is the sequence of inter-arrival times, $X_1$, $X_2$, $\cdots$. 
These are positive random variables defined in terms of the arrival epochs by 
$X_1 = S_1$ and $X_j = S_{j}- S_{j-1}$ for $j>1$. 
Hence, given the $X_j$, the arrival epochs $S_j$ are specified as
\begin{eqnarray}
S_m = \sum\nolimits_{j=1}^{m}X_j. 
\end{eqnarray}
The joint distribution of $\mathbf{X}^{(m)} = \{X_1, \cdots, X_m\}$ for all $m > 1$ is sufficient to specify
the arrival process where $X_j$ ($1\leq j \leq m$) is exponentially distributed. 
The second alternative for specifying an arrival process is the counting process $N(t)$, where
for each $t > 0$, the random variable $N(t)$ is the number of arrivals up to and including time $t$.
The counting process is an uncountably infinite family of random variables $\{N(t); t>0\}$. 
When examining a sequence of arrivals, we usually consider a truncated counting process with a fixed number of arrivals. 
We denote $\mathbf{S}^{(m)} = \{S_1,\cdots, S_m\}$ be the sequence of joining epochs, and denote 
$\mathbf{s}^{(m)}=\{s_1,\cdots,s_m\}$ be the vector of joining times of this sequence. 

The joint distribution of joining epochs is 
a known result in queueing theory \cite{poissonbook}. By convolving a number of exponential 
distributions, the joint probability density function (p.d.f.) of the sequence $\mathbf{S}^{(m)}$ is given by 
\begin{eqnarray}
f_{\mathbf{S}^{(m)}}(\mathbf{s}^{(m)}) = \lambda^m \exp(-\lambda s_m), \quad 0 {<} s_1{<}\cdots{<}s_m. 
\label{eq:open_pdf}
\end{eqnarray}
Let $f_{\mathbf{S}^{(m)}|N(t)}(\mathbf{s}^{(m)}|(m,t)) $ be the joint density of $\{S_1, \cdots, S_m\}$ conditioned on $N(t) = m$ 
and time $t$. 
Then, this density is constant over the region $0<s_1<\cdots<s_m<t$ and has the expression
 \begin{eqnarray}
f_{\mathbf{S}^{(m)}}(\mathbf{s}^{(m)}|(m,t)) = \frac{m!}{t^m}.
\label{eq:open_conditional_pdf}
\end{eqnarray}
Here, Eq. \eqref{eq:open_pdf} is the joint density of $\mathbf{S}^{(m)}$ where $S_M$ can be arbitrary in this sequence. 
Eq. \eqref{eq:open_conditional_pdf} is usually the joint density of $\mathbf{S}^{(m)}$ in a fixed duration from 0 to $t$. 

An interesting observation is that the joint p.d.f. is not explicitly related to $s_1$ until $s_{m-1}$, but is implicitly coupled 
in the conditions $0 {<} s_1{<}\cdots{<}s_m<t$. The conditional joint density is uniformly distributed. 
For this Poisson process, the probability moment function for $N(t)$ (i.e. 
the number of arrivals in $(0,t]$) is given by 
\begin{eqnarray}
\mathbb{P}_{N(t)}(m) = \frac{(\lambda t)^m \exp(-\lambda t)}{m!}. 
\label{eq:open_poisson_pmf}
\end{eqnarray}
In what follows, we will investigate whether our analytical framework for the closed crowdsensing system can be 
generalized to an open system with external arrivals.

\subsection{Earliest-n Strategy of Open Crowdsensing System}

\subsubsection{Stage-II}
If a contributor joins the contest at time $s_i$, the probability of 
having no more than $n{-}1$ arrivals is given by 
\begin{eqnarray}
\mathbb{P}_{i\in\{n\}} = \sum\nolimits_{k=0}^{n{-}1}\frac{\exp(-\lambda s_i)(\lambda s_i)^k}{k!}.
\label{eq:prob_earliest_n_open}
\end{eqnarray}
Then, the maximum reward that the $i^{th}$ joining contributor can acquire is $b\mathbb{P}_{i\in\{n\}}$ on average. 

Consider a truncated counting system with a maximum number of $M$ contributors where $M$ is reasonably large. 
The basis of this simplification is that the joining of the potential contributors after contributor $M$ is 
usually too late from the requester's angle. 
%Note that this simplification is based on the assumption 
% where $N$ can be arbitrarily large. 
The payoff of the contributor arriving at $t_i$ is the same as that of Eq. \eqref{eq:pi_earliest_n} except for 
substituting $\mathbb{P}_{i\in\{n\}}$ in Eq. \eqref{eq:earliest-n-prob} by the one in Eq. \eqref{eq:prob_earliest_n_open}.
For the $i^{th}$ joining contributor, the p.d.f. of joining time distribution is given by Eq. \eqref{eq:open_pdf}.

When the $i^{th}$ contributor joins, he has no knowledge on the joining times of all the others. After isolating the 
$i^{th}$ arrival epoch from the arrival sequence, we assume the remaining sequence as the original Poisson counting system. 
This assumption slightly deviates from the original arrival sequence, while greatly reducing the complexity of modelling the 
expected utility of the $i^{th}$ contributor. Then, for the remaining $(M-1)$ contributors, the p.d.f. of $j^{th}$ contributor is computed 
by Eq. \eqref{eq:open_pdf}. The expected utility of contributor $i$ on all the combinations of $\mathbf{S}^{(M)}$ is given by 
\begin{eqnarray}
\mathbb{E}[\pi_i([e_i, s_i], [\mathbf{e}_{-i}, \mathbf{s}_{-i}])] \!\!\!&=&\!\!\! {\int_{0}^{\infty}\int_{0}^{s_{M{-}1}}\cdots \int_{0}^{s_2}}\frac{e_ib(s_i)}{e_0{+}e_i {+} E_{-i}}\nonumber\\
&& \!\!\!\!\!\!\!\!\!\ \!\!\!\!\!\!\!\!\!\ \!\!\!\!\!\!\!\!\!\!\!\!\! \!\!\!\!\!\!\!\!\!\!\!\!\!\!\! \!\!\!\!\!\!\!\!\!\!\!\!\!\!\!
\cdot \lambda^{M{-}1} \exp(-\lambda s_{M{-}1}) \;\; ds_1 \cdots ds_{M{-}1} - e_i, 
\label{eq:complicated_integral}
\end{eqnarray}
where there are $M{-}1$ integrals and $i$ is not regarded as an item in the range $[1, M{-}1]$. 
The procedure of finding the Bayesian Nash equilibrium follows that of the closed crowdsensing system. 

\subsubsection{Stage-I} At the Bayesian Nash equilibrium, the $i^{th}$ joining contributor exerts $e_i^*$ efforts that 
yield the utility $w(s_i)e_i^*(s_i)$ to the requester. 
Because each contributor makes decision independently, the requester sees a collection of $M$ contributions each of which 
depends on its own joining time. 
For $M$ contributors with the sequence of arrival epochs 
$\{S_1, \cdots, S_M\}$, 
the expected utility of the requester at the Bayesian Nash equilibrium is
\begin{eqnarray}
E[\mathcal{U}] \!\!\!\!&=&\!\!\!\! E[\sum\nolimits_{m=1}^{M} w(S_i)e_i^*(S_i)] 
=\sum\nolimits_{m=1}^{M} E[w(S_i)e_i^*(S_i)] \nonumber\\
\!\!\!\!&=&\!\!\!\! {\int_{0}^{\infty}\int_{0}^{s_{M}}\cdots \int_{0}^{s_2}} w(s_i) e_i^*(s_i) ds_1 \cdots ds_{M},
\end{eqnarray}
and the expected reward paid to contributors is 
\begin{eqnarray}
E[R] &=& \int_{0}^{\infty}\int_{0}^{s_{M}}\cdots \int_{0}^{s_2}\frac{b\sum\nolimits_{i=1}^{n}e_i^*(s_i)}{e_0 + \sum\nolimits_{i=1}^{M}e_i^*(s_i)}\nonumber\\
&&\cdot \lambda^{M} \exp(-\lambda s_{M}) \;\;ds_1\cdots ds_M. 
\end{eqnarray}

For a vector of joining times $\mathbf{S}^{(M)}$, the efficiency of the requester is computed as 
\begin{eqnarray}
\mathcal{E} = \frac{\big(\sum\nolimits_{i=1}^{M}w(s_i) e_i^*(s_i)\big)\big(e_0 + 
\sum\nolimits_{i=1}^{M}e_i^*(s_i)\big)}{b\sum\nolimits_{i=1}^{n}e_i^*(s_i)}. 
\end{eqnarray}
Then, the expected efficiency is 
\begin{eqnarray}
\mathbb{E}[\mathcal{E}] \!\!\!&=&\!\!\!  \int_{0}^{\infty}\int_{0}^{s_{M}}\cdots \int_{0}^{s_2}  
(\sum\nolimits_{i=1}^M w(s_i) e_i^*(s_i))
\nonumber\\
\!\!\!&&\!\!\!\!\!\!\!\!\!\!\!\!\!\!\cdot \frac{e_0 {+} \sum\nolimits_{i=1}^{M}e_i^*(s_i)}{{b\sum\nolimits_{i=1}^{n}e_i^*(s_i)} }\cdot
\lambda^{M} \exp(-\lambda s_{M})\;ds_1\cdots ds_M. 
\end{eqnarray}
Compared with the closed crowdsensing system, the analytical framework of the open system remains the same. 
In fact, the open system is very similar to the closed system except that the joining time of an independent contributor is 
substituted by the joint distribution of all the incoming contributors. 

%of the contributors are heterogeneous. 
% and the 
%number of contributors is large. 

\subsection{Termination Time Strategy of Open Crowdsensing System}
\subsubsection{Stage-II}
In an open crowdsensing system, the total number of contributors joining before the deadline $T$ can be arbitrarily large. 
To avoid the analytical complexity, we suppose that there are at most $M$ contributors. 
For each contributor arriving at $S_i \leq T$, he can estimate the probability of meeting other $k$ contributors with the joining 
times before $T$. In this situation, there are $(k{+}1)$ contributors in total conditioned on the existence of at least one 
contributor joining before $T$. Then, there has
\begin{eqnarray}
\mathbb{P}(k, \infty) =  \frac{\exp(-\lambda T)(\lambda T)^{k+1}}{((k+1)!)(1-\exp(-\lambda T))}. 
\label{eq:extension1}
\end{eqnarray} 

We then substitute $\mathbb{P}(k, N-1)$ in Eq. \eqref{eq:expected_utility_T} by $\mathbb{P}(k, \infty)$ in Eq. \eqref{eq:extension1}. 
The Bayesian Nash equilibrium $e_i^*$ is solved in the same way as the closed system. 
Given the arrival sequence $\{S_1, \cdots, S_M\}$, all the contributors with $S_i \leq T$ will exert the same amount of efforts, 
and those with $S_i > T$ will not participate. In addition, $e_i^*$ is unrelated to the exact values of arrival epochs $S_1,\cdots, S_M$.
Therefore, we denote $e_i^* = e^*$ if $S_i \leq T$ and $e_i^* = 0$ otherwise.

\subsubsection{Stage-I} Given the arrival sequence $\mathbf{S}^{(M)}=\{S_1, \cdots, S_M\}$, 
the utility of the requester is:
\begin{eqnarray}
\mathcal{U} = \sum\nolimits_{i=1}^{M} w(S_i)e_i^*(S_i).  
\end{eqnarray}
When the requester designs the incentive mechanism, he will estimate the number of arrivals before $T$ using Eq. \eqref{eq:open_poisson_pmf}. Note that Eq. \eqref{eq:open_poisson_pmf} computes the probability of having $m$ arrivals before $T$ while Eq. \eqref{eq:extension1} computes the probability of ``seeing'' other $k$ arrivals before $T$ by a tagged contributor. 
Suppose that $m$ out of $M$ contributors join the crowdsensing before $T$. 
The joint distribution of $\mathbf{S}^{(m)}$ conditioned on $N(T) = m$ is obtained by Eq. \eqref{eq:open_conditional_pdf}. 
The conditional expected utility of the requester is 
\begin{eqnarray}
E[\mathcal{U}|N(T) = m] 
\!\!\!&=&\!\!\! e^*  \frac{m!}{T^m}\int_{0}^{T}\int_{0}^{s_{m}}\cdots \int_{0}^{s_2} \nonumber\\
\!\!\!&&\!\!\!\!\!\!\!\!\! (\sum\nolimits_{i=1}^{m}w(s_i)) \;\;ds_1 ds_2 \cdots ds_m . 
\label{eq:conditional_utility}
\end{eqnarray}
Considering all the possible scenarios regarding the number of arrivals before $T$, we obtain the expected utility of the requester 
\begin{eqnarray}
E[\mathcal{U}] = \sum\nolimits_{m=1}^{M} \mathbb{P}_{N(t)}(m) E[\mathcal{U} | N(T) = m].
\end{eqnarray}
The expected reward allocated to the contributors can be calculated as 
\begin{eqnarray}
E[R] = \sum\nolimits_{m=1}^{M} \mathbb{P}_{N(t)}(m) \frac{bm e^*}{e_0 + m e^*}
\end{eqnarray}
where the budget is $B$. 

When $m$ contributors join before $T$, the efficiency of the requester is given by 
\begin{eqnarray}
\mathcal{E} = \big(\sum\nolimits_{i=1}^{m}w(S_i) \big) (e_0 + m e^*) \Big/(b m),\quad m> 0.
\end{eqnarray}
The conditional expected efficiency is thus 
\begin{eqnarray}
E[\mathcal{E} | N(T) = m] &=& \frac{e_0+me^*}{bm} \frac{m!}{T^m}\int_{0}^{T}\int_{0}^{s_{m}}\cdots \int_{0}^{s_2} \nonumber\\
&& \!\!\!\!\!\!(\sum\nolimits_{i=1}^{m}w(s_i)) \;\;ds_1 ds_2 \cdots ds_m . 
\label{eq:conditional_result_ttime}
\end{eqnarray}
At the first sight, one can observe that the requester's conditional expected 
efficiency involves a complicated $N$-dimensional integral. However,  the special structure of the integral allows us 
to derive a much more succinct result.

\begin{lemma}
The conditional expected efficiency of the requester for the termination time strategy can be calculated as
\begin{eqnarray}
E[\mathcal{E} | N(T) = m] = \frac{e_0+me^*}{bT} \int_0^T w(x)dx. 
%&&\!\!\!\!\!\!\!\!\!\!E[\mathcal{E} | N(T) = m] = \frac{e_0+me^*}{bm} \frac{m!}{T^m}\int_{0}^{T} \nonumber\\
% \lambda^m\exp(-\lambda y)\nonumber\\
%&& \cdot \frac{y^{m-2}}{(m-2)!}\int_{0}^{y} w(x)dx dy . 
\label{eq:simplified_result_ttime}
\end{eqnarray}
\label{lemma:open_ttime}
\end{lemma}
\noindent\textbf{Proof:} Please refer to Appendix-I. \done % \ref{sec:appendix1}.

Eq. \eqref{eq:simplified_result_ttime} gives rise to the expected utility of the requester by traversing all possible $m$
\begin{eqnarray}
\!\!\!\!\!\!E[\mathcal{E}] \!\!\!\!\!&=&\!\!\!\!\! \sum\nolimits_{m=1}^{M} \mathbb{P}_{N(t)}(m) \cdot E[\mathcal{E} | N(T) = m] \nonumber\\
\!\!\!\!\!\!\!&=&\!\!\!\!\!\!\! \sum\nolimits_{m=1}^{M}\!\! \!\!\!\frac{\lambda^m T^{m{-}1} (e_0{+}me^*)\exp({-}\lambda T)}{b(m!)}\!\! \int_0^T\!\!\! w(x)dx.
\label{eq:optimal_ttime_open}
\end{eqnarray}
The optimal termination time $T^*$ maximizes Eq.\eqref{eq:optimal_ttime_open} where a binary search approach can be used to find $T^*$.

\noindent\textbf{Remark 6:} When the arrival of the contributors deviates far away from Poisson process, our analytical 
approach also applies, while the simple expression of the efficiency at the termination time strategy is not obtainable.

\section{Performance Evaluation} 
\label{sec:perv}

In this section, we conduct numerical simulations to validate the effectiveness of the proposed 
incentive mechanisms. Both the closed and the open systems are considered. 

%In this section, we conduct trace-driven experiments to validate the effectiveness of the proposed incentive mechanisms. 
%Both the closed and the open systems are considered. 

\subsection{Simulation Setup}

\begin{figure*}[!tb]
\begin{minipage}[t]{0.33\linewidth}
\centering
\includegraphics[width=2.5in]{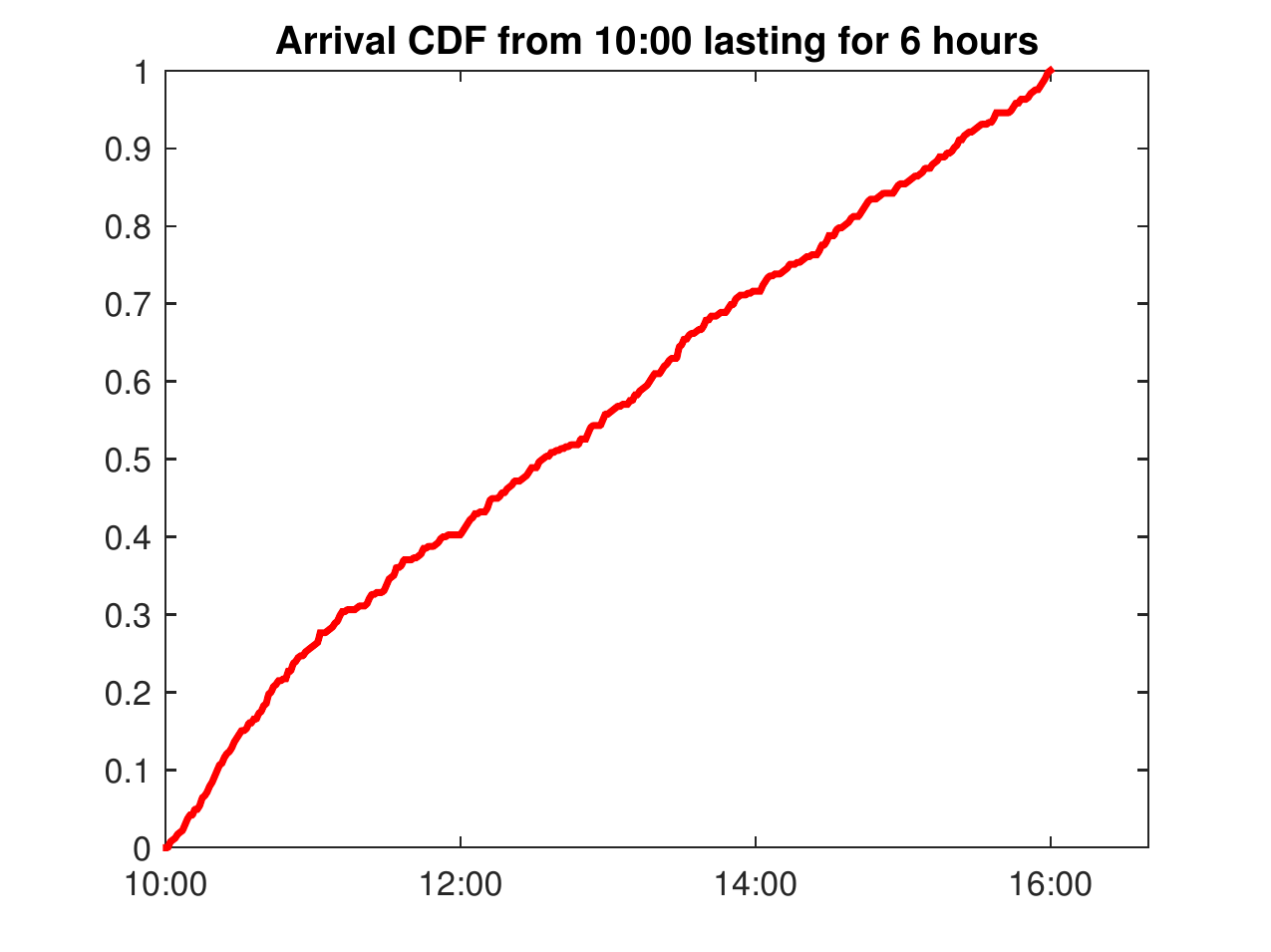}
\vspace{-0.5cm}
\caption{CDF curve of users' joining time (starting from 10:00 am)}
\label{fig:10am_cdf}
\end{minipage}
\hspace{1ex}
\begin{minipage}[t]{0.33\linewidth}
\centering
\includegraphics[width=2.5in]{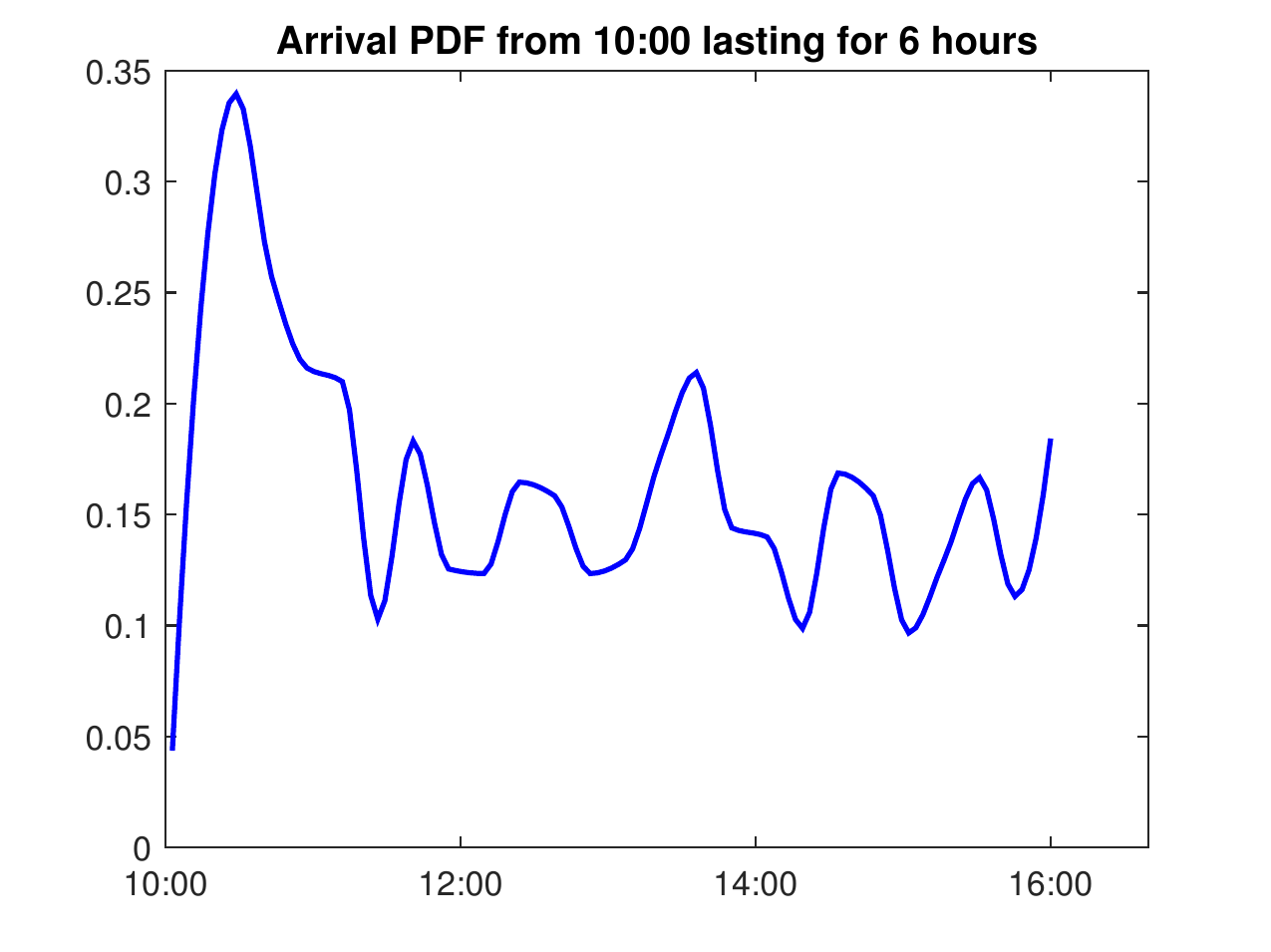}
\vspace{-0.5cm}
\caption{PDF curve of users' joining time (starting from 10:00 am)}
\label{fig:10am_pdf}
\end{minipage}
\hspace{1ex}
\begin{minipage}[t]{0.33\linewidth}
\centering
\includegraphics[width=2.5in]{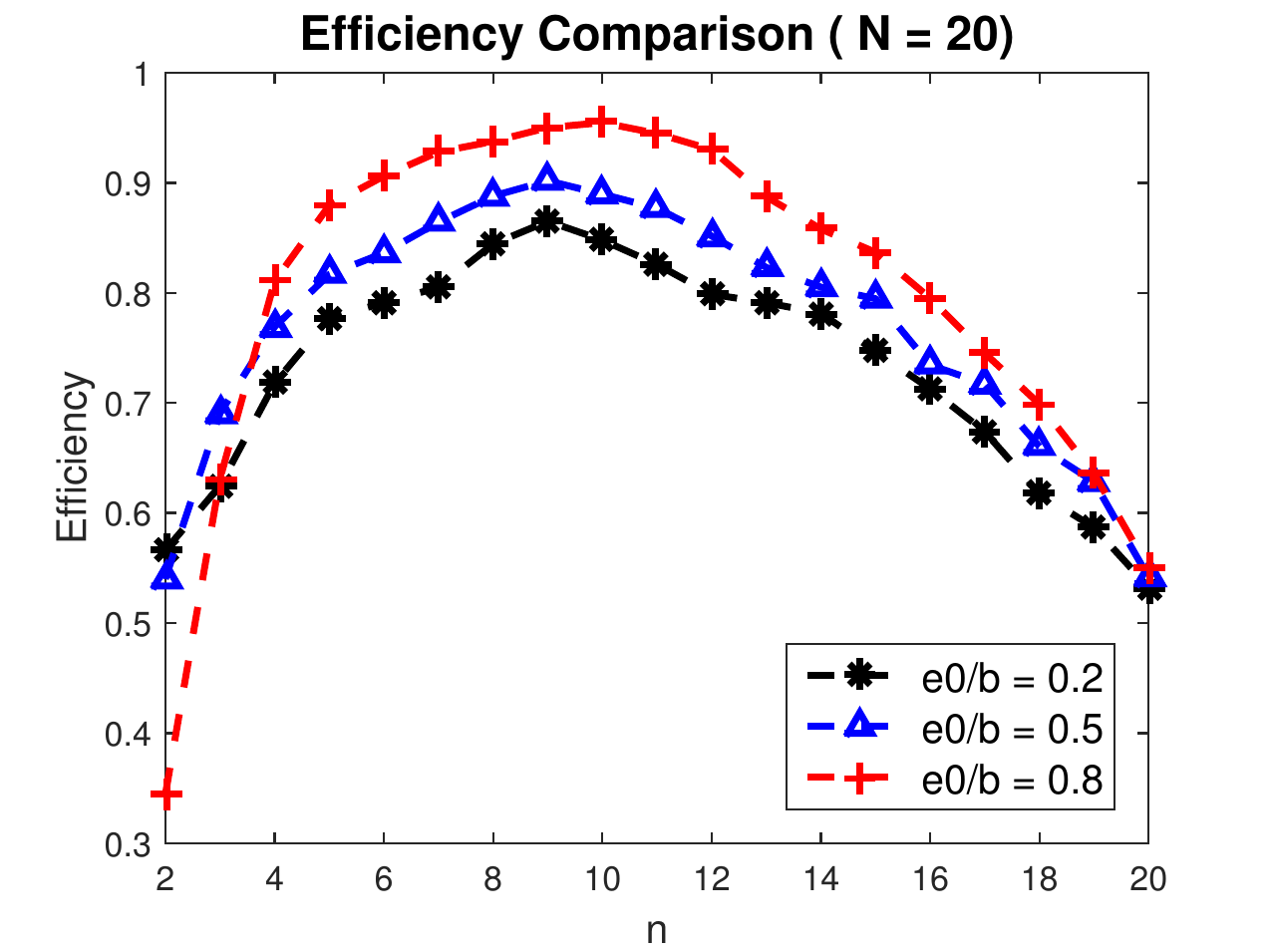}
\vspace{-0.5cm}
\caption{Efficiency of earliest-$n$ scheme: closed system with step weight function}
\label{fig:efficiency_earliestn_step_closed}
\end{minipage}
\vspace{-0.1cm}
\end{figure*}

We collect the WiFi access record of college students and faculties at Guanghua Building of Fudan University. 
This dataset spans one month, containing the anonymous 
user ID, the location of APs, the time stamp of accessing APs. 
%The records with very short accessing duration are cleansed for the anonymous 
%user may drive a car, ride a bicycle or is running around the venues. 
%These users are not regarded as potential contributors at that time because they may be too busy to undertake the 
%crowdsensing tasks. 
For the closed system, we randomly select 20 users from the top one hundred active users. Note that the active users will 
possess enough samples to generate the meaningful distribution functions of the joining time. 
Figure \ref{fig:10am_cdf} and \ref{fig:10am_pdf} display the cumulative distribution function 
and the probability distribution function of joining time where the starting point is 10:00 am and 
the ending point is 4:00 pm.  
For the open system, we set an arrival rate without using the collected data because the arrival rate is too large 
in this tall building (e.g. more than one person per minute on average). 

The timeliness property is reflected by two examples in the closed system: a step function 
$w(t) = \{1, 0.6, 0.2, 0\}$ 
with discontinuous points at $t = \{10:00, 11:30, 13:00, 16:00\}$, and 
an inverse quadratic function $w(t) = (1+\frac{1}{6}(t-10))^{-2}$ for $t\in [10, 16]$  
(i.e. in the range between 10:00 and 16:00). 
To try more choices, we let the step function be 
$w(t) = \{1, 0.6, 0.2, 0\}$ where $w(t)$ changes abruptly 
at the normalized time $\{0, 0.5, 1, 1.5\}$, and let the inverse cubic weight function be 
$w(t) = (1+t/3)^{-3}$. 
The total reward for allocation is $B = 1$ if not mentioned explicitly. 
We consider three different sets of $\{e_0, b\}$ that have $\frac{e_0}{b} = \{0.2, 0.5, 0.8\}$.

%\begin{figure}[htb]
%\centering
%\includegraphics[width=2.8in]{./simu_figures2/arrival_cdf.jpg}
%\vspace{-0.3cm}
%\caption{CDF curve of users' joining time (starting from 10:00 am)}
%\label{fig:10am_cdf}
%\end{figure}

%\begin{figure}[htb]
%\centering
%\includegraphics[width=2.8in]{./simu_figures2/arrival_pdf.jpg}
%\vspace{-0.3cm}
%\caption{PDF curve of users' joining time (starting from 10:00 am)}
%\label{fig:10am_pdf}
%\end{figure}

\subsection{Closed System}

\noindent\textbf{Earliest-$n$ Strategy.} 
We compute the BNE numerically following the approach in \cite{Wasser}. The basic idea is to 
decompose the continuous space into a number of multi-dimensional cubes. Then, the integral is substituted by 
the sum of its value on all the cubes. To reduce the complexity of traversing all of them,  
we adopt the more efficient Monte Carlo integration approach. 
When the number of contributors is large, the computational burden is relieved 
for we do not need to traverse the less important cubes.

Figure \ref{fig:efficiency_earliestn_step_closed} illustrates the efficiency of the requester with the step weight 
function as $n$ increases from 2 to 20. One can see 
that by limiting the number of rewarded contributors, the requester can even double his efficiency. 
It is wise to choose $n$ to be between 8 and 12.  
Meanwhile, a larger $\frac{e_0}{b}$ usually brings a higher efficiency. 
One exception happens at $n=2$ because the mobile users might not aways participate 
in the competition thus yielding a zero efficiency occasionally. 
When $e_0$ is set to $0.8b$ other than $0.5b$ and $0.2b$, the maximum efficiency of the requester is improved by around 
0.06 and 0.09 respectively. 
This implies that introducing a ``nature'' player can effectively enhance the requester's efficiency.

\begin{figure*}[!tb]
\begin{minipage}[t]{0.33\linewidth}
\centering
\includegraphics[width=2.6in]{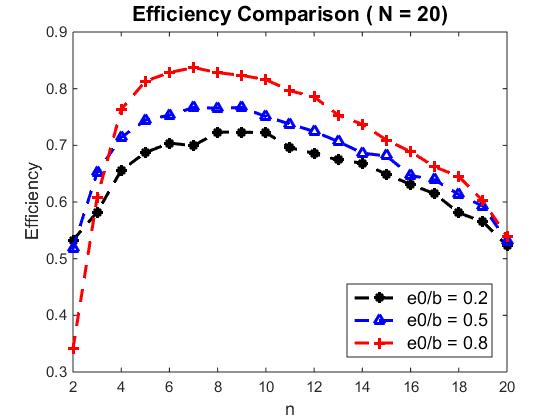}
\vspace{-0.5cm}
\caption{Efficiency of earliest-$n$ scheme: closed system with inverse weight function}
\label{fig:efficiency_earliestn_hyper_closed}
\end{minipage}
\hspace{1ex}
\begin{minipage}[t]{0.33\linewidth}
\centering
\includegraphics[width=2.6in]{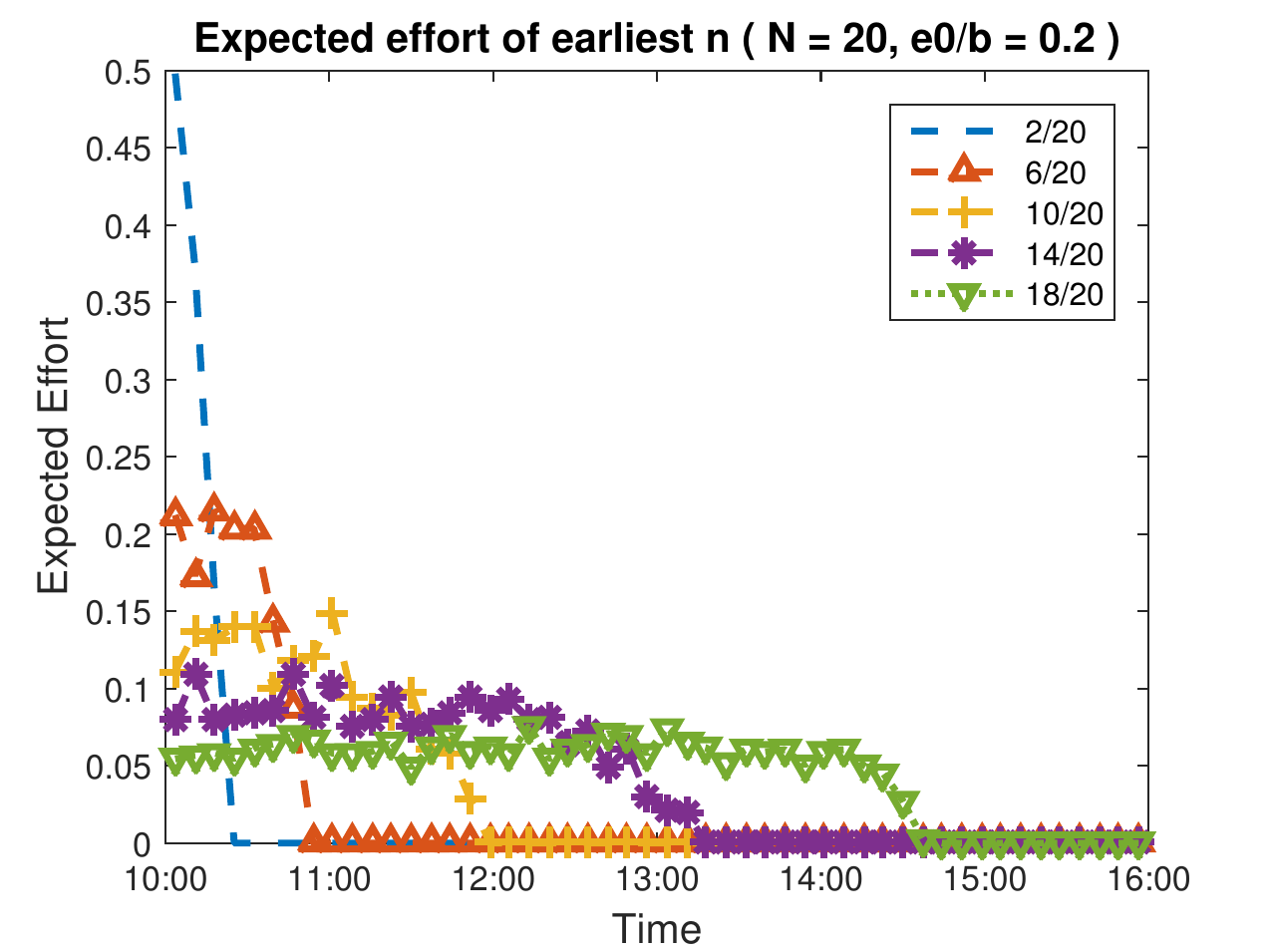}
\vspace{-0.5cm}
\caption{Efforts of a contributor of earliest-$n$ scheme: closed system}
\label{fig:efficiency_earliestn_closed}
\end{minipage}
\hspace{1ex}
\begin{minipage}[t]{0.33\linewidth}
\centering
\includegraphics[width=2.6in]{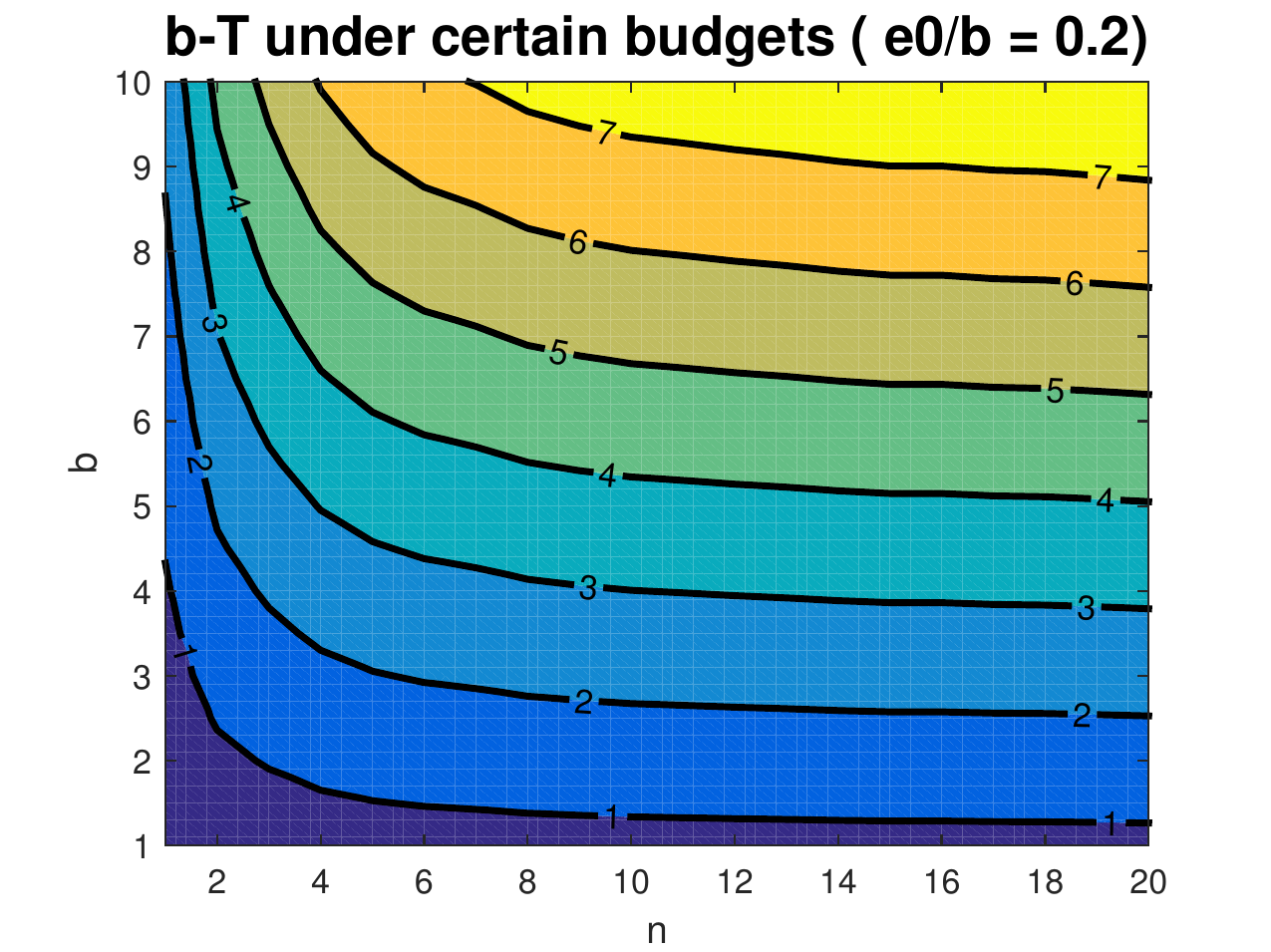}
\vspace{-0.5cm}
\caption{Contour lines of earliest-$n$ scheme: closed system}
\label{fig:efficiency_earliestn_closed}
\end{minipage}
\vspace{-0.1cm}
\end{figure*}

In Figure \ref{fig:efficiency_earliestn_hyper_closed}, we illustrate the efficiency with the 
inverse weight function. 
As $n$ increases from 2 to 20, the requester's efficiency 
increases in the beginning and decreases in the end. The optimal $n$ can be chosen to be six to 
eight.

%{\color{red} adding several sentences}

In Figure \ref{fig:efficiency_earliestn_closed}, we show the efforts exerted by a contributor at the BNE 
when he joins the crowdsensing at different times. Each curve indicates a different earliest-$n$ strategy. 
When the joining time $t$ increases, he contributes less and less efforts, and upon a certain threshold 
he does not participate. An early comer exerts more efforts than a late one. 
For different earliest-$n$ strategies, the participation threshold for a larger $n$ is also longer. 
In the proposed incentive mechanisms, $n$ and $b$ are configured simultaneously so that the 
allocated reward equals to the budget. 
Figure \ref{fig:efficiency_earliestn_closed} shows the contour lines of the pairs $(n, b)$ with different budget $B$s. 
Any point in a contour line refers to a combination of $n$ and $b$ yielding the allocated reward equal to 
the budget. We observe that $b$ is a strictly decreasing function of $n$. 
When rewarding more contributors, the maximum achievable reward for each of them will be smaller. 
Note that the efforts and the contour lines are solely determined by the 
competition at stage-I, irrelevant to the shapes of the weight function of the requester.

%\begin{figure}[!htb]
%\centering
%\includegraphics[width=2.8in]{./simu_figures2/efficiency_step_en_closed.jpg}
%\vspace{-0.3cm}
%\caption{Efficiency of earliest-$n$ scheme: closed system with step weight function}
%\label{fig:efficiency_earliestn_step_closed}
%\end{figure}

%\begin{figure}[!htb]
%\centering
%\includegraphics[width=2.8in]{./simu_figures2/efficiency_exp_en_closed.jpg}
%\vspace{-0.3cm}
%\caption{Efficiency of earliest-$n$ scheme: closed system with exponential weight function}
%\label{fig:efficiency_earliestn_exp_closed}
%\end{figure}

%\begin{figure}[!htb]
%\centering
%\includegraphics[width=2.8in]{./simu_figures2/effort_en_closed.jpg}
%\vspace{-0.3cm}
%\caption{Efforts of a contributor of earliest-$n$ scheme: closed system}
%\label{fig:efficiency_earliestn_closed}
%\end{figure}

%\begin{figure}[!htb]
%\centering
%\includegraphics[width=2.8in]{./simu_figures2/contour_en_closed.jpg}
%\vspace{-0.3cm}
%\caption{Contour lines of earliest-$n$ scheme: closed system}
%\label{fig:efficiency_earliestn_closed}
%\end{figure}

\begin{figure*}[!tb]
\begin{minipage}[t]{0.33\linewidth}
\centering
\includegraphics[width=2.5in]{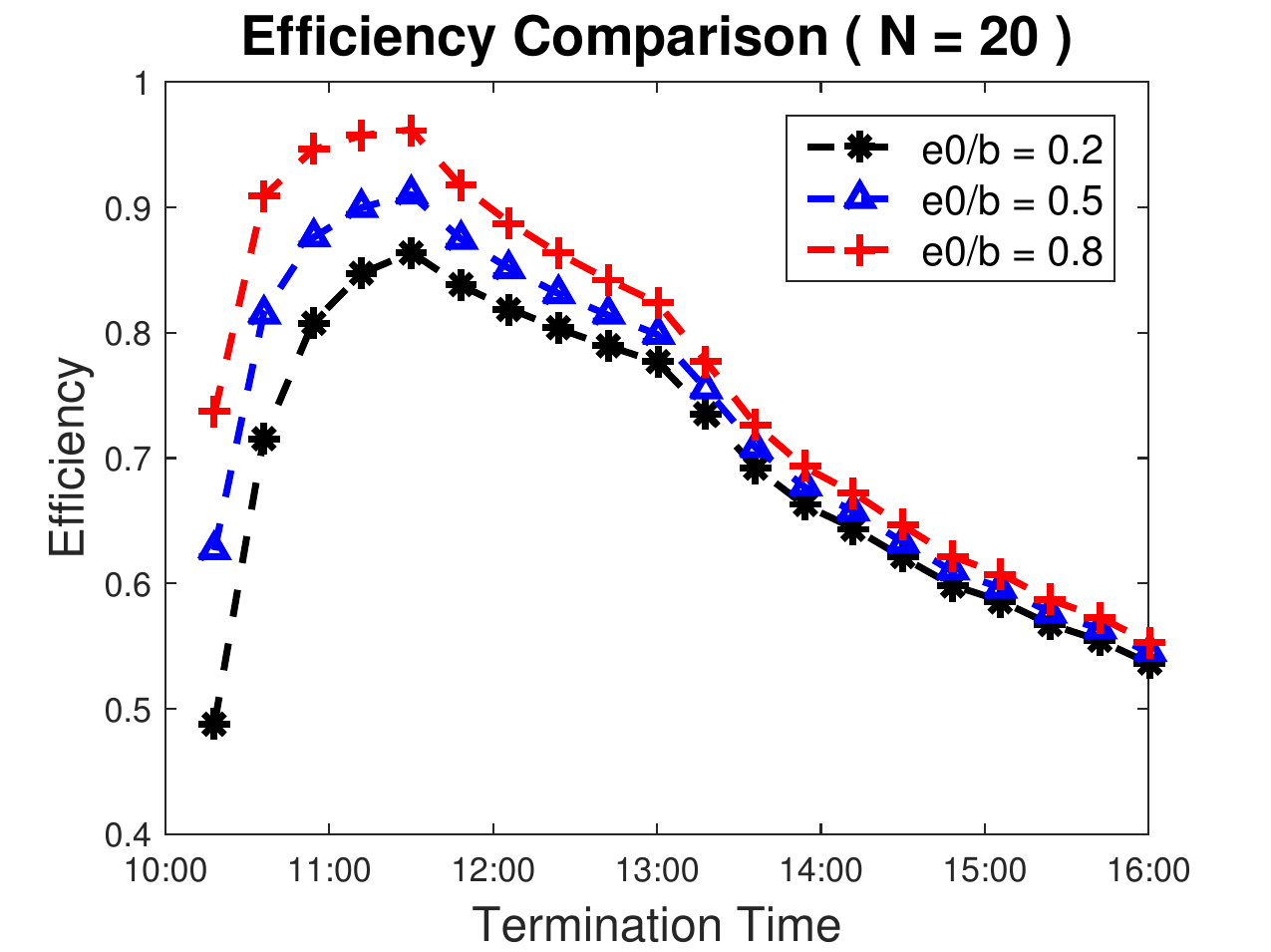}
\vspace{-0.5cm}
\caption{Efficiency of termination time scheme:  closed system with step weight function}
\label{fig:efficiency_tt_step_closed}
\end{minipage}
\hspace{1ex}
\begin{minipage}[t]{0.33\linewidth}
\centering
\includegraphics[width=2.5in]{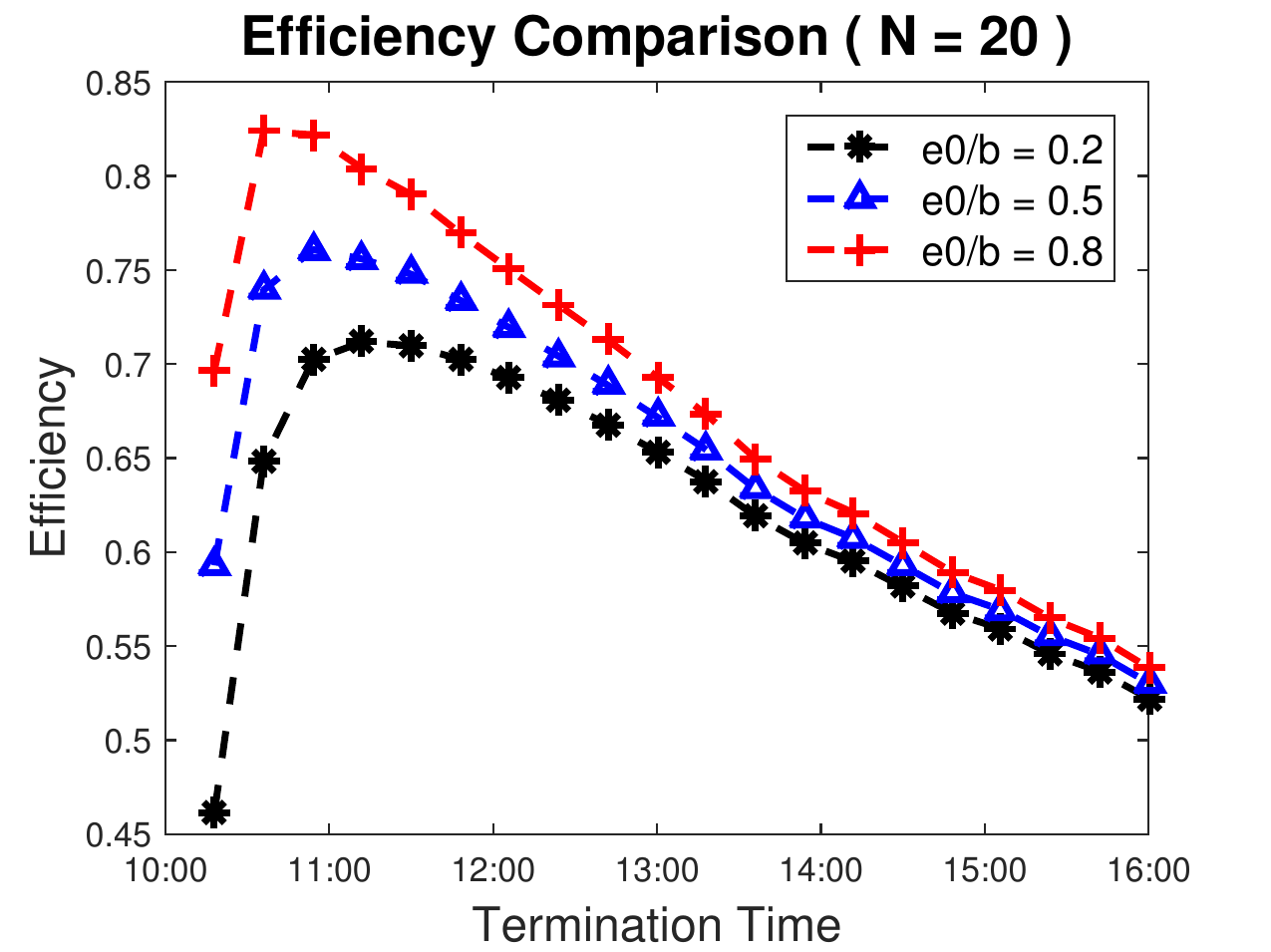}
\vspace{-0.5cm}
\caption{Efficiency of termination time scheme: closed system with inverse weight function}
\label{fig:efficiency_tt_hyper_closed}
\end{minipage}
\hspace{1ex}
\begin{minipage}[t]{0.33\linewidth}
\centering
\includegraphics[width=2.5in]{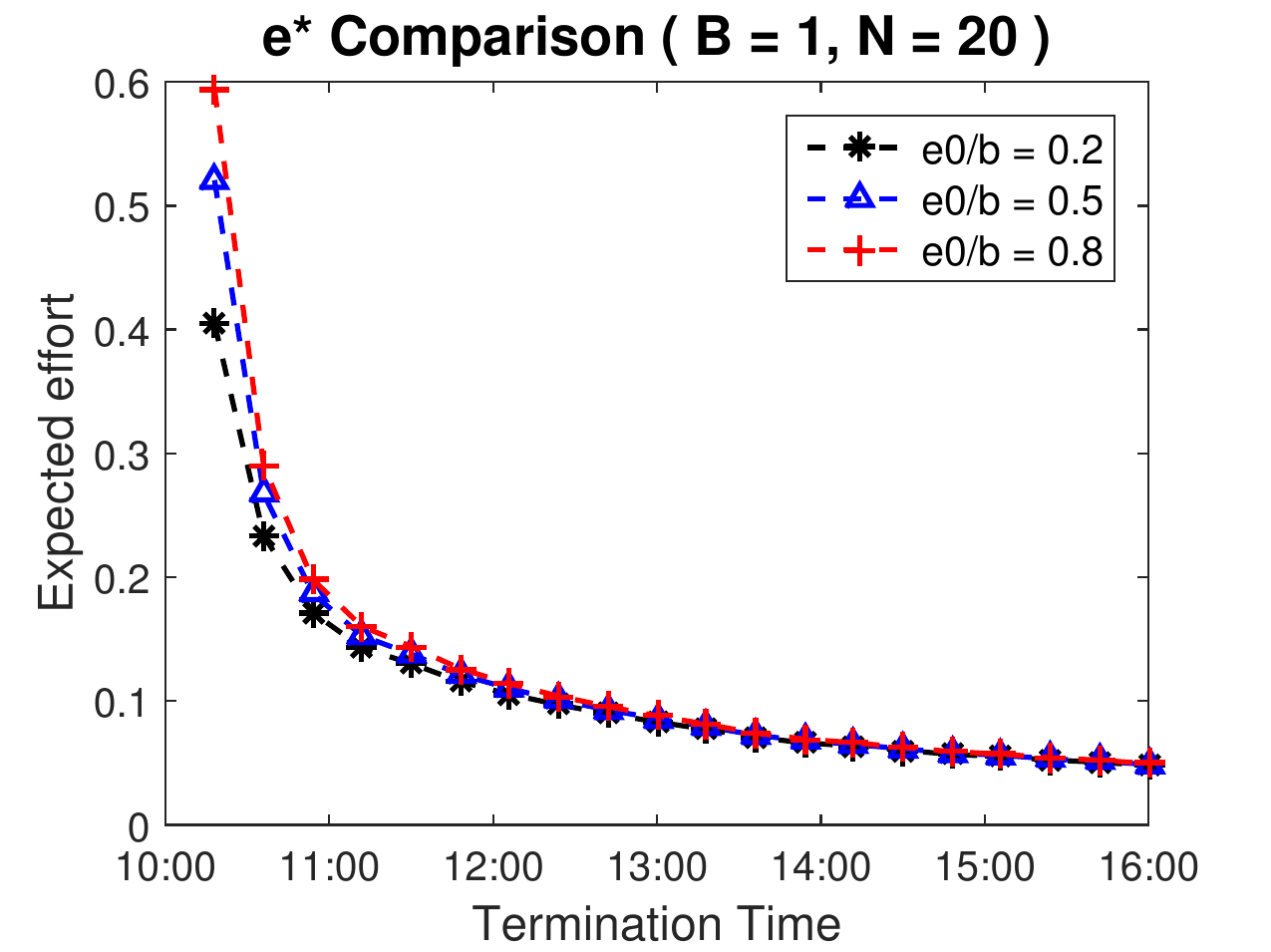}
\vspace{-0.5cm}
\caption{Efforts of a contributor of termination time scheme: closed system}
\label{fig:efforts_tt_closed}
\end{minipage}
\vspace{-0.1cm}
\end{figure*}

\noindent\textbf{Termination Time Strategy.} 
Figure \ref{fig:efficiency_tt_step_closed} plots the efficiency of the requester with the step weight function 
for the termination time strategy. The efficiency is evaluated by postponing this deadline for fifteen minutes every time. 
As the termination time increases, the efficiency increases first and then decreases where the highest value 
is achieved around 11:15 am.  
Figure \ref{fig:efficiency_tt_hyper_closed} illustrates the efficiency of the requester when the 
weight function is linearly decreasing. The optimal termination times are nearly 
11:00, 11:30 and 12:00 for $e_0$ to be $0.2b$, $0.5b$, $0.8b$ respectively. 
Both sets of experiments demonstrate that the appropriate selection of the termination time can greatly improve the 
efficiency of the requester. 
In addition, a similar observation holds when $\frac{e_0}{b}$ chooses different values. 
Compared with the case $e_0 = 0.2b$, the requester's efficiency improves by nearly 0.04 at $e_0 = 0.5b$ 
and by 0.1 at $e_0 = 0.8b$.

\begin{figure*}[!tb]
\begin{minipage}[t]{0.33\linewidth}
\centering
\includegraphics[width=2.5in]{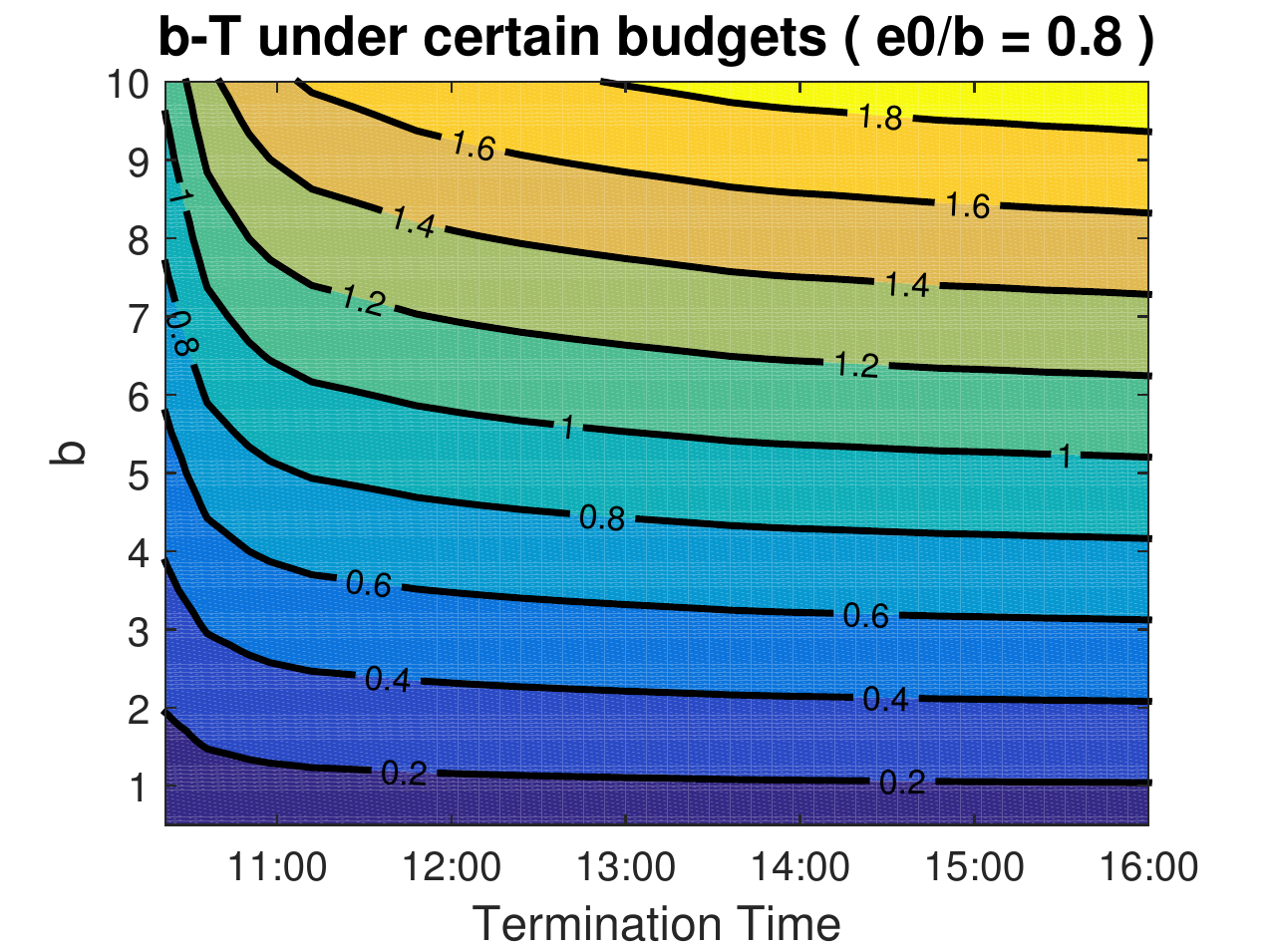}
\vspace{-0.5cm}
\caption{Contour lines of termination time scheme: closed system}
\label{fig:contour_tt_closed}
\end{minipage}
\hspace{1ex}
\begin{minipage}[t]{0.33\linewidth}
\centering
\includegraphics[width=2.5in]{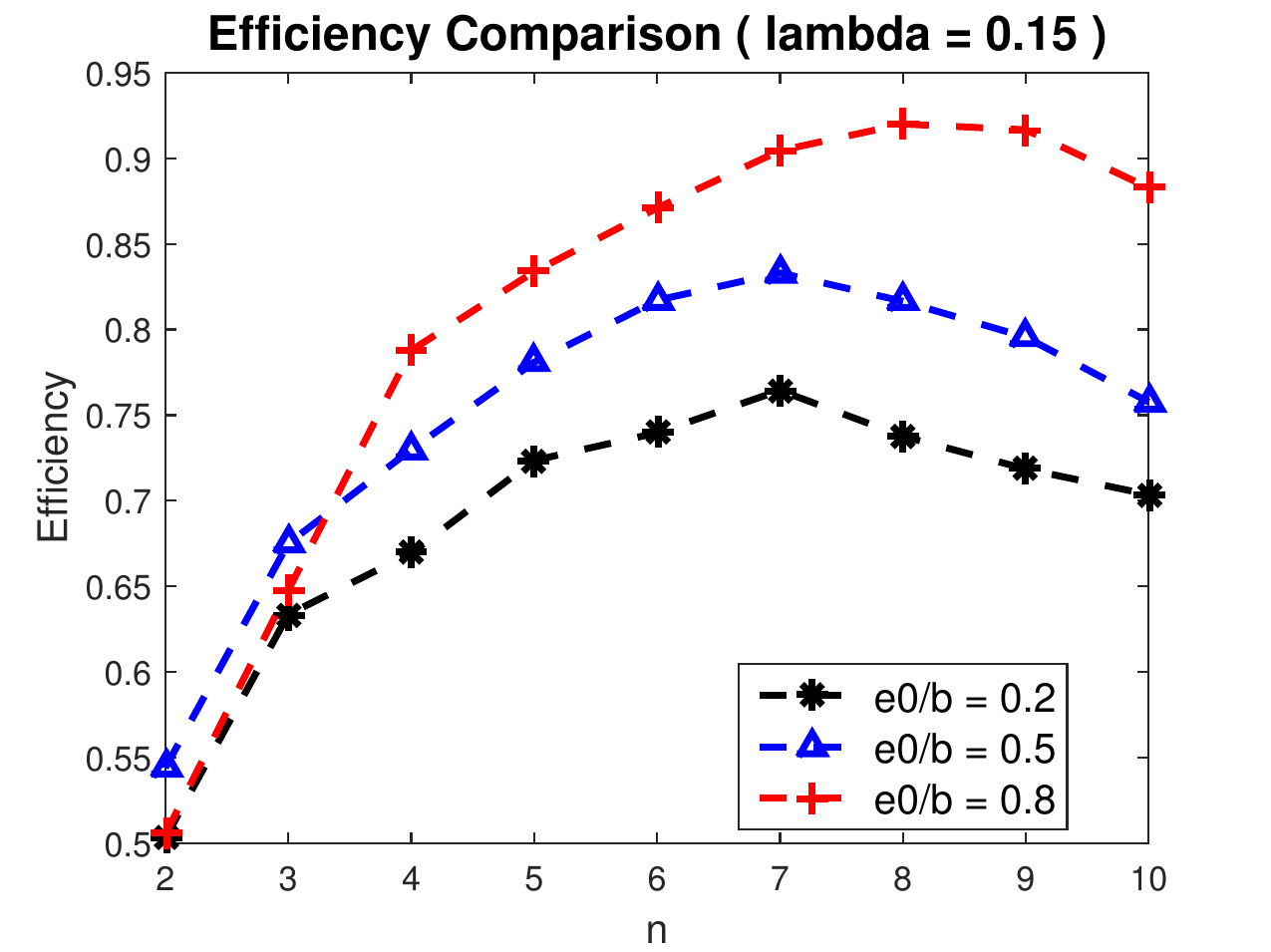}
\vspace{-0.5cm}
\caption{Efficiency of earliest-$n$ scheme: open system with step weight function}
\label{fig:efficiency_en_step_open}
\end{minipage}
\hspace{1ex}
\begin{minipage}[t]{0.33\linewidth}
\centering
\includegraphics[width=2.5in]{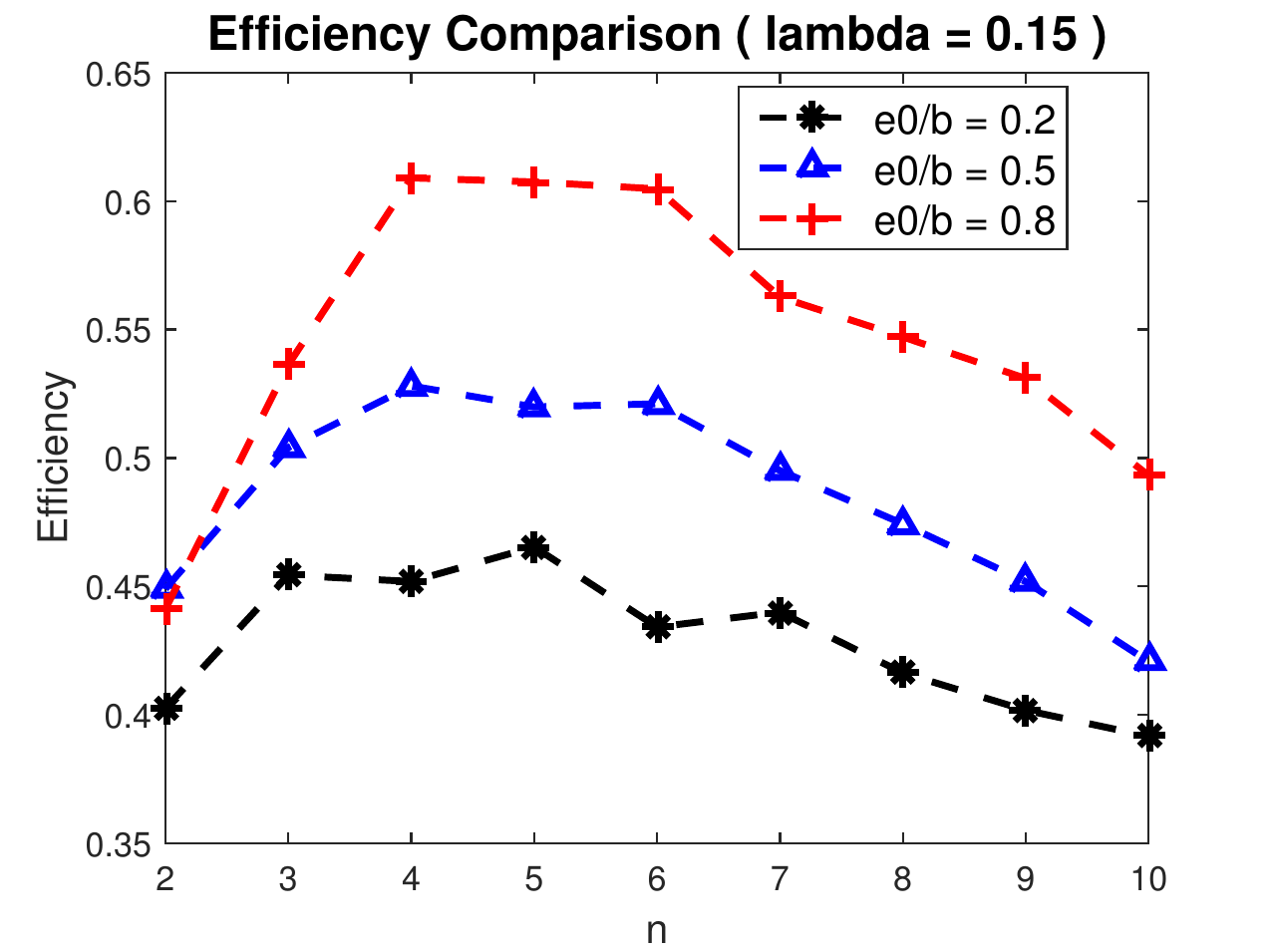}
\vspace{-0.5cm}
\caption{Efficiency of earliest-$n$ scheme: open system with inverse weight function}
\label{fig:efficiency_en_hyper_open}
\end{minipage}
\vspace{-0.6cm}
\end{figure*}

Figure \ref{fig:efforts_tt_closed} plots the amount of efforts exerted by a contributor under different termination times. 
The amount of efforts decreases with the increase of the termination time, and the rate of decline is especially large 
10:00 and 11:00. The reasons of this ``cliff effect'' are twofold: i) multiple contributors may flush into the competition; 
ii) the amount of efforts at the BNE is more sensitive to the change of the number of contributors when it is relatively small. 
As the designer of the termination time strategy, the requester needs to configure $b$ and $T$ jointly so that the 
expected expenditure equals to the budget. 
Figure \ref{fig:contour_tt_closed} shows the contour lines of $(b, T)$ under different budgets. 
When the termination time $T$ increases, more contributors will participate in the competition at the BNE. 
Therefore, the maximum achievable reward $b$ should be lower so as to balance the expenditure and the budget. 

\begin{comment}
\begin{figure}[htb]
\centering
\includegraphics[width=2.8in]{./simu_figures2/efficiency_step_tt_closed.jpg}
\vspace{-0.3cm}
\caption{Efficiency of termination time scheme:  closed system with step weight function}
\label{fig:efficiency_tt_step_closed}
\end{figure}

\begin{figure}[htb]
\centering
\includegraphics[width=2.8in]{./simu_figures2/efficiency_exp_tt_closed.jpg}
\vspace{-0.3cm}
\caption{Efficiency of termination time scheme: closed system with exponential weight function}
\label{fig:efficiency_tt_exp_closed}
\end{figure}

\begin{figure}[htb]
\centering
\includegraphics[width=2.8in]{./simu_figures2/effort_tt_closed.jpg}
\vspace{-0.3cm}
\caption{Efforts of a contributor of termination time scheme: closed system}
\label{fig:efforts_tt_closed}
\end{figure}
\end{comment}

%\begin{figure}[htb]
%\centering
%\includegraphics[width=2.8in]{./simu_figures2/contour_tt_closed.jpg}
%\vspace{-0.3cm}
%\caption{Contour lines of termination time scheme: closed system}
%\label{fig:contour_tt_closed}
%\end{figure}

\subsection{Open System}

\noindent\textbf{Earliest-$n$ Strategy.} 
Figure \ref{fig:efficiency_en_step_open} shows that 
the efficiency of a contributor increases first and then decreases 
with the increase of the number of rewarded contributors. 
In this set of experiments, the nature player has a significant impact on the requester's efficiency. 
When the earliest eight contributors are rewarded, the efficiencies at $e_0 = 0.2b, 0.5b$ and $0.8b$ 
are around 0.73, 0.82 and 0.92 respectively. 
Given the arrival rate $\lambda = 0.15$, the inter-arrival time is exponentially distributed with the mean of 
four hundred seconds. Thus, some potential contributors may arrive very late. Introducing 
the nature player may exclude the participation of late comers so that the requester's efficiency is improved. 
Figure \ref{fig:efficiency_en_hyper_open} illustrates the efficiency of the requester 
with the inverse function, which the similar patterns are demonstrated.

\begin{figure*}[!tb]
\begin{minipage}[t]{0.33\linewidth}
\centering
\includegraphics[width=2.5in]{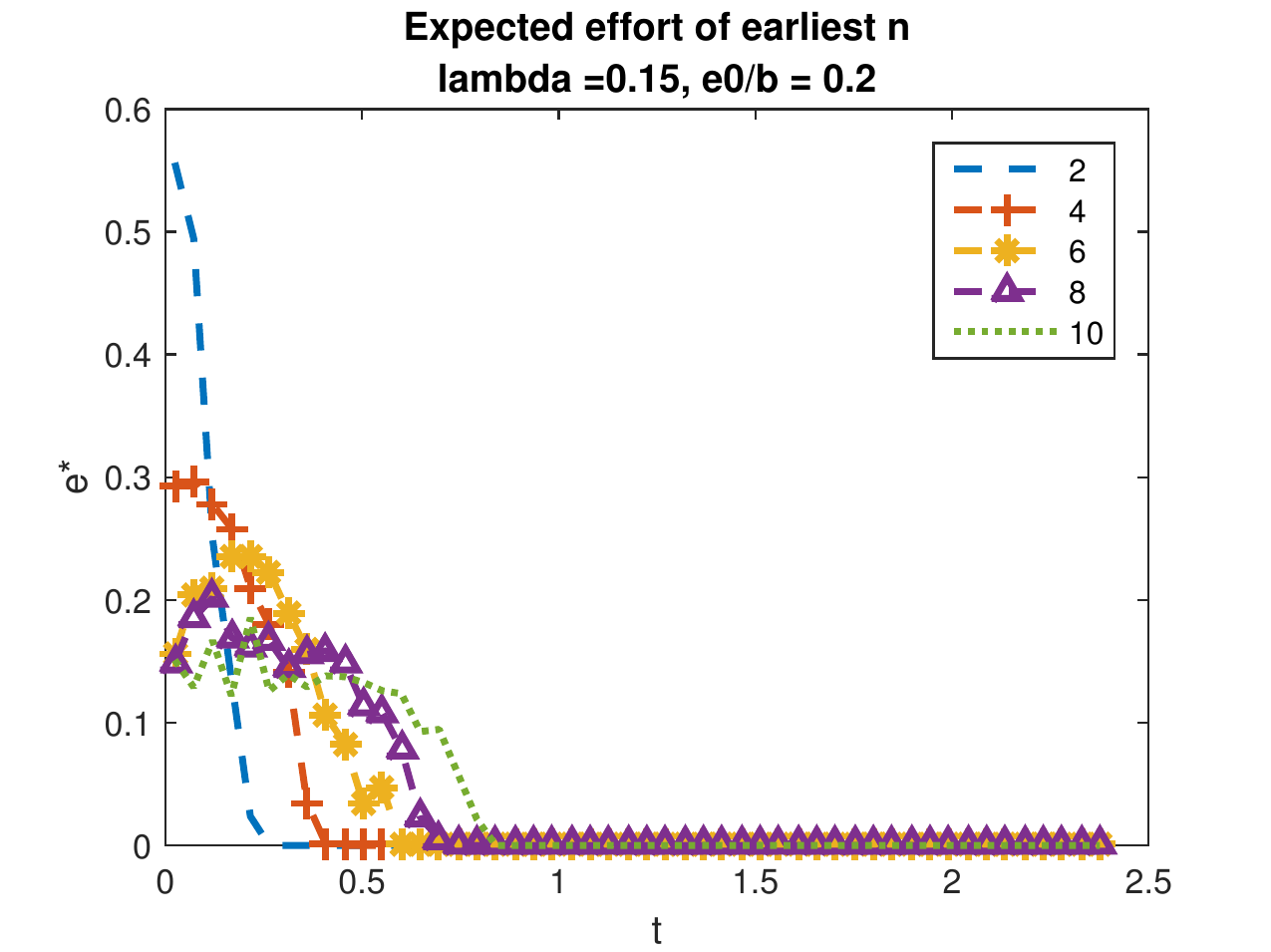}
\vspace{-0.5cm}
\caption{Efforts of a contributor of earliest-$n$ scheme: open system}
\label{fig:efforts_earliestn_open}
\end{minipage}
\hspace{1ex}
\begin{minipage}[t]{0.33\linewidth}
\centering
\includegraphics[width=2.5in]{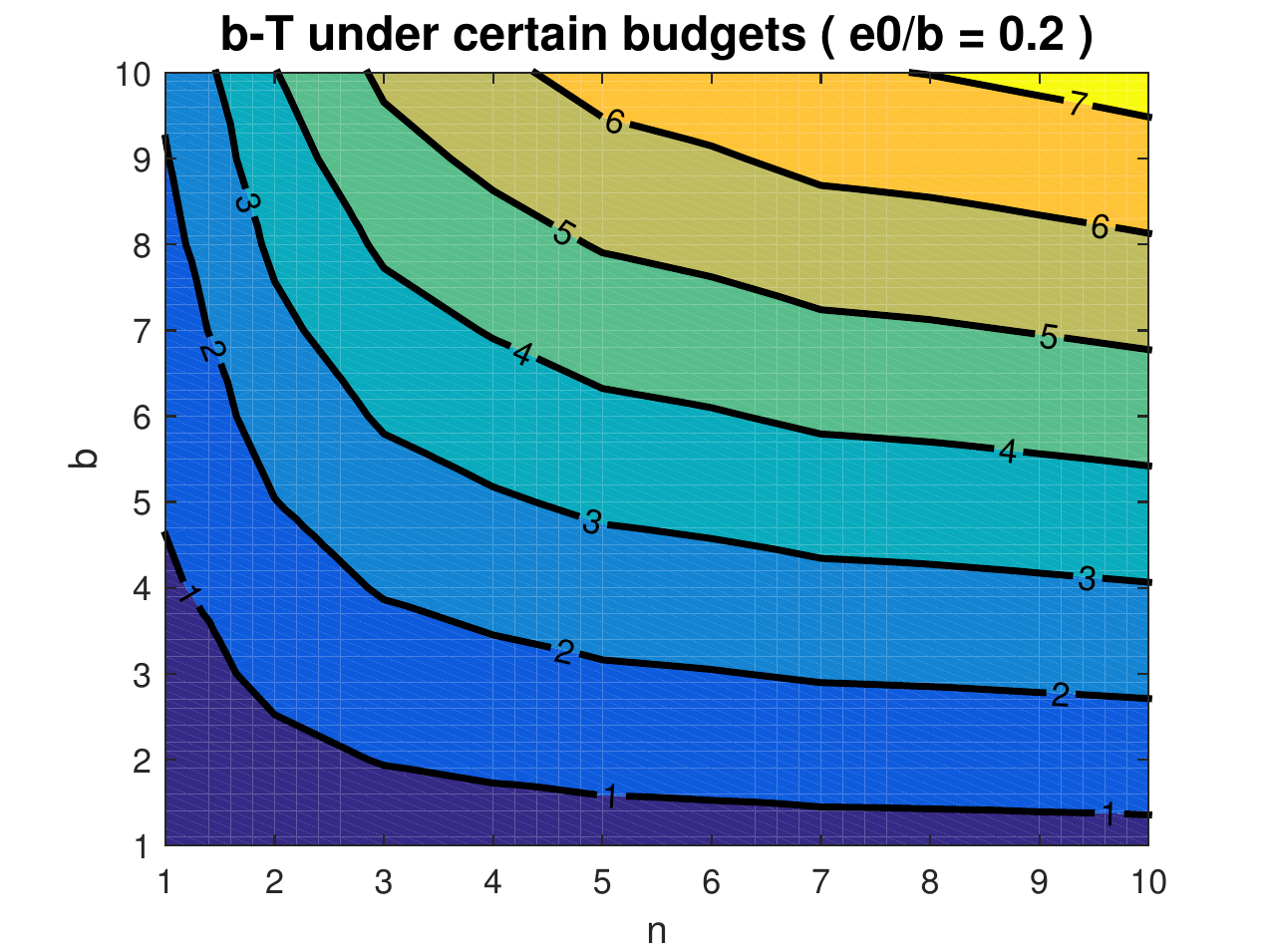}
\vspace{-0.5cm}
\caption{Contour lines of earliest-$n$ scheme: open system}
\label{fig:contour_earliestn_open}
\end{minipage}
\hspace{1ex}
\begin{minipage}[t]{0.33\linewidth}
\centering
\includegraphics[width=2.5in]{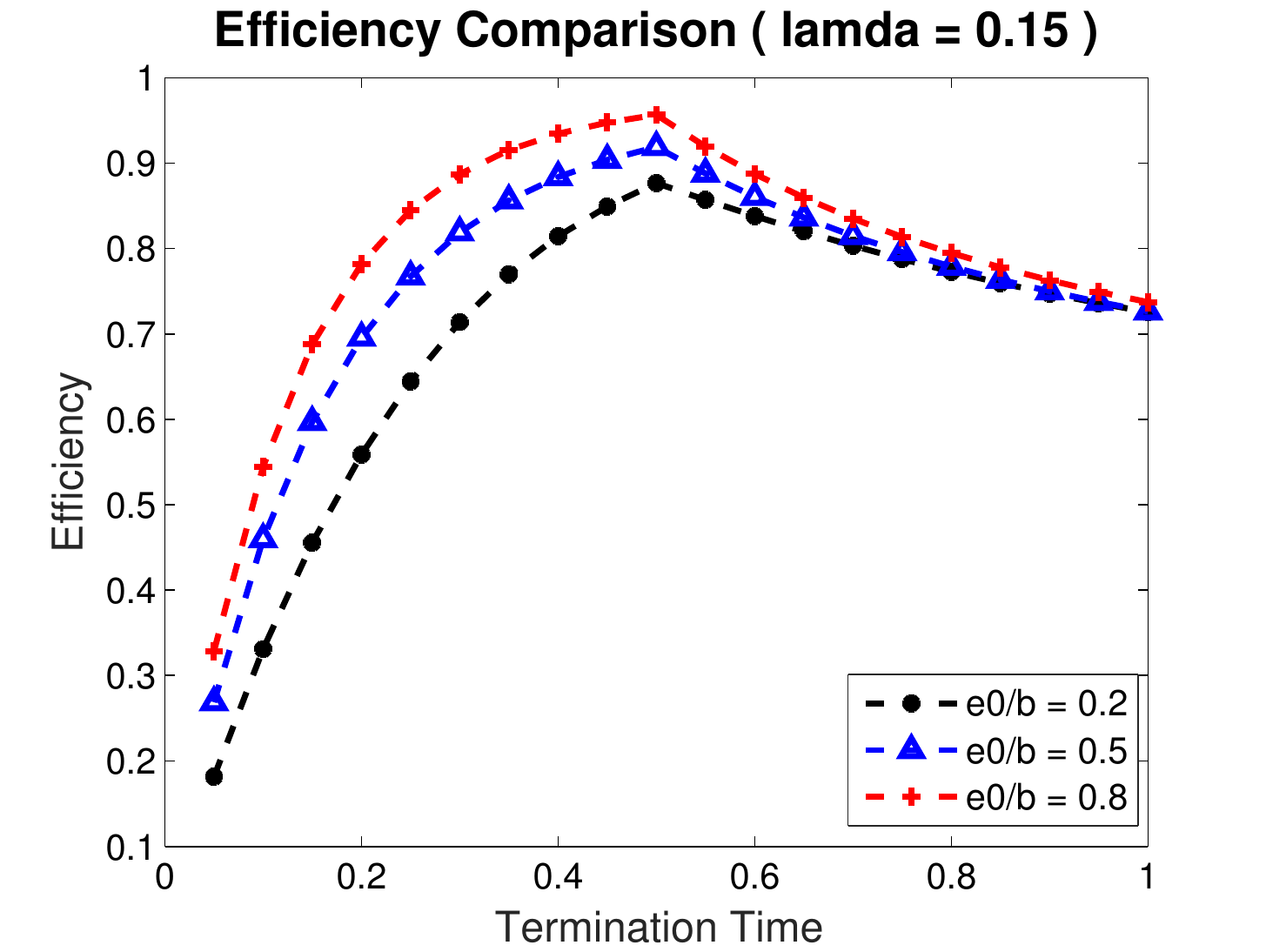}
\vspace{-0.5cm}
\caption{Efficiency of termination time scheme: open system with step weight function}
\label{fig:efficiency_tt_step_open}
\end{minipage}
\vspace{-0.5cm}
\end{figure*}

Figure \ref{fig:efforts_earliestn_open} illustrates the amount of efforts exerted by a contributor where 
the x-coordinate indicates the joining time. When the joining time becomes larger, the contributor 
is inclined to allocating less efforts at the BNE. 
Figure \ref{fig:contour_earliestn_open} plots the contour lines for the pair $(n, b)$ with different budgets. 
The larger budget $B$ means the larger $b$ for the fixed $n$, and the larger $n$ for the fixed $b$. 
Given the fixed $B$, the increases of $n$ (resp. $b$) results in the decrease of $b$ (resp. $n$). 

\begin{comment}
\begin{figure}[!htb]
\centering
\includegraphics[width=2.8in]{./simu_figures2/efficiency_step_en_open.jpg}
\vspace{-0.3cm}
\caption{Efficiency of earliest-$n$ scheme: open system with step weight function}
\label{fig:efficiency_en_step_open}
\end{figure}

\begin{figure}[!htb]
\centering
\includegraphics[width=2.8in]{./simu_figures2/efficiency_exp_en_open.jpg}
\vspace{-0.3cm}
\caption{Efficiency of earliest-$n$ scheme: open system with exponential weight function}
\label{fig:efficiency_en_exp_open}
\end{figure}
\end{comment}

\begin{comment}
\begin{figure}[!htb]
\centering
\includegraphics[width=2.8in]{./simu_figures2/effort_en_open.jpg}
\vspace{-0.3cm}
\caption{Efforts of a contributor of earliest-$n$ scheme: open system}
\label{fig:efforts_earliestn_open}
\end{figure}

\begin{figure}[!htb]
\centering
\includegraphics[width=2.8in]{./simu_figures2/contour_en_open.jpg}
\vspace{-0.3cm}
\caption{Contour lines of earliest-$n$ scheme: open system}
\label{fig:contour_earliestn_open}
\end{figure}
\end{comment}

\noindent\textbf{Termination Time Strategy.} 

\begin{figure*}[!tb]
\begin{minipage}[t]{0.33\linewidth}
\centering
\includegraphics[width=2.5in]{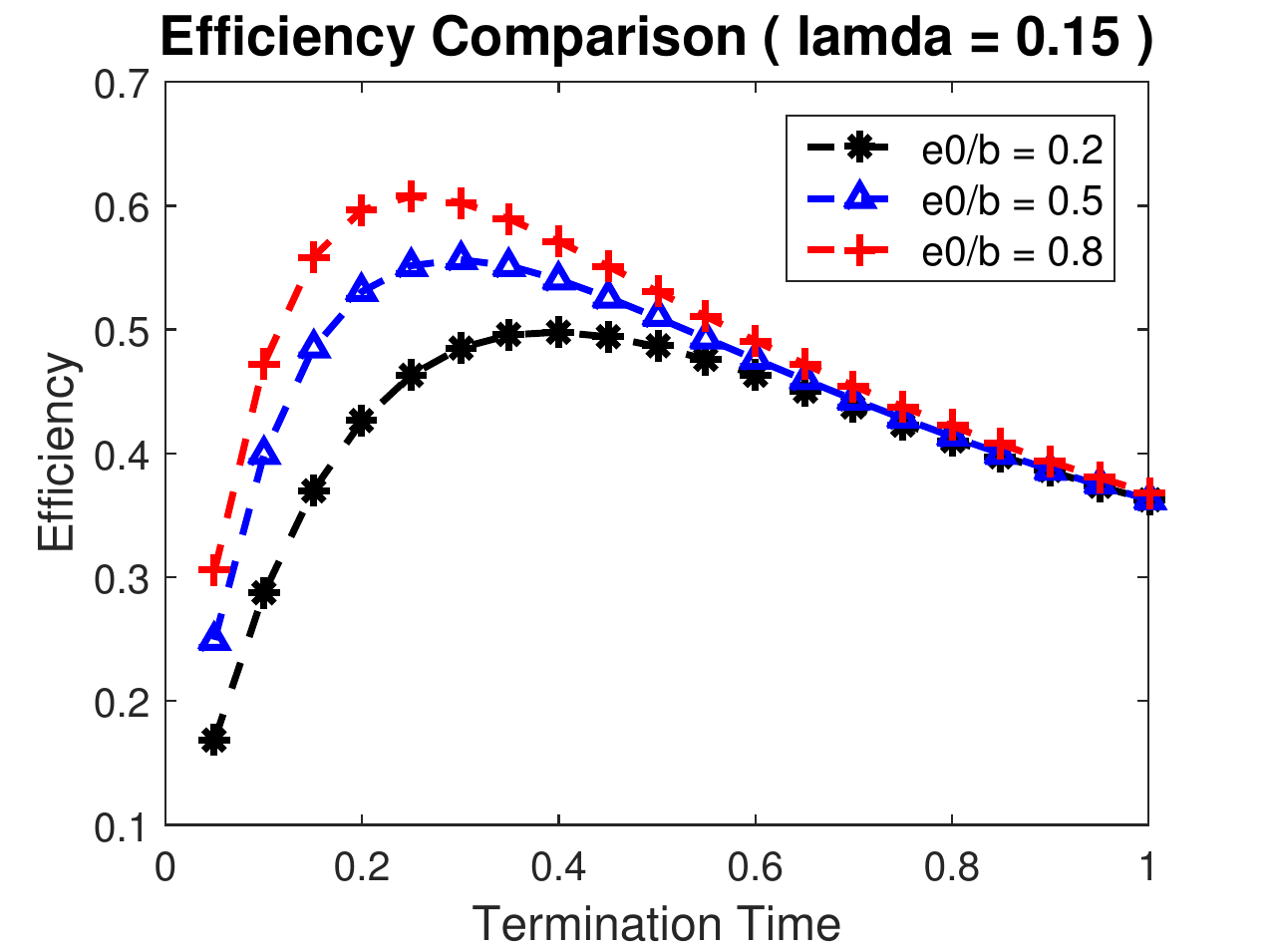}
\vspace{-0.5cm}
\caption{Efficiency of termination time scheme: open system with inverse weight function}
\label{fig:efficiency_tt_hyper_open}
\end{minipage}
\hspace{1ex}
\begin{minipage}[t]{0.33\linewidth}
\centering
\includegraphics[width=2.5in]{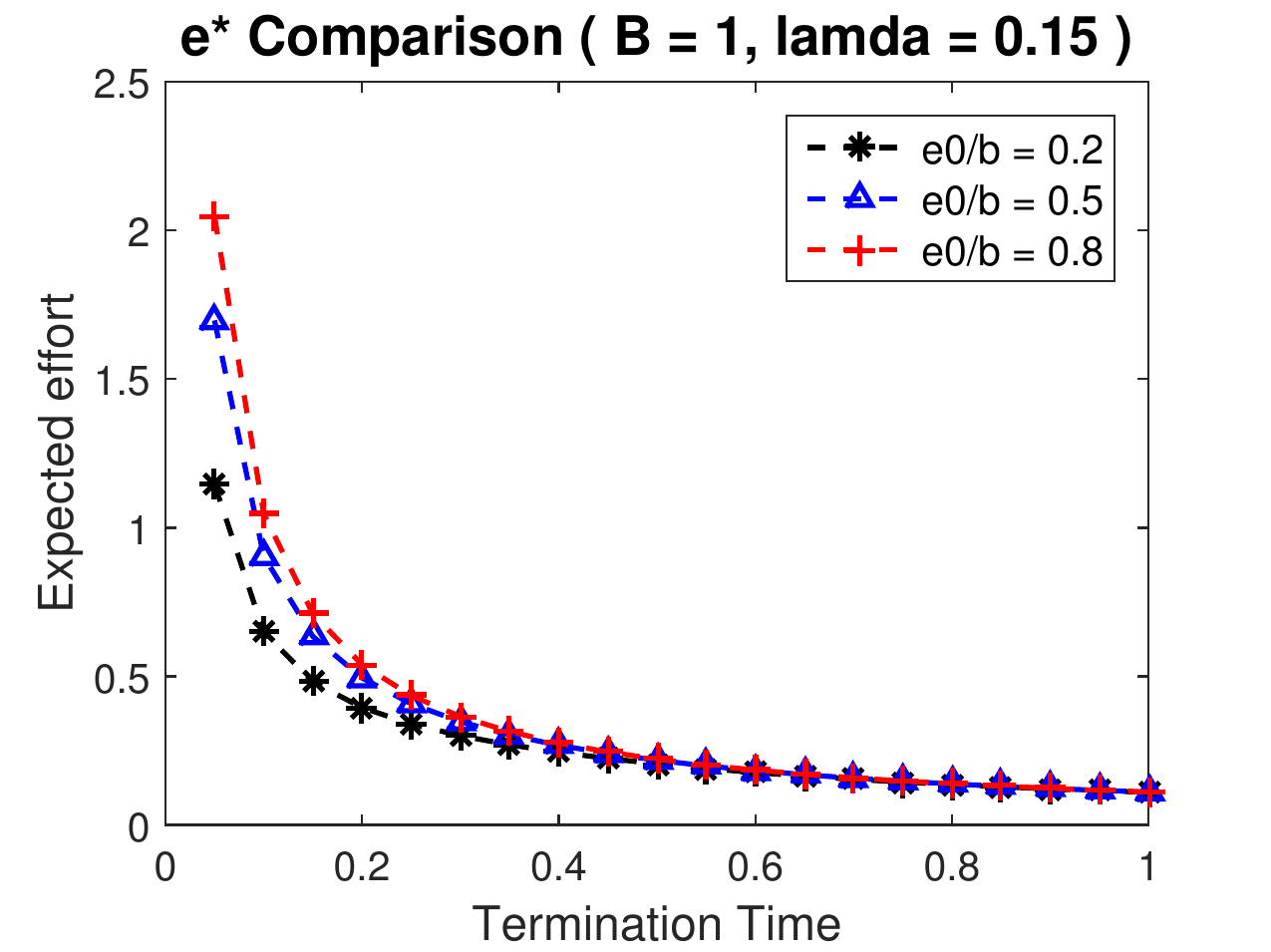}
\vspace{-0.5cm}
\caption{Efforts of a contributor of termination time scheme: open system}
\label{fig:efforts_tt_open}
\end{minipage}
\hspace{1ex}
\begin{minipage}[t]{0.33\linewidth}
\centering
\includegraphics[width=2.5in]{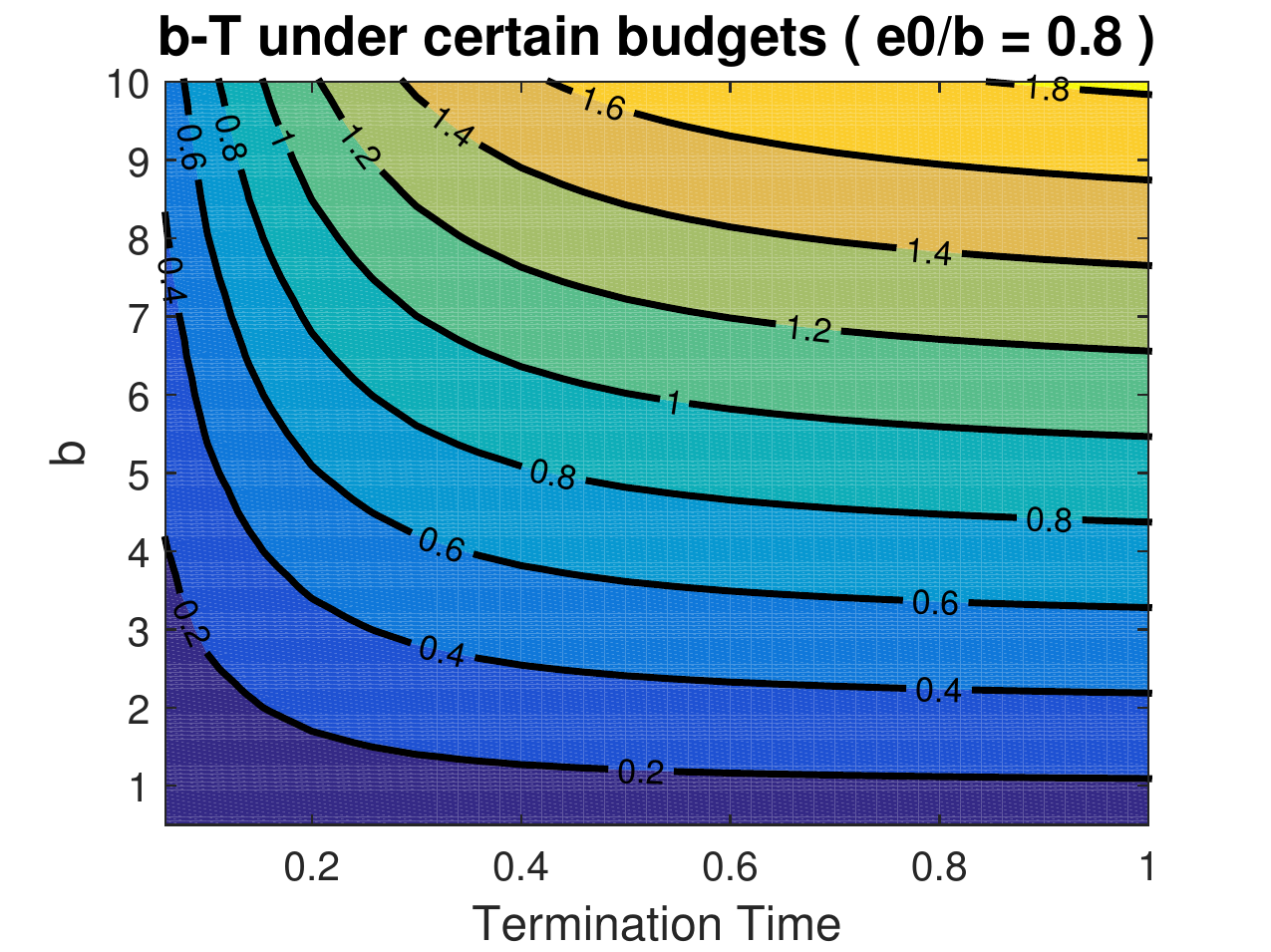}
\vspace{-0.5cm}
\caption{Contour lines of termination time scheme: open system}
\label{fig:contour_tt_open}
\end{minipage}
\vspace{-0.5cm}
\end{figure*}

Figure \ref{fig:efficiency_tt_step_open} and \ref{fig:efficiency_tt_hyper_open} show the efficiency of the requester 
with the step and the inverse weight functions. 
Considering the case of $e_0 = 0.8b$, the peak of the efficiency appears at 0.5 hour for the step weight function and 
appears at 0.3 hour for the inverse weight function. 
We have two observations regarding the efficiency of the requester. One is that the efficiency of termination time 
strategy is better than that of the earliest-$n$ strategy. 
The other is that the nature player is less influential in determining the efficiency of the requester. 
Compared with the case $e_0 = 0.2b$, the efficiency is improved by around 0.05 at the peaks for the case $e_0 = 0.5b$, 
and by around 0.1 at the peaks for the case $e_0 = 0.8b$. 
The main reason for both phenomena is that the earliest-$n$ strategy 
overly drives out the late joining contributors.

Figure \ref{fig:efforts_tt_open} illustrates the amount of efforts exerted by a contributor at different joining times. 
One can observe a similar ``cliff effect'' as that in the closed system. Figure \ref{fig:contour_tt_open} 
shows the contour lines of the pair $(b, T)$ where the expected expenditure equals to the budgets. 
For a fixed budget, choosing a larger termination time means a smaller maximum achievable budget, and vice versa. 
When $T$ (resp. $b$) is fixed, a larger budget $B$ requires the configuration of a larger $b$ (resp. $T$).

%\begin{figure}[!htb]
%\centering
%\includegraphics[width=2.8in]{./simu_figures2/efficiency_step_tt_open.jpg}
%\vspace{-0.3cm}
%\caption{Efficiency of termination time scheme: open system with step weight function}
%\label{fig:efficiency_tt_step_open}
%\end{figure}

\section{Related Work}
\label{sec:related} 

There is a growing literature on the optimal design of incentive mechanism in crowdsourcing and mobile crowdsensing applications. 
For instance, %\cite{EC09:Jain,WWW11:Ghosh,WWW12:Ghosh,EC11:Ghosh,WWW12:Ghosh2,SODA12:Chawla} 
\cite{EC09:Jain,WWW11:Ghosh} %,WWW12:Ghosh,SODA12:Chawla} 
targeted at incentivizing high quality user generated content in online question and answer forums. 
%A large body of work such as 
\cite{Infocom15:Wang,Mobihoc15:Cheung,NIPS11:Shah,Sigmetrics13:Shah} 
provided entertainment-like or monetary incentives for workers to label tasks in MTurk. 
The proliferation of mobile handheld devices triggers a variety of crowdsensing technologies. Pioneering sensing systems consist of 
NoiseTube \cite{noisetube} for noise monitoring, SignalGuru \cite{signalguru} for traffic monitoring, 
%CityExplorer \cite{cityexplorer}, TrMCD \cite{TrMCD} for 
%location based mobile games or trajectory discovery applications 
and some others \cite{Mobihoc15:Cheung}. 
%More 
%references can be found in a recent survey \cite{survey15:Zhang}.
We categorize the literature on incentive mechanism design into two groups according to their methodologies: 
one is the auction based approach and the other is the game theoretic approach. 

\textbf{Auction based approach.} DiPalantino and Vojnovic connected crowdsourcing to an all-pay auction model in 
which users selected among, and subsequently compete in, multiple contests offering various rewards \cite{EC09:DiPalantino}. 
%In \cite{SODA12:Chawla}, 
%authors presented an all-pay auction model where the crowdsourcer only benefits from the submission of the 
%highest quality. %An optimal auction design was developed with the consideration of the distribution of contestant skills and 
%the number of contestants. 
Singla and Krause \cite{WWW2013:Singla} presented a near optimal, posted-price mechanism 
for online budgeted procurement in crowdsourcing platforms. %, which was budget feasible, 
%near-optimal for the requester, incentive compatible for workers.  
Authors in \cite{Infocom15:Wang} modeled the interaction between workers and the crowdsourcer as a reverse auction for task labelling under 
strict budget constraint. 
Yang et al. designed an auction-based incentive mechanism for mobile phones to collect and analyze data \cite{Mobicom12:Yang}. 
%The proposed mechanism was computationally efficient, individually rational, profitable, and truthful. 
Jin et al. designed incentive mechanisms based on reverse combinatorial auctions that approximately maximized the social welfare 
with a guaranteed approximation ratio \cite{Mobihoc15:Jin}. Luo et al. studied an incentive mechanism based on all-pay auction with 
more realistic factors such as information asymmetry, population uncertainty and risk aversion \cite{Infocom:Luo}.%, TMC:Luo}. 

\textbf{Game theoretic approach.} 
Authors in \cite{EC09:Jain} studied the question of designing incentives for online Q\&A forums. 
%, in which the NEs of 
%three representative scoring rules were analyzed. 
Ghosh and McAfee modeled the economics of incentivizing high-quality user generated content 
using noncooperative game, and investigated the highest quality at the NE under an elimination mechanism \cite{WWW11:Ghosh}. 
%They further generalized their work to consider the entry cost of all the participants in \cite{WWW12:Ghosh}. 
%Ghosh and Hummel analyzed the equilibrium behavior of a rank-order mechanism which possessed a symmetric mixed 
%equilibrium strategy \cite{EC11:Ghosh}. They also extended their work to incorporate the abilities of contestants in \cite{WWW12:Ghosh2}. 
Yang et al. \cite{Mobicom12:Yang} studied a platform-centric incentive mechanism for maximizing the total effort from contributors in 
mobile crowdsensing that utilizes a Stackelberg game framework. 
%A market-based framework for participatory sensing was presented with a new metric comprising both the 
%information quality and timeliness of a specific real-time sensed quality \cite{TMC:Luo2}. 
Authors in \cite{Infocom:Luo2} presented an optimal 
Tullock contest model for crowdsensing with incomplete information. 

%for crowdsourcing where the cost of per-unit of effort is the private information to each contestant. 

The studies most relevant to ours are \cite{Mobicom12:Yang,Infocom:Luo2} that utilized Tullock-like contest models. 
In the pioneering work \cite{Mobicom12:Yang}, the crowdsensing was modeled by a complete information Tullock game. \cite{Infocom:Luo2}
investigated the same contest model with incomplete information that admited a BNE, and presented an 
optimal prize function to maximize the total efforts. 
Both studies do not consider the timeliness of contributions. % in which the crowdsourcer incentives early contestants to 
%contribute more effort. 
We rigorously show how an appropriate form of contest model can be chosen, and propose 
two novel Stackelberg Bayesian contest mechanisms to incentivize early joining contestants to 
contribute more efforts. 

\section{Conclusion}
\label{sec:conclusion}

In this paper, we address the incentive mechanism deign for timeliness 
sensitive crowdsensing using the Tullock contest framework. Each contributor is featured by 
his joining time, and the requester has a higher valuation toward an 
early contribution than a late one with the same effort. 
The core issue faced by the requester is how to offer different preferences 
to the contributors regarding their joining times, given the fixed budget. 
Two representative reward discrimination schemes, the earliest-$n$ and 
the termination time schemes, are proposed. 
We develop a Stackelberg Bayesian game framework to capture the competition of the contributors without the
complete information, and to characterize the orchestration of the competition 
by the requester to maximize his crowdsensing efficiency. 
We prove the existence and uniqueness of  BNE, and present the optimal parameter configuration 
for each scheme. Our game framework is powerful in its applicability to both the closed crowdsensing system with a fixed 
number of players and the open system whose arrival of contributors follows a stochastic process.  
Numerical simulations validate the effectiveness of the proposed incentive mechanisms.

%In this paper, we present the first attempt to design incentive mechanisms for timeliness sensitive crowdsensing using the Tullock 
%contest framework. Each contributor is featured by his joining time, and the requester has a higher valuation toward an 
%early contribution than a late one with the same effort. 
%The core issue faced by the requester is how to offer different preferences to the contributors regarding their joining times, 
%given the fixed budget. 
%We develop a Stackelberg Bayesian game framework to capture the competition of the contributors without the
%complete information of the joining times, and to characterize the orchestration of the competition 
%by the requester to achieve the maximum crowdsensing efficiency. 
%Two representative reward discrimination schemes, the earliest-$n$ and 
%the termination time schemes, are proposed where the joining time of a contributor is private knowledge. 
%We prove the existence and uniqueness of Bayesian NE in the competition, and present the optimal parameter configuration 
%for each scheme. Our game framework is powerful in that it is applicable to both the closed crowdsensing system with a finite 
%number of players and the open system whose arrival of contributors follows a stochastic process.  
%Numerical simulations validate the effectiveness of the proposed incentive mechanisms. 

\bibliographystyle{abbrv}
%\bibliography{sigproc}  % sigproc.bib is the name of the Bibliography in this case

%\input{sec_appendix}

%\newpage
%.
\newpage

\section*{Appendix-I: Proofs and Derivations}
\label{sec:appendix1}

\subsection*{\textbf{A. Proof of Lemma \ref{lemma:n_contributors}}}
\noindent\textbf{Proof:} i) It is easy to prove by contradiction. Suppose $0< e_i^*<e_j^*$ if $b_i\geq b_j$. 
According to the expression in \eqref{eq:n_ne_conditions}, the left hand for $i$ is greater than that for $j$. 
The right hand for $i$ is the same as that for $j$, which equality does not hold. Thus, if $e_i^* > e_j^* >0$, there must have $b_i > b_j$. 

ii) When $e_0$ is greater than $b_i$, the left hand of the expression in \eqref{eq:n_ne_conditions} is always less than the right hand. 
Hence, $e_i^*$ is 0 all the time. For the same reason, the number of participating contributors decreases if $e_0$ increases. \done

\subsection*{B. Optimal Reward Discrimination}

With the complete information on $t_i$ for all $i$, the requester is able to steer the configuration of $b(t_i)$ (i.e. $b_i$ for 
brevity) to maximize his efficiency (equivalent to utility maximization when there is no uncertainty): 
\begin{eqnarray}
&\max & \sum\nolimits_{i=1}^{N} w_i e_i^*  \nonumber\\
&\textrm{s.t.} &  \sum\nolimits_{i=1}^{N}r_i([\mathbf{e}^*_{i},\mathbf{t}^*_{i}]) = B.
\end{eqnarray}
The optimal design falls in the scope of nonlinear programming. The individual rationality rule makes the optimization problem 
not differentiable anywhere. Solving the optimal reward vector $\mathbf{b}^*$ involves complicated operations on some 
implicit functions. For clarity, we hereby examine the optimal incentive mechanism in the case $e_0 = 0$. When $e_0$ is 
nonzero, all the procedures remain the same, and are omitted henceforth. 

The optimization problem with $e_0 = 0$ is simplified as
\begin{eqnarray}
&\max & \sum\nolimits_{i=1}^{n^*} w_i \big(\frac{n^*-1}{\sum_{j=1}^{n^*}\frac{1}{b_j}} - \frac{1}{b_i}(\frac{n^*-1}{\sum_{j=1}^{n^*}\frac{1}{b_j}})^2\big) \nonumber\\
&\textrm{s.t.} &  \sum\nolimits_{i=1}^{n^*} b_i - \frac{n^*(n^*-1)}{\sum_{i=1}^{n^*}\frac{1}{b_i}}= B, \label{eq:equality_constraint}
\end{eqnarray}
where $n^*$ is the number of players with nonzero efforts at the NE. 
Since $n^*$ is an inter coupled with the NE, this optimization problem is inherently non-differentiable on $b_i$. 
To bypass this difficulty, we treat each $n^*$ as a constant in the range $[1, N]$, and compute 
the optimal reward discrimination vector $\{b_1^*, b_2^*, \cdots, b_{n^*}^*\}$. Then, the vector $\{b_1^*, b_2^*, \cdots, b_{n^*}^*\}$ 
is feed into the original game model so as to check the individual rationality condition. If individual rationality is satisfied, 
this vector maximizes the requester's utility owing to the uniqueness of the NE.  

We use the Lagrangian method to solve the constrained optimization problem. Define $\mathcal{L}(\mathbf{b}, \lambda)$ as
\begin{eqnarray}
\mathcal{L}(\mathbf{b}, \lambda) \!\!\!&=&\!\!\! \sum\nolimits_{i=1}^{n^*} w_i \big(\frac{n^*-1}{\sum_{j=1}^{n^*}\frac{1}{b_j}} - \frac{1}{b_i}(\frac{n^*-1}{\sum_{j=1}^{n^*}\frac{1}{b_j}})^2\big) \nonumber\\
\!\!\!& &\!\!\!+ \lambda\big( \sum\nolimits_{i=1}^{n^*} b_i - \frac{n^*(n^*-1)}{\sum_{i=1}^{n^*}\frac{1}{b_i}} - B\big)
\label{eq:lagrangian}
\end{eqnarray}
where $\lambda$ is the Lagrangian multiplier. The optimal $\mathbf{b}^*$ maximizes $\mathcal{L}(\mathbf{b}, \lambda)$ while 
satisfying the equality constraint in Eq.\eqref{eq:equality_constraint}.
The standard approach of searching $\mathbf{b}^*$ is to compute the derivatives of $\mathcal{L}(\mathbf{b}, \lambda)$ on each 
$b_i$ and let $\frac{d\mathcal{L}(\mathbf{b}, \lambda)}{d b_i} =0$ for each $i$, $\forall \; 1\leq i \leq n^*$. 
Then, we express each $b_i$ as a function of $\lambda$, submit all the $b_i(\lambda)$ to Eq.\eqref{eq:equality_constraint}, and 
compute $\lambda^*$ and $\mathbf{b}^*$ subsequently. 
Due to the complexity of the implicit function $b_i(\lambda)$, a practical approach is to use a binary search of $\lambda^*$. 
The iteration stops when the left-hand of Eq.\eqref{eq:equality_constraint} is in the vicinity of $B$. 

\subsection*{\textbf{C. Proof of Theorem \ref{lemma:E_U_earliest_n}}}
\noindent\textbf{Proof:} We apply results in \cite{Wasser} to prove the existence of a BNE, and results in \cite{Cornes} 
to prove the uniqueness. For consistency with \cite{Wasser} and \cite{Cornes}, it is useful to perform the change of 
variables $c_i := -\frac{1}{b(t_i)}$. Then, contributor $i$'s payoff is equivalent to
\begin{eqnarray}
\pi_i(\mathbf{e}, c_i) = b(t_i)\big(\frac{e_i}{e_0 + e_i +E_{-i}} + c_i e_i\big)
\end{eqnarray}
where $e_i\in [0, b]$ and $c_i\in (-\infty, -\frac{1}{b}]$. Consider a constant $c\in (-\infty, -\frac{1}{b}]$, 
the probability distribution of $c_i$ is obtained by 
\begin{eqnarray}
\mathbb{P}(c_i\leq c) \!\!\!&=&\!\!\! \mathbb{P}(\frac{-1}{b(t_i)}\leq  c) = \mathbb{P}(b(t_i) \leq -\frac{1}{c}) \nonumber\\
\!\!\!&=&\!\!\! \mathbb{P}\big(t_i \geq b^{-1}(-\frac{1}{c})\big) = 1 - F\big(b^{-1}(-\frac{1}{c})\big), \nonumber
\end{eqnarray}
where $b^{-1}(\cdot)$ is an inverse function for $b(t_i) = bP_{i\in\{n\}}$. Hence, each $c_i$ is drawn from the 
distribution $\hat{F}_i(c_i) = 1 - F_i(b^{-1}(-\frac{1}{c_i}))$. 
Given a strategy $x_j:[-\infty, -\frac{1}{b}] \rightarrow [0, b]$ for each $j\neq i$, contributor $i$'s expected payoff 
amounts to 
\begin{eqnarray}
\mathbb{E}[\pi_i(\mathbf{e}, c_i)] = b(t_i) \Big(\int_{-\infty}^{-\frac{1}{b}}\cdots\int_{-\infty}^{-\frac{1}{b}}
\frac{e_i}{e_0+e_i+E_{-i}}\nonumber\\
\prod_{j=1, \neq i}^Nd\hat{F}_j(c_j) + c_ix_i\Big). \nonumber
\end{eqnarray}
We take the partial derivative of $\mathbb{E}[\pi_i(\mathbf{e}, c_i)]$ over $e_i$ and $c_i$ so to have 
$\frac{d^2\mathbb{E}[\pi_i(\mathbf{e}, c_i)]}{de_idc_i} = 1 >0$. 
According to \cite{Wasser}, the \emph{single crossing condition} for an incomplete
information game is satisfied. In addition, $\mathbb{E}[\pi_i(\mathbf{e}, c_i)]$ is continuous as long as $e_0>0$, 
which satisfies the requirements of uniqueness.  
Hence, there exists a pure strategy Bayesian Nash equilibrium. 

The uniqueness of pure strategy BNE can be directly proved because $\mathbb{E}[\pi_i(\mathbf{e}, c_i)]$ satisfies 
the conditions from U1 to U3, and from D1 to D2 mention in \cite{Wasser}. \done

\subsection*{\textbf{D. Proof of Theorem \ref{theorem:bounds}}}

\noindent\textbf{Proof:} i) It is direct to see that all the contributors may participate in the contest for the special case $n=N$. 
The left-hand side (LHS) of Eq.\eqref{eq:bayesian_NE_condition} is no larger than 1 if $b\leq e_0$, and is greater than 1 otherwise. 
Hence, either all the contributors do not exert effort, or exert the same amount of effort at the NE. 

ii) If $n$ is less than $N$, b(t) is infinitely close to 0 as $t$ approaches infinity. Then, $b(t)$ is in the range $(0, b]$. 

The left-hand side of Eq.\eqref{eq:bayesian_NE_condition} is strictly decreasing in $e_i^*$. 
Hence, if $e_i^*(\hat{t})$ is greater than 0 for some joining time $\hat{t}$, then $e_i^*(t)  \geq e_i^*(\hat{t})$ for all 
$t < \hat{t}$ because $b(t)$ is a strictly decreasing function of $t$. Therefore, a constant $\bar{t}$ must exist such that 
$e_i^*(t) =0$ for $t\geq \bar{t}$ while $e_i^*(t)$ is positive and strictly decreasing for $t>\bar{t}$.
Note that the LHS of Eq.\eqref{eq:bayesian_NE_condition} is maximized when $E_{-i}^*$ is 0. 
This implies $b(\bar{t}) \geq e_0$, or equivalently $t\leq b^{-1}(e_0)$.

iii) Assuming $t_i < \bar{t}$. We multiply both sides of Eq.\eqref{eq:bayesian_NE_condition} with $e_i^*(t)$ and obtain
\begin{eqnarray}
\mathbb{E}\Big[\frac{e_0 + E_{-i}^*}{(e_0 + e_i^*(t_i) + E_{-i}^*)^2} \Big] e_i^*(t_i)b(t_i) = e_i^*(t_i). 
\label{eq:proof_bound_earliestn_no1}
\end{eqnarray}
Denote a random variable $\Omega$ with $E[\Omega] = e_i^*(t_i)/b(t_i)$. The above equation yields
\begin{eqnarray}
\Omega = \frac{e_i^*(t_i)}{e_0 + e_i^*(t_i) + E_{-i}^*} \cdot\big(1 - \frac{e_i^*(t_i)}{e_0 + e_i^*(t_i) + E_{-i}^*} \big) \leq \frac{1}{4}. \nonumber
\end{eqnarray}
Thus, we obtain 
\begin{eqnarray}
e_i^*(t_i) \leq \frac{1}{4}b(t_i). \nonumber
\end{eqnarray}
The equality holds at $e_i^*(t_i) = e_0 + E_{-i}^*$ which implies $e_0\leq \frac{1}{4}b(t_i)$. 

When $e_0$ is larger than $\frac{1}{4}b(t_i)$, the upper bound of $e_i^*(t_i)$ can be tighter. In what follows, we present 
the upper bound of $e_i^*(t_i)$ for $e_0\geq \frac{1}{4}b(t_i)$. We take the derivative of $\Omega$ over $E_{-i}^*$ 
and obtain
\begin{eqnarray}
\frac{d\Omega}{dE_{-i}^*} = \frac{e_i^*(t_i)\big(e_i^*(t_i) - e_0 - E_{-i}^*\big)}{(e_0+e_i^*(t_i)+E_{-i}^*)^3} \leq 0. \nonumber
\end{eqnarray}
Thus, the maximum $\Omega$ is obtained on the curve with $E_{-i}^* = 0$. 
Under this situation, $\Omega$ is simplified as 
\begin{eqnarray}
\Omega \big |_{E_{-i}^*=0} = \frac{e_0e_i^*(t_i)}{(e_0+e_i^*(t_i))^2}. \nonumber
\end{eqnarray}
We further differentiate $\Omega|_{E_{-i}^*=0}$ over $e_i^*(t_i)$ and obtain 
\begin{eqnarray}
\frac{d\Omega}{de_{i}^*(t_i)} \Big |_{E_{-i}^*=0}= \frac{e_0(e_0-e_i^*(t_i))}{(e_0+e_i^*(t_i))^3} \geq  0. \nonumber
\end{eqnarray}
The maximum $\Omega$ is obtained at $e_i^*(t_i)$ is chosen to be $\frac{1}{4}b(t_i)$. Hence, the following inequality holds
\begin{eqnarray}
\Omega \leq \frac{\frac{1}{4}b(t_i)e_0}{(e_0+\frac{1}{4}b(t_i))^2}. \nonumber
\end{eqnarray}
According to Eq.\eqref{eq:proof_bound_earliestn_no1}, there has
\begin{eqnarray}
e_i^*(t_i) \leq \frac{\frac{1}{4}b(t_i)^2 e_0}{(e_0+\frac{1}{4}b(t_i))^2}. \nonumber
\end{eqnarray}
This concludes the proof. \done

\subsection*{\textbf{E. Proof of Theorem \ref{theorem:symmetric_BNE}}}

\noindent\textbf{Proof:} To prove the existence and uniqueness of BNE, we need to show the unique fixed point solution 
to the set of best response functions. 
By taking the derivative of $\mathbb{E}[\pi_i]$ over $e_i$, we obtain
\begin{eqnarray}
\frac{d\mathbb{E}[\pi_i]}{de_i} = \sum_{k=0}^{N-1}\mathbb{P}(k,N{-}1) \cdot \frac{b\sum\nolimits_{j=0,\neq i}^{k+1}e_j}{(\sum\nolimits_{j=0}^{k+1}e_j)^2} - 1 \leq 0, \;\;\; \forall i.
\label{eq:uniqueness_tts_no1}
\end{eqnarray}
The equality holds for the circumstance $e_i \geq 0$. Since the requester does not discriminate the contributors 
who join before $T$, all of them exerts the same amount of effort according to the above optimality condition. 

Next, we will show the uniqueness of this symmetric equilibrium. In Eq.\eqref{eq:uniqueness_tts_no1}, when the equality 
holds, the right-hand is a strictly decreasing function of $e_i$. Then, $e_i$ is unique to enable $\frac{d\mathbb{E}[\pi_i]}{de_i} =0$ 
for all $i$ with $t_i \leq T$. Thus, there exists a unique BNE in the crowdsourcing contest with termination time strategy. \done

\begin{comment}
\subsection*{Proof}

Differentiating the above equation over $e_0$, there exists
\begin{eqnarray}
\frac{de^*}{de_0} = \frac{-4e^*-2e_0+vb(e^*)^{v-1}}{8e^*+4e_0+vb(1-v)(e^*)^{v-2}e_0-v^2b(e^*)^{v-1}}.
\label{eq:symmetric_ne_no2}
\end{eqnarray}
Denote by $e^{*\dag}$ as the solution to make $\frac{de^*}{de_0} = 0$ if it exists. For $v\in(0,1)$, 
$e^{*\dag}$ is actually the unique solution to the expression $-4e^{*\dag}-2e_0+vb(e^{*\dag})^{v-1}=0$.
When $e^*$ is less than $e^{*\dag}$, $e^*$ is a strictly increasing function of $e_0$; when $e^*$ is above 
$e^{*\dag}$, it is a strictly decreasing function of $e_0$. 
When $v$ is 1, the derivative $\frac{de^*}{de_0}$ may not achieve 0. If $e_0\geq \frac{b}{2}$, there has 
$e^{*\dag} = \frac{2e_0-b}{4}$. Otherwise, $\frac{de^*}{de_0}$ is always negative. 
Therefore, as $e_0$ increases, $e^*$ decreases on the contrary. 
\end{comment}

\subsection*{\textbf{F. Derivation of Equation \eqref{eq:expected_eff_termination}}}

Equation \eqref{eq:expected_eff_termination} is obtained by the following steps:
\begin{eqnarray}
\mathbb{E}[\mathcal{E}] \!\!\!&=&\!\!\! \sum_{k=1}^{N}\mathbb{P}(k, N)  \frac{\mathbb{E}[\sum_{l=1}^{k}w(\tilde{t}_l)]e_i^*(e_0+ke_i^*)}{bke_i^*} \nonumber\\
\!\!\!&=&\!\!\! \sum_{k=1}^{N}\mathbb{P}(k, N)  \frac{\sum_{l=1}^{k}\mathbb{E}[w(\tilde{t}_l)](e_0+ke_i^*)}{bk} \nonumber\\
\!\!\!&=&\!\!\! \frac{1}{bp}\sum_{k=1}^{N}\mathbb{P}(k, N) (e_0{+}ke_i^*)\int_0^T \!\!\!w(t) f(t)dt \nonumber\\
\!\!\!&=&\!\!\! \big(\frac{e_0}{bp}\sum_{k=1}^{N}\mathbb{P}(k, N) +  \frac{e_i^*}{bp}\sum_{k=1}^{N}k\mathbb{P}(k, N) \big)\int_0^T \!\!\!w(t) f(t)dt \nonumber\\
\!\!\!&=&\!\!\! \big(\frac{e_0}{bp}(1-(1-p)^N) +  \frac{Ne_i^*}{b}\big)\int_0^T \!\!\!w(t) f(t)dt .
\end{eqnarray}

\subsection*{\textbf{G. Proof of Lemma \ref{lemma:open_ttime}}}
\noindent \textbf{Proof:} We prove this lemma by mathematical induction. 

First of all, we claim that the integral part of Eq. \eqref{eq:conditional_result_ttime} is 
\begin{eqnarray}
&&\!\!\!\!\!\!\!\int_{0}^{T}\int_{0}^{s_{m}}\cdots\int_{0}^{s_2}(\sum\nolimits_{k=1}^{m}w(s_k)) \;\;ds_1 \cdots ds_m \nonumber\\
&=& \frac{T^{m{-}1}}{(m-1)!} \int_0^Tw(s_m)ds_m. 
\label{eq:proof_induction1}
\end{eqnarray}

i) When $m$ equals to 2, the integral part of $E[\mathcal{E} | N(T) = m]$ is 
\begin{eqnarray}
&&\int_{0}^{T}\int_{0}^{s_{2}}(w(s_1)+w(s_2)) \;\;ds_1 ds_2 \nonumber\\
&=& \int_{0}^{T}\int_{0}^{s_{2}}w(s_1)\;\;ds_1 ds_2 + \int_{0}^{T}\int_{0}^{s_{2}}w(s_2) \;\;ds_1 ds_2 \nonumber\\
&=& \int_{0}^{T}\int_{0}^{s_{2}}w(s_1)\;\;ds_1 ds_2 + \int_{0}^{T}s_2 w(s_2) \;\;ds_2\nonumber\\
&=& \int_{0}^{T}\int_{0}^{s_{2}}w(s_1)\;\;ds_1 ds_2 + \big[s_2\int_0^{s_2}w(x)dx\big]_{0}^{T} \nonumber\\
&& - \int_0^{T}\int_0^{y}w(x)dxdy\nonumber\\
&=& T\int_0^{T}w(x)dx.
\end{eqnarray}

ii) We assume that Eq. \eqref{eq:simplified_result_ttime} holds for $m>2$.

iii) For $m+1$, the integral part of Eq. \eqref{eq:conditional_result_ttime} is 
\begin{eqnarray}
&&\!\!\!\!\!\!\!\int_{0}^{T}\int_{0}^{s_{m{+}1}}\cdots\int_{0}^{s_2}(\sum\nolimits_{k=0}^{m{+}1}w(s_k)) \;\;
ds_1 ds_2\cdots s_mds_{m{+}1} \nonumber\\
&=& \!\!\int_{0}^{T}\frac{(s_{m{+}1})^{m}}{m!} w(s_{m{+}1})ds_{m{+}1}+ \nonumber\\
&& + \int_{0}^{T} \frac{(s_{m{+}1})^{m{-}1}}{(m-1)!}
\int_{0}^{s_{m{+}1}}w(s_m)ds_m ds_{m{+}1} \nonumber\\
&=& \big[\frac{(s_{m{+}1})^{m}}{m!}\!\!\int_0^{s_{m{+}1}}\!\!\!\!\!w(s_m) ds_m\big]_0^T \nonumber\\
&=& \frac{T^m}{m!}\int_0^Tw(x)dx. 
\end{eqnarray}
Therefore, our claim is proved by mathematical induction.
Submitting Eq. \eqref{eq:proof_induction1} to Eq. \eqref{eq:conditional_result_ttime}, we obtain
\begin{eqnarray}
E[\mathcal{E} | N(T) = m] = \frac{e_0+me^*}{bT} \int_{0}^{T} w(x)dx,
\label{eq:proof_induction5}
\end{eqnarray}
which concludes the proof. \done

\newpage
.
\newpage

\section*{Appendix-II: Justification of Contest Success Function}
%\section{Paradoxes of Timeliness Sensitive Incentive Mechanism: Two-Contributor Case}
\label{sec:paradox}

Define $\mathcal{E}$ as the requester's efficiency that is the utility brought by per-unit payment, conditioned on the fixed budget $B$ allocated to 
the contributors, 
i.e. $\mathcal{E} = \frac{\mathcal{U}}{B}\big|_{R=B}$. 
The requester is able to optimize $\mathcal{E}|_{R=B}$ by properly discriminating the contributors with different joining times. 
An auxiliary metric is named as ``discrimination gain'' denoted by 
$\mathcal{G}= \mathcal{E}_{d}/\mathcal{E}_{nd}$ where 
$\mathcal{E}_{nd}$ and  $\mathcal{E}_{d}$ are the requester's efficiencies without and with discrimination.

\subsection*{A. Generalized Tullock Contest Success Function}

Let $\mathbf{a} = \{a_i\}_{i=1}^N$,  $\mathbf{b} = \{b_i\}_{i=1}^N$ and $\mathbf{v} = \{v_i\}_{i=1}^N$ be the set of non-negative 
coefficients that are related to $\{t_i\}_{i=1}^{N}$. For notational simplicity, we do not display $\{t_i\}_{i=1}^{N}$ unless 
necessary. We define $\mathbf{e}_{-i} =  \{e_j\}_{j=1,\neq i}^N$ and 
$\mathbf{t}_{-i} =  \{t_j\}_{j=1,\neq i}^N$. 
Let $r_i([e_i, t_i], [\mathbf{e}_{-i}, \mathbf{t}_{-i}])$ be the reward obtained by the $i^{th}$ contributor, given the sets of 
joining time $\mathbf{t}$ and effort level $\mathbf{e}$. There has
\begin{eqnarray}
r_i([e_i, t_i], [\mathbf{e}_{-i}, \mathbf{t}_{-i}]) = \frac{a_i(e_i)^{v_i}}{e_0 {+} a_i(e_i)^{v_i}{+}\sum_{j{=}1,\neq i}^{N}a_j(e_j)^{v_j}}b_i,
\label{eq:basicincentive}
\end{eqnarray}
where $v_i$ is in the range $(0, 1]$ for all $i$. According to contest theory \cite{contest_survey}, this reward sharing rule coincides 
all the five axioms for contest success functions (CSFs). 
The physical interpretation is that a contributor acquires a 
larger reward if he contributes a higher effort, and a smaller reward if any of his opponent spends more effort. 
The rate of increased reward is shrinking as this contributor exerts more and more effort. 
The constant $e_0$ is introduced by the requester that if it is positive, the reward will not be  completely assigned to the
contributors. A larger $e_0$ means that the requester retains a higher percentage of reward. Actually, a non-zero 
$e_0$ is equivalent to introducing a \emph{NATURE} player, where $e_0$ influences the crowdsensing effort of 
all the players. Throughout this work, if not mentioned explicitly, 
we suppose $e_0$ to be 0, i.e. the whole reward will be shared among the contributors. 
The vectors of coefficients, $\mathbf{a}$, $\mathbf{b}$ and $\mathbf{v}$, are introduced to offer discriminations 
to different contributors. They are placed at different ``positions'' in the contest success function so to modify 
crowdsensing contest in different ways. 
\begin{itemize}

\item \textbf{$\{a_i\}_{i=1}^N$ - Weight Discrimination.}  Given identical $\mathbf{v}$ and $\mathbf{b}$, for the 
same effort $e_i = e_j$, contributor $i$ obtains a higher reward than contributor $j$, i.e. $r_i > r_j$ if $a_i > a_j$.

\item \textbf{$\{v_i\}_{i=1}^N$ - Exponent Discrimination.} Given identical $\mathbf{a}$, $\mathbf{b}$ and 
$e_i = e_j$, we endow the early contribution a higher exponent, that is, $v_i > v_j$ if $i < j$.

\item \textbf{$\{b_i\}_{i=1}^N$ -  Reward Discrimination.} The maximum possible rewards obtained by different contributors 
can be distinct. Given identical $\mathbf{a}$, $\mathbf{v}$ and $e_i = e_j$, there exists $r_i > r_j$ if $b_i > b_j$.
\end{itemize}
The weight discrimination changes the priority and the exponent discrimination changes the elasticity of effort in bringing 
reward. Denote by $R$ the total reward paid to the contributors that has $R = \sum\nolimits_{i=1}^{N} r_i([\mathbf{e}, \mathbf{t}])$. 
In the incentive mechanism design, the requester needs to make $R$ equal to $B$. 
Once $\mathbf{b}$ are homogeneous and $e_0$ is zero, these two discrimination schemes do not alter 
the total reward paid by the requester. 
However, the reward discrimination scheme provides different penalties to the maximum achievable reward, 
henceforth changing the aggregate reward paid by the requester. 
Our main purpose is to investigate 
whether these discriminate schemes can incentive more timely effort from the contributors. To this goal, when we 
manipulate one vector of coefficients, the other two vectors of coefficients are supposed to be identical, thus effectively 
simplifying our analysis.

In what follows, we analyze the NE properties of Game \textbf{G1}. 
We reveal a couple of paradoxes in the discrimination schemes and provide important insights 
on incentive mechanism design. 

\emph{Challenges:} A deep understanding of this incentive mechanism is usually intractable 
for two reasons: firstly, the Nash equilibrium of a Tullock contest does not admit a close-form solution in general; 
secondly, all the parameters are coupled in a complicated way such that there does not yield any 
insight without proper separation of them. 

To overcome these difficulties, we simplify our analysis in two aspects. One is to begin with the  
scenario of two contributors, and then to generalize our observations to an arbitrary population. 
The other is to assume that only one set of parameters in $\{\mathbf{a}, \mathbf{b}, \mathbf{v}\}$ are heterogeneous 
among the contributors when each of the discrimination rules is studied. 
With above simplifications, we can gain important insights on what form an appropriate incentive mechanism should take on. 
Our analysis starts from a simple scenario with $N=2$ if not mentioned explicitly. 

%In this section, we let $N$ be 2 if not mentioned explicitly. 
%It is very difficult to analyze the performance of incentive mechanism for two reasons: 

\subsection*{B. Nash Equilibrium with Weight Discrimination}
\medskip
\boxed{
	\begin{minipage}{3.1in}
		{\ensuremath{\mbox{\sc  Weight Discrimination:}}}
		$b_1 = b_2 = b$, $v_1=v_2=v$, $e_0 = 0$, $w_1\geq w_2$, $a_1 = a$ and $a_2 = 1$.
	\end{minipage}
}
\medskip

For the special case $N=2$, we can solve the unique NE in explicit form. We suppose that 
the first contributor joins the contest earlier than the second. 
The extent of preference to the first contributor is reflected by the coefficient $a$. 
By letting the derivatives $\frac{d\pi_1}{d e_1}$ and  $\frac{d\pi_2}{d e_2}$ be 0, we obtain 
two equations that yield the following solution
\begin{eqnarray}
e_1^* = e_2^* = \frac{abv}{(1+a)^2} \;. 
\end{eqnarray}
The derivatives satisfy $\frac{de_i^*}{d a} < 0$ for $a >1$ and $\frac{de_i^*}{d a} > 0$ for $a <1$, 
which implies that the maximum $e_i^*$ is obtained at $a = 1$. 
The efficiency of the requester is given by $\mathcal{E} = \frac{abv(w_1{+}w_2)}{(1+a)^2}$ whose maximum value is reached at $a=1$. 
This yields the following paradox.

\noindent\textbf{Paradox 1:} The weight discrimination scheme adopted by the requester will diminish his efficiency on the contrary.

%For the special case $N=2$, we can solve the unique NE in explicit form. Recall that $e_0$ is 0 if not mentioned explicitly. 
%We let $b_i = 1$ and 
%$v_i = v$ for $i=1, 2$. Player 1 joins the contest earlier than player 2. Then, we let $a_1 = a \geq 1$ 
%and $a_2 = 1$. The extent of preference to player 1 is reflected by the coefficient $a$. 

%\begin{eqnarray}
%\begin{cases}
%e_1^* = \\
%e_2^* = 
%\end{cases}
%\end{eqnarray}

\subsection*{C. Nash Equilibrium with Exponent Discrimination}

\medskip
\boxed{
	\begin{minipage}{3.1in}
		{\ensuremath{\mbox{\sc  Exponent Discrimination:}}}
		$b_1 = b_2 = b$, $a_1=a_2=a$, $e_0 = 0$, $v_1 \geq v_2 $ and $w_1\geq w_2$.
	\end{minipage}
}
\medskip

Following the same approach, it is easy to yield $e_1^*: e_2^* = v_1:v_2$. 
The unique NE of \textbf{G1} is a feasible solution to the following equation
\begin{eqnarray}
\big(\frac{v_1}{v_2}\big)^{2v_1}(e_2^*)^{(v_1-v_2+1)} + (e_2^*)^{(v_2-v_1+1)} + 2\big(\frac{v_1}{v_2}\big)^{v_1}e_2^* \nonumber\\
= bv_2\big(\frac{v_1}{v_2}\big)^{v_1}. 
\label{eq:2palyers_ed}
\end{eqnarray} 
Let us take a look at a special case with $b=1$. Any reasonable strategy of a contributor that leads to a positive
payoff should satisfy $e_1 \leq 1$ and $e_2 \leq 1$. Suppose that the both contributors exert the same mount of effort.
Due to $v_1\geq v_2$, there has $e_1^{v_1} \leq e_2^{v_1}$. 
The late contributor obtains a higher share of the reward, and hence a larger payoff. 
On the contrary, if the exponents satisfy $v_1 < v_2$, the early contributor acquires a smaller share of reward with 
$e_1 = e_2 \geq 1$, and then a lower payoff.  This is in conflict with the original discrimination purpose to 
offer more benefits to the early contributor. 

The exponent discrimination scheme may degrade the requester's efficiency. Formally, the following 
lemma holds.
\begin{lemma} 
In the exponent discrimination with $v_i {>} v_j$, the NE of game \textbf{G1} satisfies: (1) $e_i^* > e_j^*$; 
(2) if all the parameters are fixed except $v_j$, increasing $v_j$ also improves the effort of both contributors. 
\label{lemma:exponent}
\end{lemma}

\noindent\textbf{Paradox 2:} If the requester favors the earlier contributor by configuring 
a larger exponent, the payoff of the early one could be worse than that of the late one. 
A large difference between two exponents (i.e. sufficient discrimination) leads to the reduction of 
effort in the both contributors.

\subsection*{D. Nash Equilibrium with Reward Discrimination}

\medskip
\boxed{
	\begin{minipage}{3.1in}
		{\ensuremath{\mbox{\sc  Reward Discrimination:}}}
		$a_1 = a_2 = a$, $v_1=v_2=v$, $e_0 = 0$, $w_1 = w \geq w_2 = w/u$, $b_1 = b>b_2=b/\beta$.
	\end{minipage}
}
\medskip

Similarly, we suppose that the first contributor joins the contest earlier than the second one, i.e. $w_1 \geq w_2$. 
It is reasonable to set $b_1 \geq b_2$ in the reward discrimination. 
%For notational simplicity, the coefficients are set to $a_1=a_2=a$, $v_1=v_2=v$, $b_1 = b$ and $w_1 = w$. 
We denote two new variables, $\beta:=\frac{b_1}{b_2}\geq 1$ and $u:=\frac{w_1}{w_2}\geq 1$. 
By differentiating $\pi_i$ over $e_i$, and letting the derivatives be 0, we obtain 
$e_1^*=\beta e_2^*$ at the unique NE. This NE is subsequently solved by
\begin{eqnarray}
e_1^* = \frac{vb\beta^{v}}{(\beta^v + 1)^2} \;\;\; \textrm{and} \;\;\;
e_2^* = \frac{vb\beta^{v-1}}{(\beta^v + 1)^2}.
\end{eqnarray}

\noindent\textbf{Observation 1:} If the requester prefers the early contribution by setting 
a larger maximum achievable reward, the early contributor exerts more effort than the late one at the NE. 

\textbf{Analysis on Requester's Efficiency.}
The total reward paid by the requester is 
\begin{eqnarray}
R = \frac{b(\beta^v + \beta^{-1})}{\beta^v + 1}  \leq b.
\end{eqnarray} 
The above expression yields the efficiency of requester as
\begin{eqnarray}
\mathcal{E}= \frac{w_1e_1^*+w_2e_2^*}{R}
= \frac{vw(\beta^{v+1}+\beta^v/u)}{(1+\beta^{v+1})(1+\beta^v)}.
\end{eqnarray}
We next investigate the impacts of $u$, $v$ and $\beta$ on the efficiency. 
%The main results are summarized as the following Lemma.
\begin{lemma} 
The interplays between the efficiency of requester, $\mathcal{E}_d$, and the parameters $u$, $v$ and $\beta$ are summarized 
as follows:
\begin{itemize}

\item $\mathcal{E}$ is a decreasing function of $u$.

\item There exists a constant $\hat{\beta}_1$ such that $\mathcal{E}$ is an increasing function of $\beta$ if $\beta \leq \hat{\beta}_1$, 
and a decreasing function of $\beta$ if $\beta > \hat{\beta}_1$. As $\beta$ approaches infinity, the requester's efficiency becomes 0.

%$\mathcal{E}_d$ is an increasing function of $\beta$ when $\beta$ is slightly greater than 1, and is strictly decreasing when 
%$\beta$ is large. As $\beta$ approaches infinity, the requester's efficiency becomes 0. 

\item There exists a constant $\hat{\beta}_2$ such that $\mathcal{E}$ is a strictly increasing function of $v$ if $\beta \leq \hat{\beta}_2$. 
Otherwise, as $v$ increases from 0 to 1, $\mathcal{E}$ increases until reaching its maximum at the extreme point $\hat{v}$, 
and decreases afterwards. 
\end{itemize}
\label{lemma:reward_discrimination}
\end{lemma}

%\noindent\textbf{Observation 2:} In the reward discrimination scheme, if the early contributor is offered enough 
%preference (i.e. large $\beta$), the efficiency of requester is low. 

\begin{figure}[!htb]
	\begin{minipage}{0.45\linewidth}
		\centering
		\includegraphics[width=1.8in, height = 1.7in]{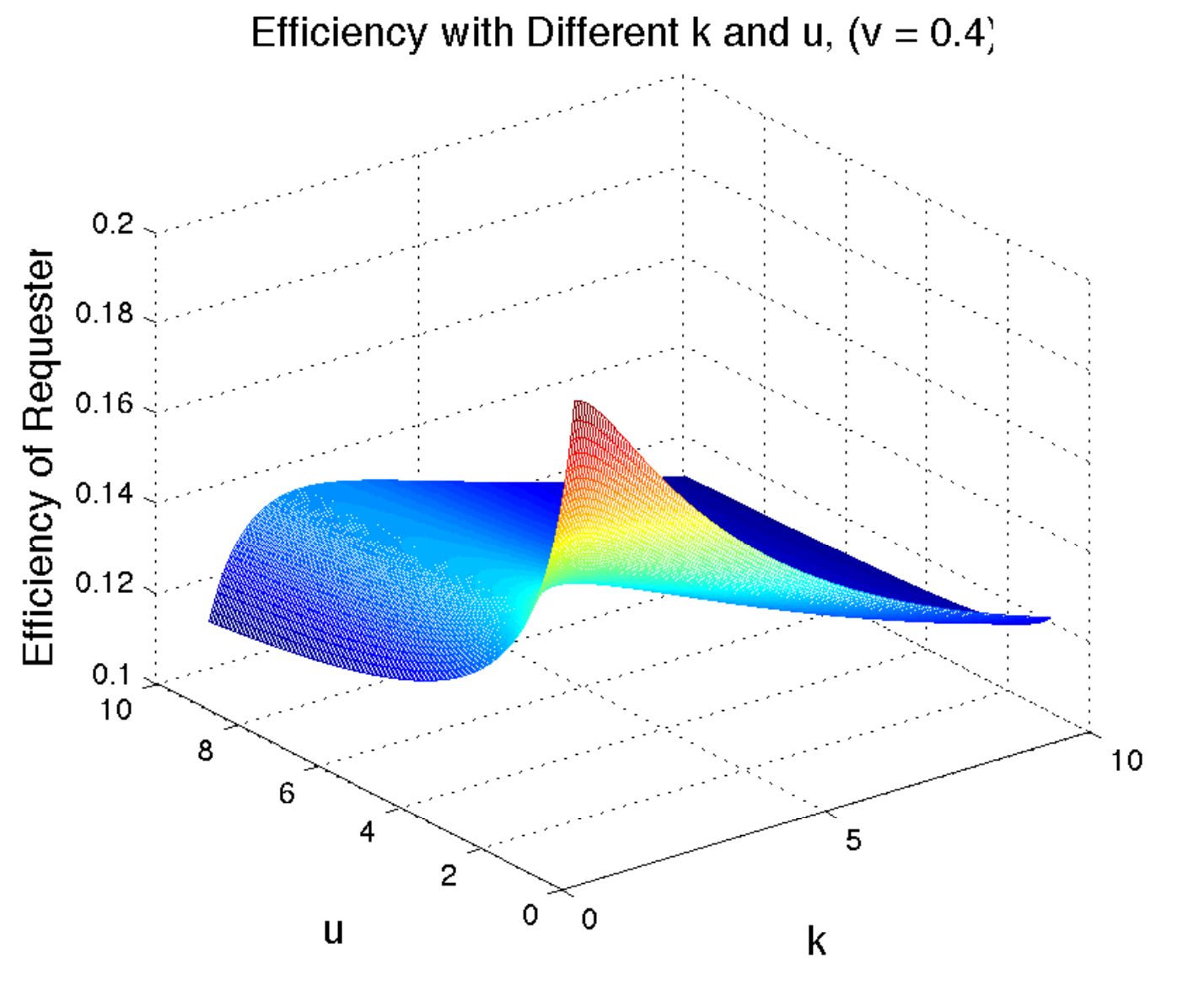}
%		\vspace{-0.8cm}
%		\caption{Discrmination gain ($v = 0.5$}
		\label{fig:eff_no1}
	\end{minipage}
	\hspace{0.1cm}
	\begin{minipage}{0.45\linewidth}
		\centering
		\includegraphics[width=1.8in, height = 1.7in]{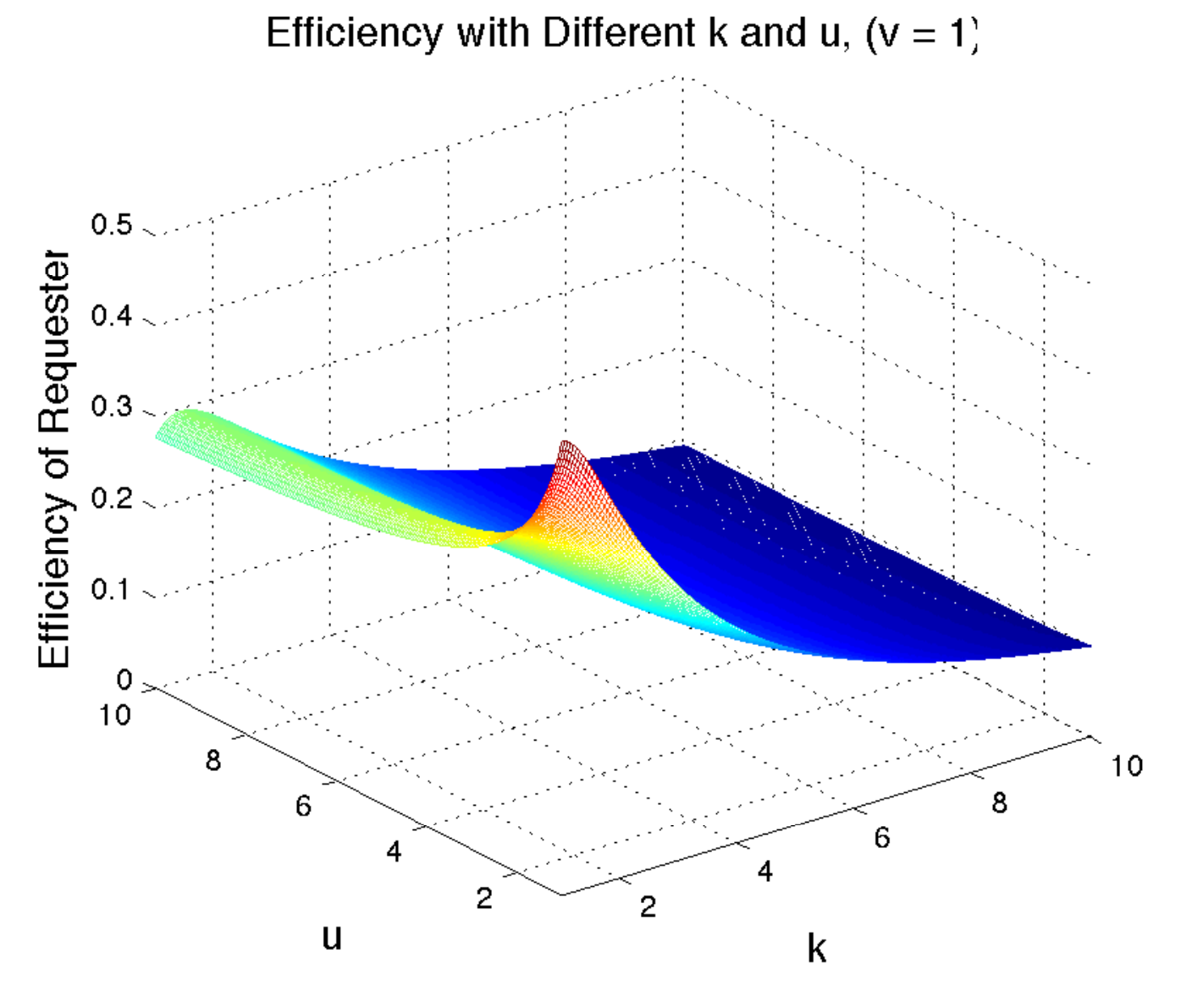}
%		\vspace{-0.8cm}
%		\caption{Discrimination gain with $v {=} \frac{1}{4}$}
		\label{fig:eff_no2}
	\end{minipage}
\vspace{-0.3cm}
\caption{Efficiency of requester with $v = 0.4$ (left) and $v = 1$ (right)}
\label{fig:eff_illustration}
\end{figure}

For the second property, if the early contributor is offered enough preference (i.e. large $\beta$), the efficiency of requester is low.
As an example of the third property, when $\beta$ is less than 3.9026, $\max_v\mathcal{E}$ is obtained at $v=1$. 
This implies that a reasonable tradeoff is to let $v$ be 1 and $\beta$ be relatively small. 
We plot the relationship between $\mathcal{E}$ 
and $u$ as well as $w$ for $v=0.4$ and $v=1.0$ in Fig.\ref{fig:eff_illustration}. 
The efficiency of the requester is intuitively higher in the case $v=1$.   

\noindent\textbf{Observation 2:} To acquire a high efficiency of the requester,  he needs to choose 
a large exponent $v$ and a relatively small reward ratio $\beta$ in the contest function. 

%the requester should NOT 
%configure a much larger maximum achievable reward to the early joining contributor than the late one. 
%Meanwhile, it is recommended to adopt a large exponent (e.g. $v=1$) in the contest function. 

\textbf{Analysis on Discrimination Gain.}
When there is no reward discrimination, $\beta$ equals to 1 so to have
$\mathcal{E} = \frac{vw(u+1)}{4bu}.$ Comparing $\mathcal{E}_{d}$ (with discrimination) with $\mathcal{E}_{nd}$ (with no discrimination), 
we obtain
\begin{eqnarray}
\mathcal{G} = \frac{\mathcal{E}_d}{\mathcal{E}_{nd}} = \frac{4(u\beta^{v+1} + \beta^v)}{(1+\beta^{v+1})(1+\beta^v)(1+u)},
\end{eqnarray}
which clearly shows that the discrimination gain is jointly determined by $u$, $v$ and $\beta$. 
Recalling that contributor 1 joins the contest earlier than contributor 2, any feasible reward 
discrimination scheme must have $\beta\geq 1$ and $u\geq 1$. 
The impacts of $v$, $u$ and $\beta$ on $\mathcal{G}$ is provided below. 

\begin{lemma}
	The discrimination gain $\mathcal{G}$ satisfies the following properties.
	\begin{itemize}
		\item $\mathcal{G}$ is a decreasing function of $v$ for all $v\in(0, 1]$.
		\item $\mathcal{G}$ is an increasing function of $u$. 
		\item There exists an optimal $\beta = \hat{\beta}_3$ to maximize $\mathcal{G}$.
	\end{itemize}
	\label{lemma:discrimination_gain}
\end{lemma}

\begin{figure}[!htb]
	\begin{minipage}{0.45\linewidth}
		\centering
		\includegraphics[width=1.8in, height = 1.7in]{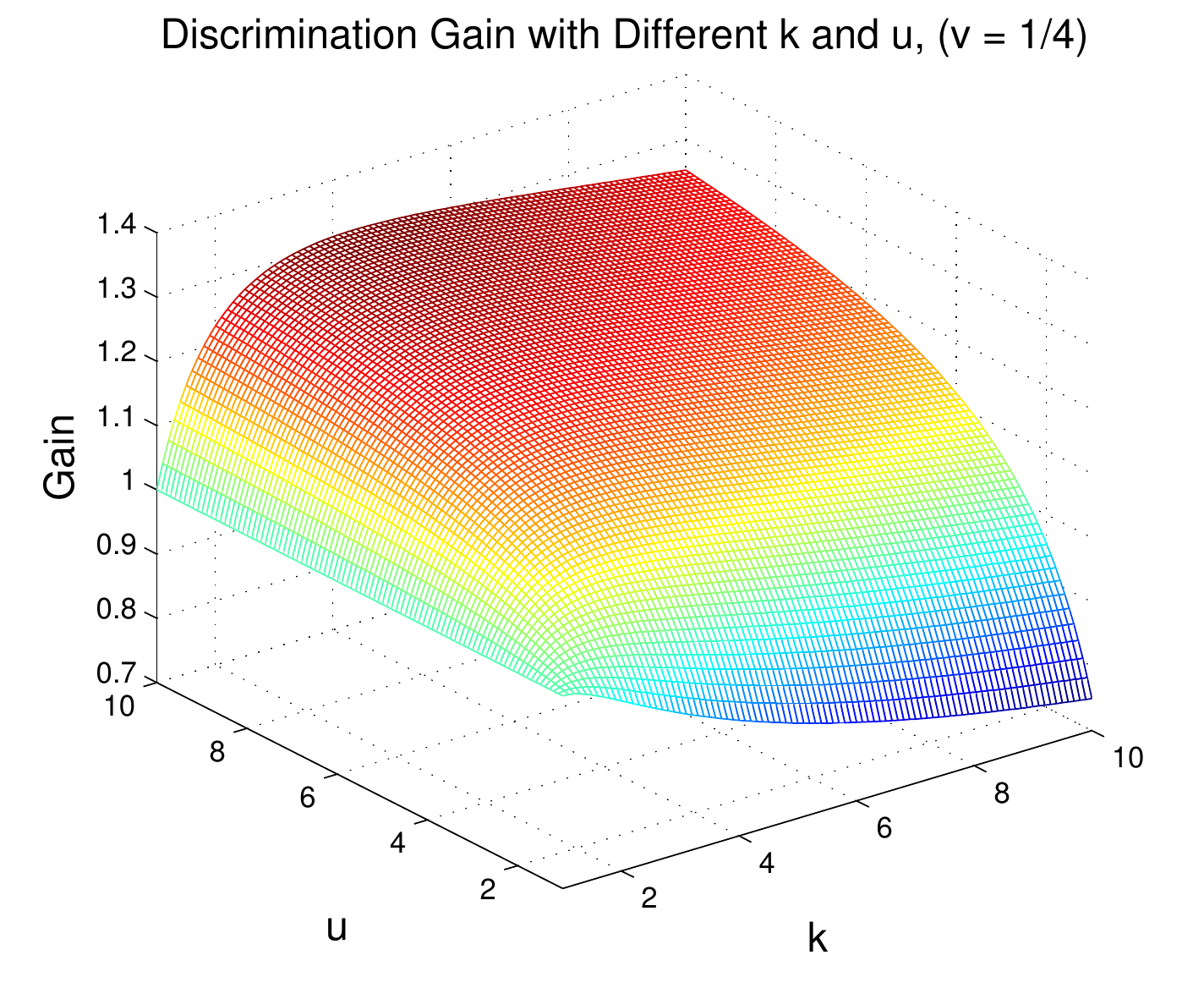}
	\end{minipage}
	\hspace{0.1cm}
	\begin{minipage}{0.45\linewidth}
		\centering
		\includegraphics[width=1.8in, height = 1.7in]{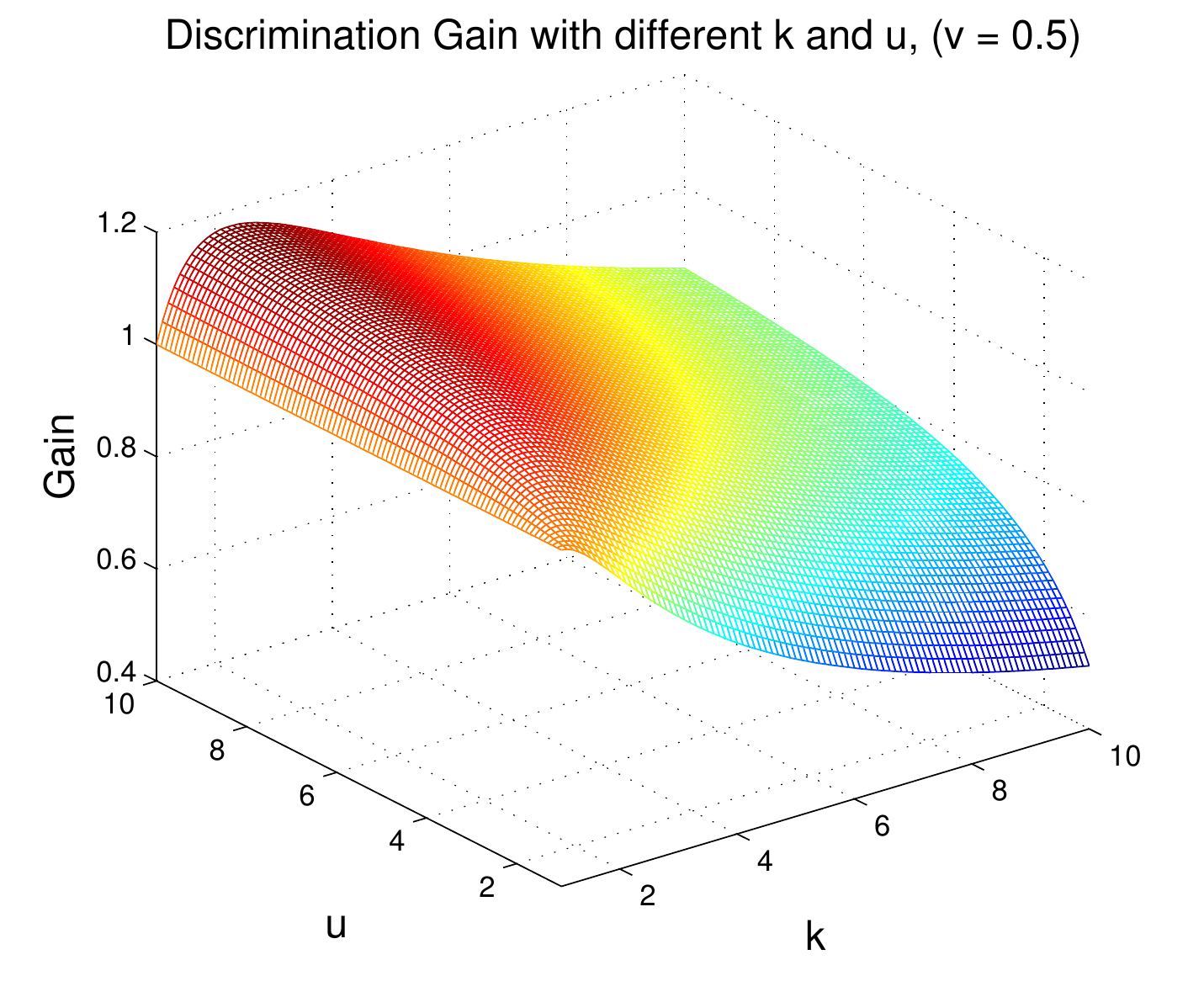}
	\end{minipage}
\vspace{-0.3cm}
\caption{Discrimination gain with $v = \frac{1}{4}$ (left) and $v = \frac{1}{2}$ (right)}
\label{fig:discrim_gain}
\end{figure}

The relationship between $\mathcal{G}$ and $v$ as well as $u$ is now clear. However, it is not intuitive 
to show the impact of $\beta$ on $\mathcal{G}$. Here, a simple example is presented. Suppose that $v$ 
is 1, the maximum $\mathcal{G}$ is obtained at $\beta=1.5214$. In Fig.\ref{fig:discrim_gain}, 
we illustrate the relationship between $\mathcal{G}$ and $u$ as well as $\beta$ in 
two different cases, $v=\frac{1}{4}$ and $\frac{1}{2}$.

\noindent\textbf{Observation 3:} To achieve a high discrimination gain, the requester needs to 
configure a small exponent in the contest function. 

\noindent\textbf{Paradox 3:} From observations 2 and 3, we find that the reward discrimination scheme is hardly to 
achieve both high efficiency and high discrimination gain. 

So far, we have identified three paradoxes in the weight, exponent and reward discrimination schemes respectively. 
Though our study is limited to the case of two competing contributors, these paradoxes usually 
exist in the crowdsourcing with more than two contributors. Our next purpose is to find an appropriate 
contest architecture that incentives more effort from the early contributor. 

\subsection*{E. Nature-as-a-player in Reward Discrimination Scheme}

\medskip
\boxed{
	\begin{minipage}{3.1in}
		{\ensuremath{\mbox{\sc  Nature Player:}}}
		$a_1 = a_2 = a$, $v_1=v_2=v$, $b_1=b_2=b$, $w_1 = w \geq w_2 = w/u$,  $e_0 \geq 0$.
	\end{minipage}
}
\medskip

In the previous analyses, the variable $e_0$ is 0 by default. Hence, the reward is completely allocated 
to the contributors, no matter what their effort is. To avoid adversary scenario that all the contributors exert 
little effort, the requester can introduce \emph{nature} as a player so to retain a fraction of his reward. 
This nature player ``exerts'' a fixed amount of nonzero \emph{virtual} effort $e_0$, while not harvesting any real reward. 
%For simplicity of notations, we let 
%$a_1=a_2=1$, $v_1 = v_2 = v$, $w_1=w$, $b_1=b$, $u=\frac{w_1}{w_2}$ and $k = \frac{b_1}{b_2}$. 

After the nature player is introduced, it is plausible that the contributors might not participate (i.e. 
exerting 0 effort). To solve the NE, we take the derivative of $\pi_i$ over $e_i$. The optimality conditions result in 
the following (in)equalities at the NE,
\begin{eqnarray}
\frac{vb_i(e_i^*)^{v-1}(e_0 {+} \sum\nolimits_{j=1,\neq i}^{N}(e_j^*)^v)}{(e_0 + \sum\nolimits_{j=1}^{N} (e_j^*)^v)^2} \;\;
\left\{
\begin{aligned}
 = 1 && \textrm{ if } e_i^* \geq 0,\\
< 1&& \textrm{ if } e_i^* = 0
\end{aligned}
\right..
\label{eq:opt_conditions}
\end{eqnarray}
From the above (in)equalities, both contributors will exert positive amount of effort at the NE when $v<1$. 
When $v=1$ and $e_0 \geq b_i$, contributors from $1$ to $i$ always obtain negative payoffs if they 
allocate any amount of effort. Their optimal strategy is not to participate in the crowdsensing, which is 
a \emph{sufficient} condition. 
Hence, the requester can use $e_0$ to control how many rewards are retained and which contributors are 
awarded. In general, Eqs.\eqref{eq:opt_conditions} do not admit a close-form solution to the NE so that we 
need to search for $e_i^*$ iteratively. In the beginning, a vector of values are endowed to $\mathbf{e}$. 
Given $\mathbf{e}_{-i}$, we compute $e_i$ using Eq.\eqref{eq:opt_conditions} for all $i$. Then, we repeat 
the above step until eventually $\mathbf{e}$ converge to a very small vicinity of the equilibrium.  

The analysis of reward discrimination scheme demonstrates that an appropriate configuration of 
reward ratio is able to bring certain discrimination gain. However, the reward discrimination impairs the 
efficiency of the requester under many circumstances. Here, our purpose is to investigate whether the reduction of efficiency can be 
mitigated by introducing the nature player. For analytical tractability, we let $b_1 = b_2 = b$. Then, there exists a
unique symmetric NE that solves the following equation with $e_1^*=e_2^*=e^*$,
\begin{eqnarray}
%(e_0 + 2e_1^*)^2 = vb(e_1^*)^{v-1}(e_0 + e_1^*). 
(e_0 + 2(e^*)^v)^2 = vb(e^*)^{v-1}(e_0 + (e^*)^v). 
\label{eq:nature_symmetric_ne_no1}
\end{eqnarray} 
After certain transformations, the solution $e^*$ is given by 
\begin{eqnarray}
(e^*)^v = \frac{1}{2}e_0\big( \frac{bv}{bv-4e^*} - 1  \big).
\label{eq:nature_symmetric_ne_no2}
\end{eqnarray}
The above equation yields the relationship between $e^*$ and $e_0$. 
\begin{lemma}
When nature is a player in the contest along with homogeneous ordinary players (i.e. $a_i=a$, $b_i=b$ and $v_i=v$ for all $i$), 
the NE strategy of ordinary players, $e^*$, is a decreasing function of $e_0$. 
\label{lemma:e_star_vs_e0}
\end{lemma}

The total reward paid to the contributors is given by
\begin{eqnarray}
R = 2b(e^*)^v \big(e_0 + 2(e^*)^v \big)^{-1}. \nonumber
\end{eqnarray}
The efficiency of the requester is computed as
\begin{eqnarray}
\mathcal{E} &\!\!\!\!\!=\!\!\!\!\!& \frac{w(1{+}u)}{2ub} ((e^*)^{1{-}v}(e_0 + 2(e^*)^v)) \nonumber\\
&\!\!\!\!\!=\!\!\!\!\!&\frac{vw(1{+}u)}{2ub} \frac{e_0 + (e^*)^v}{e_0 {+} 2(e^*)^v} = \frac{vw(1{+}u)}{4u}\big( 1{+}  \frac{e_0}{e_0{+}2(e^*)^v}  \big)
%\!\!&=&\!\! \frac{vw(1{+}u)}{4u}\big( 1+  \frac{e_0}{e_0{+}2(e^*)^v}  \big).
\label{eq:efficiency_withe0}
\end{eqnarray} 
according to Eqs.\eqref{eq:nature_symmetric_ne_no1} and \eqref{eq:nature_symmetric_ne_no2}. 
The relationship between the efficiency and nature's effort is the following. 

\begin{lemma}
When nature is a player in the contest, the efficiency of 
requester $\mathcal{E}$ is an increasing function of $e_0$. The asymptotic efficiency w.r.t. $e_0$ is 
given by 
\begin{eqnarray}
\left\{
\begin{aligned}
&\forall \;0<v<1& \lim_{e_0\rightarrow \infty} \mathcal{E}\\
&v = 1& \lim_{e_0\rightarrow b^{-}} \mathcal{E} 
\end{aligned}
\right\} = \frac{vw(1{+}u)}{2u}.
\end{eqnarray}
For a fixed $e_0$, the asymptotic efficiency w.r.t. $b$ is given by 
\begin{eqnarray}
\lim_{b\rightarrow \infty} \mathcal{E} = \frac{vw(1+u)}{4u}.
\end{eqnarray}
\label{lemma:asymptotic_efficiency}
\end{lemma} 

\noindent\textbf{Observation 4:} By introducing nature as a player, the efficiency of requester can be improved at most 
two times. To pursue the best efficiency, the requester is reasonable to configure $e_0$ as large as possible. 
However, a large $e_0$ leads to less effort from the contributors at the NE. 

These small effort may incur some challenges 
in practical applications when the observation of the effort contain noise. 
To improve the efficiency and 
to obtain a large utility, the requester needs to configure $e_0$ and $b_i$ $(\forall i)$ jointly so that the total reward paid to 
the contributors equals to the budget $B$.

\subsection*{F. Brief Summary}

We have analyzed the total utility, efficiency and discrimination gain under three schemes:  
weight discrimination, exponent discrimination and 
reward discrimination. Our observations can be summarized as below.

\begin{itemize}

\item The weight and exponent discrimination schemes are not suitable for timeliness sensitive crowdsourcing contests. 

\item The reward discrimination scheme can achieve certain discrimination gain, but risks the potential reduction of the requester's efficiency. 

\item By introducing a \emph{nature} player, the requester's efficiency can be improved, while incurring great difficulty of configuring the 
maximum achievable rewards and the nature player's effort jointly. 

\end{itemize}
The optimal incentive mechanism is very difficult to obtain due to its freedom in choosing the shape of contest success function 
and configuring coupled parameters. 
Based on our observations, the appropriate incentive mechanism should bear the form
\begin{eqnarray}
r_i([e_i, t_i], [\mathbf{e}_{-i}, \mathbf{t}_{-i}]) = \frac{b_ie_i}{e_i + \sum_{j{=}0,\neq i}^{N}e_j}, \;\;\; \forall 1{\leq} i {\leq} N,
\label{eq:simplerincentive}
\end{eqnarray}
which is named as ``standard contest function''.
In contrast to Eq.\eqref{eq:basicincentive}, though the reward allocation function in Eq.\eqref{eq:simplerincentive} is much simpler, 
there are still $N+1$ parameters to be configured. Especially, $e_0$ and $b_i$ are coupled, which adds great difficulty to the 
incentive mechanism design.

%\newpage

\section*{Appendix-III: Proofs in Justifying Contest Success Function}

\subsection*{A. Proof of Lemma \ref{lemma:exponent}}
\noindent\textbf{Proof:} Recap that the prerequisite is $0<v_2\leq v_1 \leq 1$. 
We take the derivative of the right-hand of Eq.\eqref{eq:2palyers_ed} over $v_2$ and obtain
\begin{eqnarray}
\frac{d(v_2\big(\frac{v_1}{v_2}\big)^{v_1})}{d v_2} = (1-v_1)\cdot \big(\frac{v_1}{v_2}\big)^{v_1} \geq 0. 
\end{eqnarray}
This implies that the increase of $v_2$ causes the increase of the right-hand in Eq.\eqref{eq:2palyers_ed}. 

We split the left-hand of Eq.\eqref{eq:2palyers_ed} into two parts as follows.
\begin{eqnarray}
Part_1 \!\!\!&=&\!\!\! \Big(\big(\frac{v_1}{v_2}\big)^{2v_1}-1\Big)(e_2^*)^{(v_1-v_2+1)}  + 2\big(\frac{v_1}{v_2}\big)^{v_1}e_2^*,\nonumber\\
Part_2 \!\!\!&=&\!\!\! (e_2^*)^{(v_1-v_2+1)} + (e_2^*)^{(v_2-v_1+1)} \nonumber.
\end{eqnarray}
Given a fixed $e_2^*$, as $v_2$ increases, $Part_1$ becomes smaller for sure. For a fixed $e_2^*$, the first expression in 
$Part_2$ is a decreasing function of $v_2$, while the second one is an increasing function of $v_2$. 
We take the derivative of $Part_2$ over $v_2$,
\begin{eqnarray}
\frac{d Part_2}{dv_2} = \big((e_2^*)^{(v_2-v_1+1)} - (e_2^*)^{(v_1-v_2+1)}\big) \log(e_2^*) > 0
\end{eqnarray}
no matter whether $e_2^*$ is greater than 1 or not. 
When $v_2$ increases, the right-hand of Eq.\eqref{eq:2palyers_ed} increases, 
while the left-hand decreases if $e_2^*$ does not change. In order to make the equality hold, 
$e_2^*$ must increase. Therefore, increasing $v_j$ will improve the effort exerted by the both contributors. \done

\subsection*{B. Proof of Lemma \ref{lemma:reward_discrimination}}
\noindent \textbf{Proof:} i) It is easy to see that $\mathcal{E}$ is a decreasing function of $u$.

ii) We take the first-order derivative of $\mathcal{E}$ over $\beta$, 
\begin{eqnarray}
\frac{d\mathcal{E}}{d\beta} = \frac{-v\beta^{v-1}\big(-u \beta^{1+v}+uv\beta^{2+2v}-uv\beta -u\beta}{u(1+\beta^v)^2(1+\beta^{1+v})^2} \nonumber\\
+\frac{\beta^{1+v} + v\beta^{1+2v}+\beta^{1+2v} - v\big)}{u(1+\beta^v)^2(1+\beta^{1+v})^2}
\end{eqnarray}
where $Denominator$ is a positive expression.
We further subtract the numerator denoted by $\mathcal{F}_{\beta}$ from the above equation, 
$\mathcal{F}_{\beta} = u\beta^{1+v} - uv\beta^{2+2v} + uv\beta + u\beta - \beta^{1+v} - v\beta^{1+2v}-\beta^{1+2v} + v$. 
When $\beta$ is slightly above the minimum value (i.e. 1), $\mathcal{F}_{\beta}$ is 
approximated by $2u-2$ which is greater than 0. Then, $\mathcal{E}$ is an increasing function of 
$\beta$ in this situation. When $\beta$ is sufficiently large, $\mathcal{F}_{\beta}$ is negative so that 
$\mathcal{E}$ is a decreasing function of $\beta$. Hence, due to the continuity of $\mathcal{E}$, 
there may exist one or more local extreme points. In what follows, we will show that there exists 
a unique extreme point to let $\mathcal{F}_{\beta} = 0$. 

Let $\mathcal{G}_{\beta}$ be the derivative of $\mathcal{F}_{\beta}$ over $\beta$. There has 
\begin{eqnarray}
\mathcal{G}_{\beta} = (1+v)(-2uv\beta^{1+2v}- (1+2v)\beta^{2v}-\beta^v + u\beta^v + u)
\end{eqnarray}
where it is 0 at the extreme points. Suppose $\mathcal{F}_{\hat{\beta}} = 0$ at the point 
$\beta = \hat{\beta}_1$. Subtracting $\mathcal{G}_{\beta}$ from $\mathcal{F}_{\beta}$ at the extreme 
point $\hat{\beta}_1$, we obtain 
\begin{eqnarray}
\frac{\mathcal{G}_{\hat{\beta}_1}}{1+v} - \mathcal{F}_{\hat{\beta}_1} = v(1+u\hat{\beta}_1)(1-(\hat{\beta}_1)^{1+2v}) \leq 0
\end{eqnarray}
due to $\hat{\beta}_1 \geq 1$, and the equality holds at $\hat{\beta}_1 = 1$. 
Thus, the above inequality yields $\mathcal{G}_{\hat{\beta}_1} < 0$ for any $\hat{\beta}_1 > 1$. 
In the small vicinity of $\hat{\beta}_1$, there must have $\mathcal{F}_{\beta} > 0$ 
if $\beta<\hat{\beta}_1$ and $\mathcal{F}_{\beta} < 0$ if $\beta> \hat{\beta}_1$. 
We next prove the uniqueness of $\hat{\beta}_1$. Suppose that there are two extreme points. 
Then, one must be the local maximum and the other must be the local minimum. 
Denote by $\hat{\beta}_1^{(a)}$ the local maximum and by $\hat{\beta}_1^{(b)}$ the local minimum. 
In the small vicinity of $\hat{\beta}_1^{(b)}$, $\mathcal{F}_{\beta} > 0$ 
if $\beta<\hat{\beta}_1^{(b)}$ and $\mathcal{F}_{\beta} < 0$ if $\beta> \hat{\beta}_1^{(b)}$. 
This contradicts to the assumption that there are two extreme points. 
The uniqueness of $\hat{\beta}_1$ means that there is only one extreme point in $\mathcal{E}$.

iii) We take the derivative of $\mathcal{E}$ over $v$,
\begin{eqnarray}
\frac{d\mathcal{E}}{dv} &\!\!\!=\!\!\!& (\beta^{1+v} + \beta^{1+2v} - v\beta^{1+2v}\log(\beta) +\beta^v + v\log(\beta)+1) \nonumber\\
&&\cdot \frac{(u\beta+1)\beta^v}{u(\beta^v+1)^2(\beta^{1+v}+1)^2}.
\end{eqnarray}
We denote by $\mathcal{F}_v = \beta^{1+v} + \beta^{1+2v} - v\beta^{1+2v}\log(\beta) +\beta^v + v\log(\beta)+1$. 
When $v$ is small enough (i.e. $v$ is approaching 0), $\mathcal{F}_v$ is approximated $2(1+\beta)$ which is greater than 0. 
Thus, $\mathcal{E}$ is an increasing function of $v$ when it is small. 
When $v$ approaches 1 and $\beta$ is sufficiently large, $\mathcal{F}_v$ is 
negative so that $\mathcal{E}$ is a decreasing function of $v$. Due to the continuity of $\mathcal{F}_v$, 
it is plausible that $\mathcal{F}_v$ crosses 0 as $v$ increases from 0 to 1. 
Suppose that there exists a constant $\hat{v}$ so to have $\mathcal{F}_{\hat{v}} = 0$. 
Denote a new expression $\mathcal{G}_v$ that is the derivative of $\mathcal{F}_v$ over $v$,
\begin{eqnarray}
\mathcal{G}_v = (\beta^{1+v} + \beta^{1+2v} - 2v\beta^{1+2v}\log(\beta) +\beta^v +1)\log(\beta). 
\end{eqnarray}
Because of $\mathcal{F}_{\hat{v}} = 0$, there must have $\mathcal{G}_{\hat{v}} = -v(\log(\beta))^2(1+\beta^{1+2v}) < 0$. 
Hence, the extreme point $\hat{v}$ can only be the local maximum of the efficiency function $\mathcal{E}$. 
As $v$ increases from 0 to $\hat{v}$, $\mathcal{E}$ increases first, and decreases after $v$ continue to increase beyond $\hat{v}$. 
When $\beta$ is not large enough, $\mathcal{F}_v$ may be always positive so that $\mathcal{E}$ always increases with regard to $v$. 
\done

\subsection*{C. Proof of Lemma \ref{lemma:discrimination_gain}}
	\noindent \textbf{Proof:} i) We first take the derivative of $\mathcal{G}$ over $v$ and obtain 
	\begin{eqnarray}
	\frac{d \mathcal{G}}{dv} = \frac{4(\log \beta)(u\beta^{v+1} + \beta^v)(1-\beta^{2v+1})}{(1+\beta^{v+1})^2(1+\beta^v)^2(1+u)} < 0.
	\end{eqnarray}
	Therefore, the discrimination gain is a decreasing function of the exponent $v$ when other coefficients are fixed. When $v=1$, the minimum $\mathcal{G}$ is obtained by $\frac{4(u\beta^2+\beta)}{(1+\beta^2)(1+\beta)(1+u)}$; when $v\rightarrow 0$, the maximum $\mathcal{G}$
	is approximated by $\frac{2(1+u\beta)}{(1+\beta)(1+u)}$. 
	
	ii) The derivative of $\mathcal{G}$ over $u$ is given by
	\begin{eqnarray}
	\frac{d\mathcal{G}}{d u} = \frac{4(\beta^{v+1}-\beta^v)}{(1+\beta^{v+1})(1+\beta^v)(1+u)^2} \geq 0
	\end{eqnarray}
	and the equality holds at $\beta=1$. Hence, as $u$ grows, the discrimination gain $\mathcal{G}$ increases correspondingly, which 
	strengthens the effect of reward discrimination scheme. The asymptotic gain is expressed as 
	\begin{eqnarray}
	\lim_{u \rightarrow \infty} \mathcal{G} &\!\!\!=\!\!\!& \frac{4\beta^{v+1}}{(1+\beta^{v+1})(1+\beta^v)}, \nonumber\\
	\lim_{u \rightarrow 1} \mathcal{G} &\!\!\!=\!\!\!& \frac{2(\beta^v+\beta^{1+v})}{(1+\beta^v)(1+\beta^{1+v})}. \nonumber
	\label{eq:rdgain_asymptotic}
	\end{eqnarray}
	
	iii) We further take the derivative of above expression over $\beta$ and compute the optimal 
	$\hat{\beta}_3$ to maximize the discrimination gain. 
	Here, $\hat{\beta}_3$ is the \emph{unique} feasible solution to an implicit equation,
	\begin{eqnarray}
	v\beta^{2v+1} - \beta^v - (1+v) = 0. 
	\label{eq:k_uniquesolution}
	\end{eqnarray} 
	Therefore, according to Eqs.\eqref{eq:rdgain_asymptotic} and \eqref{eq:k_uniquesolution}, the asymptotic 
	discrimination gain increases with $\beta$ first until reaching the maximum value, and then decreases 
	as $\beta$ further increases.
	\done
	
\subsection*{D. Proof of Lemma \ref{lemma:e_star_vs_e0}}
\noindent\textbf{Proof:} We prove this lemma by contradiction. 
We take the derivative of $e^*$ over $e_0$ in Eq.\eqref{eq:nature_symmetric_ne_no1}, $\frac{de^*}{de_0} = $
\begin{eqnarray}
\frac{-4(e^*)^v-2e_0+vb(e^*)^{v{-}1}}{8v(e^*)^{2v{-}1}{+}4v(e^*)^{v{-}1}e_0{+}bv(1{-}v)(e^*)^{v{-}2}e_0 {-} bv(2v{-}1)(e^*)^{2v{-}2}}.
\end{eqnarray}
Suppose $\frac{de^*}{de_0}  = 0$ at a point $e^{**}\geq 0$. Therefore, there exists
\begin{eqnarray}
4(e^{**})^v +2e_0 = vb(e^{**})^{v-1}. 
\end{eqnarray}
Submitting the above equation to Eq.\eqref{eq:nature_symmetric_ne_no1}, we obtain 
\begin{eqnarray}
\!\!\!\!\!\!&&(e_0{+}2(e^{**})^v)^2 = \frac{1}{4}v^2b^2(e^{**})^{2v{-}2}, \nonumber\\
\!\!\!\!\!\!&&vb(e^{**})^{v{-}1}(e_0 {+} (e^{**})^v) = \frac{1}{4}v^2b^2(e^{**})^{2v-2} {+} \frac{1}{2}vb(e^{**})^{v{-}1}e_0. \nonumber
\end{eqnarray}
Hence, for any non-negative $e^{**}$ to enable $\frac{de^*}{de_0} = 0$, there can only have $e_0$ = 0.
Recall that we assume $\frac{de^*}{de_0} =0$ at the point $e^{**}\geq 0$ for any $e_0>0$. This contradiction 
manifests that $e^*$ is a monotone function of $e_0$ for $e_0\geq 0$. 

We next consider a special case $e^*=(e_0)^{\frac{1}{v}}$. According to Eq.\eqref{eq:nature_symmetric_ne_no1}, there has 
$e^* = \frac{2vb}{9}$. The nominator of $\frac{de^*}{de_0}$ is
$-\frac{3}{2}vb(e^*)^{v-1}$, and the denominator is greater than $\frac{5}{3}bv^2(e^*)^{2v-2}$. Hence, 
$e^*$ is a strictly decreasing function of $e_0$ for $e^*\geq 0$ and $e_0\geq 0$. \done

\subsection*{E. Proof of Lemma \ref{lemma:asymptotic_efficiency}}
\noindent\textbf{Proof:}   
The first argument is a direct result of Lemma \ref{lemma:e_star_vs_e0}. 
When $e_0$ increases, $e^*$ decreases such that $\mathcal{E}$ becomes larger and larger. 
For $v\in (0,1)$, as $e_0$ grows to infinity, $e^*$ approaches 0. For $v=1$, when $e_0$ is infinitely close to 
$b^{-}$, $e^*$ also approaches 0. Then, the asymptotic efficiency w.r.t. $e_0$ is $\frac{vw(1{+}u)}{2u}$. 
When $e_0$ is fixed and $b$ increases to infinity, $e^*$ also increases to infinity. Then, $\mathcal{E}$
decreases until $\frac{vw(1+u)}{2u}$ according to Eq.\eqref{eq:efficiency_withe0}. \done

\end{document}